\documentclass[%
 aip,
 amsmath,amssymb,
reprint,%
]{revtex4-1}

\usepackage{graphicx}

\usepackage{dcolumn}
\usepackage{bm}

\usepackage[utf8]{inputenc}
\usepackage[T1]{fontenc}
\usepackage{mathptmx}
\usepackage{etoolbox}

\usepackage{subcaption}
\usepackage{multirow}
\usepackage{booktabs}

\makeatletter
\def\@email#1#2{%
 \endgroup
 \patchcmd{\titleblock@produce}
  {\frontmatter@RRAPformat}
  {\frontmatter@RRAPformat{\produce@RRAP{*#1\href{mailto:#2}{#2}}}\frontmatter@RRAPformat}
  {}{}
}%
\makeatother
\begin{document}

\preprint{AIP/123-QED}

\title[Validation of a Comprehensive First-Principles-Based Framework for Predicting the Performance of Future Stellarators]{Validation of a Comprehensive First-Principles-Based Framework for Predicting the Performance of Future Stellarators}

\author{D.L.C. Agapito Fernando}
\email{don.fernando@ipp.mpg.de}
\affiliation{Max-Planck-Institut f\"ur Plasmaphysik, Boltzmannstra$\beta$e 2, 85748 Garching bei M\"unchen, Germany}

\author{A. Ba{\~n}{\'o}n Navarro}
\affiliation{Max-Planck-Institut f\"ur Plasmaphysik, Boltzmannstra$\beta$e 2, 85748 Garching bei M\"unchen, Germany}

\author{D. Carralero}
\affiliation{Laboratorio Nacional de Fusion, CIEMAT, 28040 Madrid, Spain}

\author{A. Alonso}
\affiliation{Laboratorio Nacional de Fusion, CIEMAT, 28040 Madrid, Spain}

\author{A. Di Siena}
\affiliation{Max-Planck-Institut f\"ur Plasmaphysik, Boltzmannstra$\beta$e 2, 85748 Garching bei M\"unchen, Germany}

\author{J.L. Velasco}
\affiliation{Laboratorio Nacional de Fusion, CIEMAT, 28040 Madrid, Spain}

\author{F. Wilms}
\affiliation{Max-Planck-Institut f\"ur Plasmaphysik, Boltzmannstra$\beta$e 2, 85748 Garching bei M\"unchen, Germany}

\author{G. Merlo}
\affiliation{Max-Planck-Institut f\"ur Plasmaphysik, Boltzmannstra$\beta$e 2, 85748 Garching bei M\"unchen, Germany}

\author{F. Jenko}
\affiliation{Max-Planck-Institut f\"ur Plasmaphysik, Boltzmannstra$\beta$e 2, 85748 Garching bei M\"unchen, Germany}
\affiliation{Institute for Fusion Studies, The University of Texas at Austin, Austin, Texas 78712, USA}

\author{S.A. Bozhenkov}
\affiliation{Max-Planck-Institut f\"ur Plasmaphysik, Wendelsteinstra$\beta$e 1, 17491 Greifswald, Germany}

\author{E. Pasch}
\affiliation{Max-Planck-Institut f\"ur Plasmaphysik, Wendelsteinstra$\beta$e 1, 17491 Greifswald, Germany}

\author{G. Fuchert}
\affiliation{Max-Planck-Institut f\"ur Plasmaphysik, Wendelsteinstra$\beta$e 1, 17491 Greifswald, Germany}

\author{K. J. Brunner}
\affiliation{Max-Planck-Institut f\"ur Plasmaphysik, Wendelsteinstra$\beta$e 1, 17491 Greifswald, Germany}

\author{J. Knauer}
\affiliation{Max-Planck-Institut f\"ur Plasmaphysik, Wendelsteinstra$\beta$e 1, 17491 Greifswald, Germany}

\author{A. Langenberg}
\affiliation{Max-Planck-Institut f\"ur Plasmaphysik, Wendelsteinstra$\beta$e 1, 17491 Greifswald, Germany}

\author{N.A. Pablant}
\affiliation{Princeton Plasma Physics Laboratory, 100 Stellarator Road, Princeton, New Jersey 08543, USA}

\author{T. Gonda}
\affiliation{Auburn University Physics Department, 380 Duncan Dr. Auburn, Alabama 36832, USA}

\author{O. Ford}
\affiliation{Max-Planck-Institut f\"ur Plasmaphysik, Wendelsteinstra$\beta$e 1, 17491 Greifswald, Germany}

\author{L. Van{\'o}}
\affiliation{Max-Planck-Institut f\"ur Plasmaphysik, Wendelsteinstra$\beta$e 1, 17491 Greifswald, Germany}

\author{T. Windisch}
\affiliation{Max-Planck-Institut f\"ur Plasmaphysik, Wendelsteinstra$\beta$e 1, 17491 Greifswald, Germany}

\author{T. Estrada}
\affiliation{Laboratorio Nacional de Fusion, CIEMAT, 28040 Madrid, Spain}

\author{E. Maragkoudakis}
\affiliation{Max-Planck-Institut f\"ur Plasmaphysik, Wendelsteinstra$\beta$e 1, 17491 Greifswald, Germany}

\author{the Wendelstein 7-X Team}
\affiliation{For the complete member list, please refer to O. Grulke et al., Nucl. Fusion 64, 112002 (2024)}

\date{\today}

\begin{abstract}
This paper presents the validation of the \texttt{GENE-KNOSOS-Tango} framework for recovering both the steady-state plasma profiles in the considered radial domain and selected turbulence trends in a stellarator. This framework couples the gyrokinetic turbulence code \texttt{GENE}, the neoclassical transport code \texttt{KNOSOS}, and the transport solver \texttt{Tango} in a multi-timescale simulation feedback loop. Ion-scale kinetic-electron and electron-scale adiabatic-ion flux-tube simulations were performed to evolve the density and temperature profiles for four OP1.2b W7-X scenarios. The simulated density and temperature profiles showed good agreement with the experimental data using a reasonable set of boundary conditions. Equally important was the reproduction of observed trends for several turbulence properties, such as density fluctuations and turbulent heat diffusivities. Key effects were also touched upon, such as electron-scale turbulence and the neoclassical radial electric field shear. The validation of the \texttt{GENE-KNOSOS-Tango} framework enables credible predictions of physical phenomena in stellarators and reactor performance based on a given set of edge parameters.
\end{abstract}

\maketitle

\section{\label{sec:introduction}Introduction}

Within the field of nuclear fusion by magnetic confinement, stellarators are a promising alternative to tokamaks, offering several notable advantages. To name a few, stellarators are inherently capable of operating at steady-state, have higher plasma density limits, exhibit lower peak heat flux exhaust, and avoid major disruptions. \cite{Xu2024} With improved optimization for MHD stability and neoclassical transport, turbulence remains a significant obstacle for stellarators today. For instance, turbulence-driven transport surpasses the neoclassical contribution in Wendelstein 7-X (W7-X), the most advanced stellarator to date.\cite{Klinger2019} Thus, understanding plasma turbulence is crucial for designing stellarators and advancing fusion reactor development. \\

With the world's most powerful supercomputers, turbulence can now be analyzed across the entire plasma volume using gyrokinetic codes. Although gyrokinetic codes are capable of simulating the effect of turbulence on plasma temperature and density profiles up to energy confinement times, this task demands an extremely large amount of computational resources. The challenge stems from the large timescale separation between turbulence and transport phenomena. Despite this, profile prediction remains a key research goal in fusion. Developing simulation tools capable of doing so for a given magnetic configuration using only the power and particle sources is essential for designing operational scenarios with optimized confinement and turbulence properties and forecasting reactor performance. \\

An effective strategy for bridging the timescale gap is to couple a gyrokinetic code and a transport code in a feedback loop. Instead of being hindered by the timescale gap, this coupling exploits it instead, leading to a significant reduction in the computational cost of simulating steady-state plasma profiles. The turbulence code computes the turbulent fluxes for input plasma profiles using gradient-driven simulations, while the transport code evolves the profiles according to the difference between the fluxes and sources. This approach has been successfully applied and validated for tokamaks, using both flux-tube \cite{Candy2009, Barnes2010, Honda2018, Honda2019} and global \cite{DiSiena2022, DiSiena2024} gyrokinetic simulations. \\

To extend this capability for stellarators, we further developed the existing \texttt{GENE-Tango} suite \cite{DiSiena2022, DiSiena2024} to create the \texttt{GENE-KNOSOS-Tango} framework. This framework couples the gyrokinetic turbulence code \texttt{GENE} \cite{Jenko2000}, the neoclassical transport code \texttt{KNOSOS} \cite{Velasco2020, Velasco2021}, and the one-dimensional transport solver \texttt{Tango} \cite{Shestakov2003, Parker2018, DiSiena2022} in a multiple-timescale simulation feedback loop. It self-consistently incorporates turbulence, neoclassical transport, and the neoclassical radial electric field shear in each iteration. The framework can be extended to include three-dimensional effects by using \texttt{GENE-3D}. \cite{Maurer2020, Wilms2021} \\

For stellarators, the \texttt{GENE-KNOSOS-Tango} framework has previously been used to study the confinement degradation, as evidenced by the ion-temperature clamping, in electron-heated W7-X plasmas during the machine's first experimental campaign. \cite{Beurskens2021, Carralero2021, BanonNavarro2023} Additionally, it has been used to compare the core confinement properties of the HSX stellarator and the HSK and QSTK, two quasi-helically symmetric stellarator configurations optimized for ion temperature gradient (ITG) stability. \cite{BanonNavarro2024} \\

Although it has already been successfully applied to stellarator studies, the framework's validation remains necessary to ensure that simulation predictions reproduce experimental results and turbulence features across a wide range of operating conditions. Accurate reproduction of plasma temperature and density profiles, given the boundary values and heating and particle sources, and experimental trends is essential in such validation studies. This is a prerequisite for developing operation scenarios, enabling profile predictions, and designing future reactors. The latter is of utmost importance in the current context, where the increasing number of nuclear fusion start-ups is leveraging simulation codes to design power-generating machines for the near future. To date, the only other coupled code validation study for the core that is being performed for stellarators is for the gyrokinetic code \texttt{GX}\cite{Mandell2018,Mandell2024}, transport code \texttt{Trinity3D}\cite{Barnes2010,Qian2022}, and neoclassical codes \texttt{KNOSOS} and \texttt{SFINCS}\cite{Landreman2014}. \cite{Mandell2023, Zarnstorff2024} \\

In this study, we present the validation of the \texttt{GENE-KNOSOS-Tango} framework using four scenarios from the OP1.2b W7-X experimental campaign. \cite{Carralero2021, Carralero2022} These cases cover a wide range of turbulence properties, effectively adding another constraint to the validation task. In addition to matching the targeted experimental results for each scenario, the framework should also model selected trends among different turbulence regimes. This ensures the robustness and applicability of the framework across diverse experimental conditions. Improvements on previous work include the addition of kinetic-electron gyrokinetic simulations in each iteration, neoclassical effects, and varying density profiles. \\

The paper is structured as follows. An overview of the four experimental cases is presented in Section \ref{sec:w7x_scenarios}. Next, in Section \ref{sec:framework_description}, a more in-depth description on the \texttt{GENE-KNOSOS-Tango} framework is provided. In Section \ref{sec:simulation_setup}, the simulation set-up and methodology are discussed. The simulation results are presented in Section \ref{sec:results}. Finally, a summary of the results, findings, conclusion, and outlook are given in Section \ref{sec:conclusions_outlook}.

\section{\label{sec:w7x_scenarios}W7-X Scenarios}

\begin{figure*}
    \centering
    \begin{subfigure}{0.40\textwidth}
        \centering
        \includegraphics[width=\linewidth]{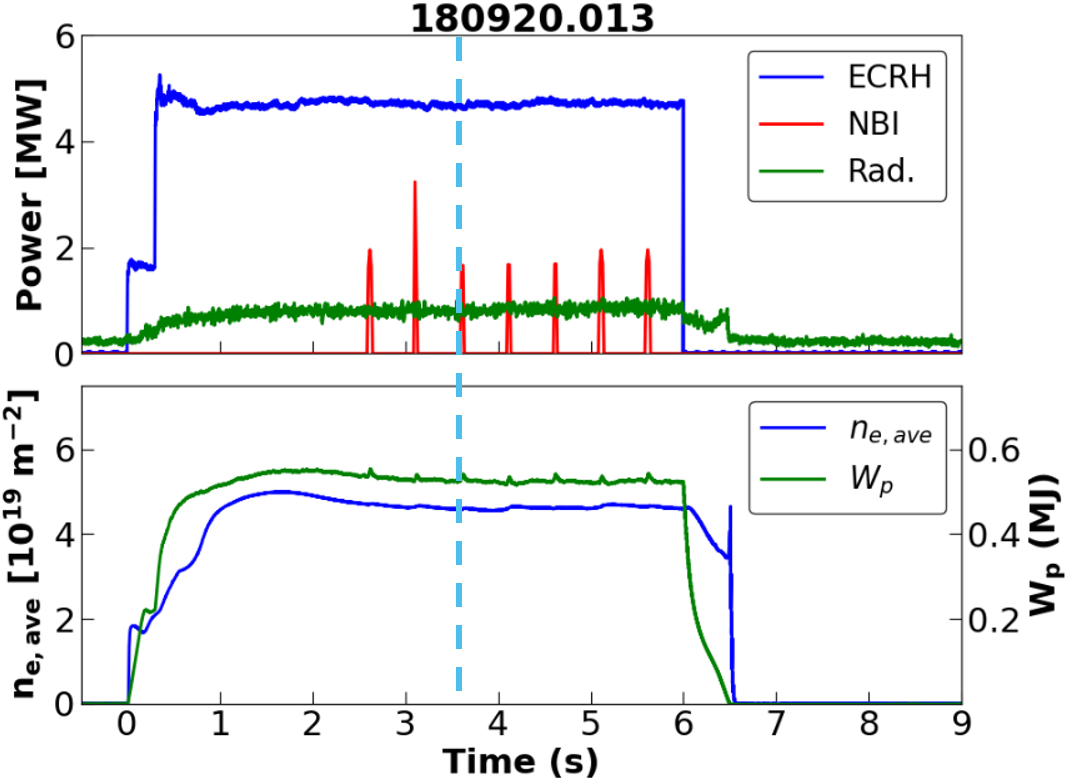}
        \caption{}
        \label{fig:Elow_shot}
    \end{subfigure}
    \hspace{0.7cm}
    \begin{subfigure}{0.40\textwidth}
        \centering
        \includegraphics[width=\linewidth]{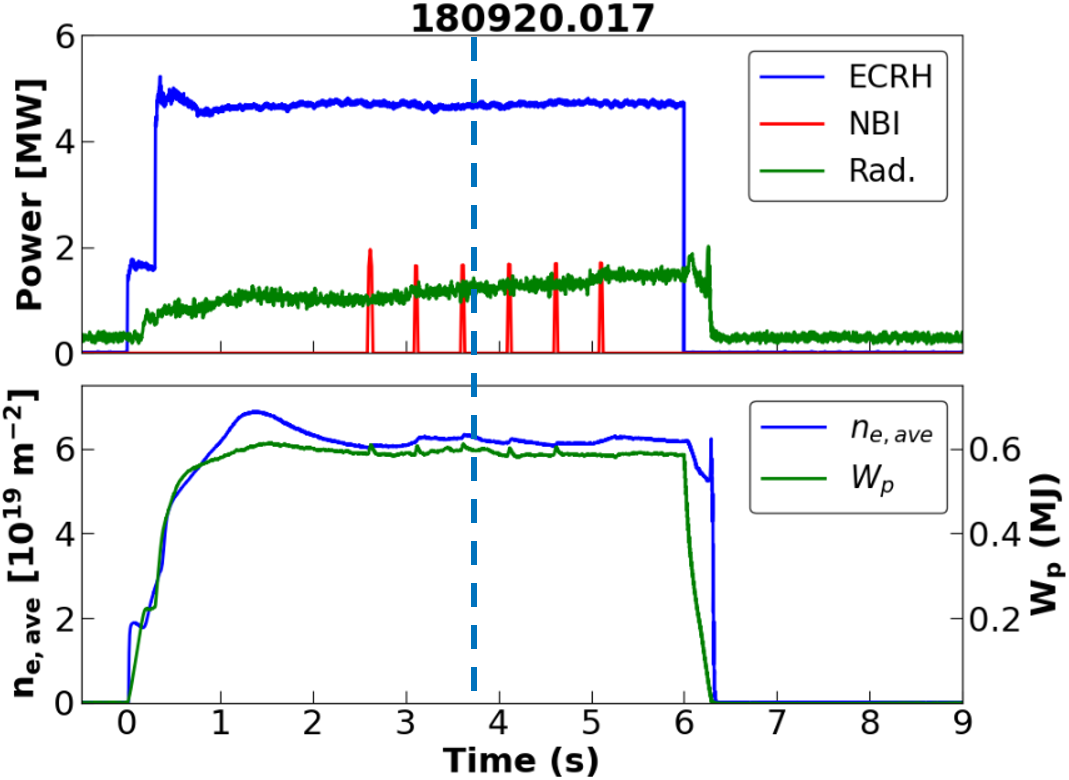}
        \caption{}
        \label{fig:Ehigh_shot}
    \end{subfigure}
    \vspace{0.5cm}
    \begin{subfigure}{0.40\textwidth}
        \centering
        \includegraphics[width=\textwidth]{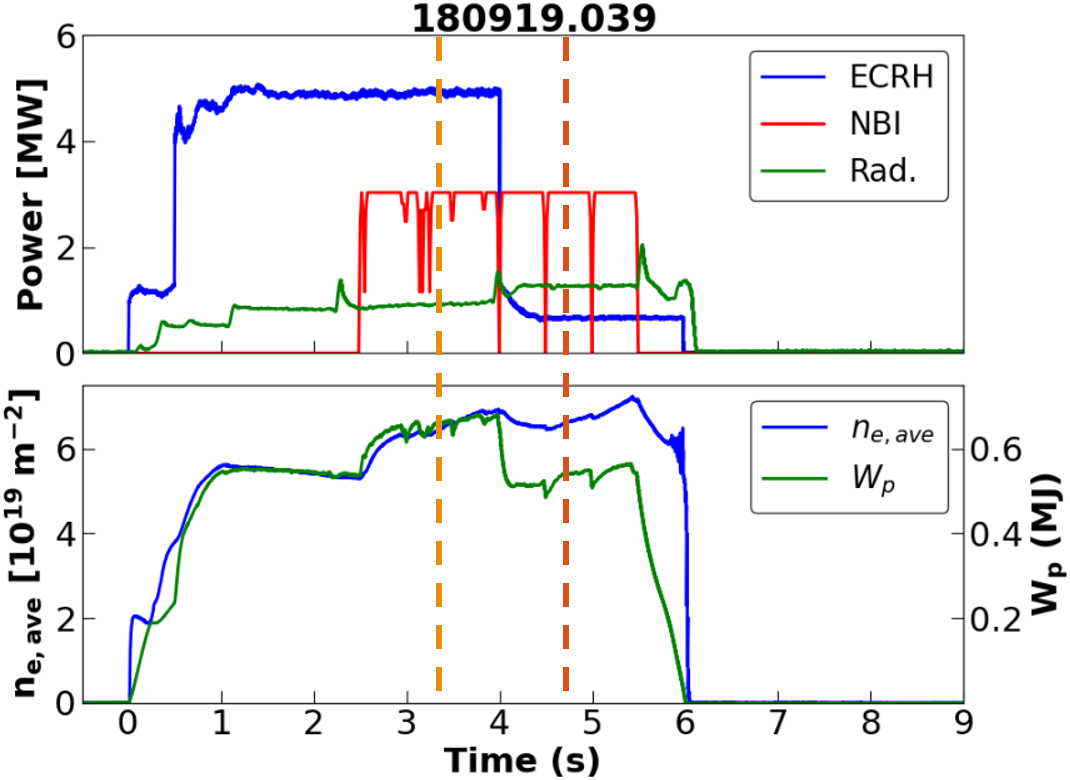}
        \caption{}
        \label{fig:NBI_shot}
    \end{subfigure}
    \caption{The three OP1.2b W7-X discharges used for this validation study: (a) Low-density with ECRH, (b) High-density with ECRH, (c) NBI with different ECRH power. Upper plots show the time traces for the input and radiated power, while the lower plots show the average densities and stored energies. Dashed lines indicate the timestamps when the experimental data was obtained for each scenario. \cite{Carralero2021}}
    \label{fig:3_discharges}
\end{figure*}

Three W7-X discharges from OP1.2b were used for this study, and these give way to four different scenarios. Fig. \ref{fig:3_discharges} shows the time traces for the heating power, average densities, and stored energies of these discharges. The naming convention for the four scenarios used in Refs. \citenum{Carralero2021} and \citenum{Carralero2022} will be retained in this paper for consistency. The first two scenarios are characterized by a core ion temperature $T_i$ below $1.5 \pm 0.2$ keV, which is typical of W7-X electron-heated plasmas\cite{Beurskens2021}, while the last two exceed this threshold value. \\

The first two cases are shots 180920.013 and 180920.017, which were purely heated by electron cyclotron resonance (ECRH) and fueled with gas puffing. Case 1 is the "low-density ECRH" scenario while case 2 is the "high-density ECRH" case, which have average densities of about 4 and 6 $\times 10^{19} \, \mathrm{m}^{-3}$ respectively. The same heating power of about 4.7 MW was applied to both. For these cases, the ion species is hydrogen. \\

From shot 180919.039, two NBI heating cases are derived based on different time stamps during the shot. Case 3, which is the "NBI + ECRH" scenario, was taken at a timestamp of about 3.5 s and had a higher ECRH power than case 4, the "NBI" scenario, which was taken at 4.5 s. It should be emphasized that even though case 4 is referred to as the "NBI" scenario, about 0.7 MW of ECRH was still present. The total heating power for case 3 was approximately 8 MW, while case 4 only received about 4 MW due to the significant reduction of applied ECRH power.

\section{\label{sec:framework_description}Framework Description}

To predict the core temperature and density profiles for these scenarios, the state-of-the-art \texttt{GENE-KNOSOS-Tango} framework is used. The gyrokinetic turbulence code \texttt{GENE} simulates the turbulent transport of each plasma species, while the neoclassical transport code \texttt{KNOSOS} calculates the neoclassical fluxes and the background neoclassical radial electric field $E_r$. Then, the transport solver \texttt{Tango} generates the new plasma profiles by comparing the total fluxes with the injected sources. The code accomplishes this by solving one-dimensional radial transport equations for pressure and density. A relaxation scheme is applied to the fluxes and profiles, wherein values from two successive iterations are linearly combined. The magnitude of the weighting parameter $\alpha$ determines the level of relaxation. This prevents large swings in the fluxes and profiles, thereby keeping the numerical method stable. The new profiles are read by \texttt{GENE} and \texttt{KNOSOS} in the next iteration, and the feedback loop continues as needed. Convergence is manually determined based on the fulfillment of the radial power and particle balances. The full framework showing the coupling and dependencies between the codes is depicted in Fig. \ref{fig:framework_phase4}. \\

\begin{figure*}
    \centering
    \begin{subfigure}{0.31\textwidth}
        \centering
        \includegraphics[width=\linewidth]{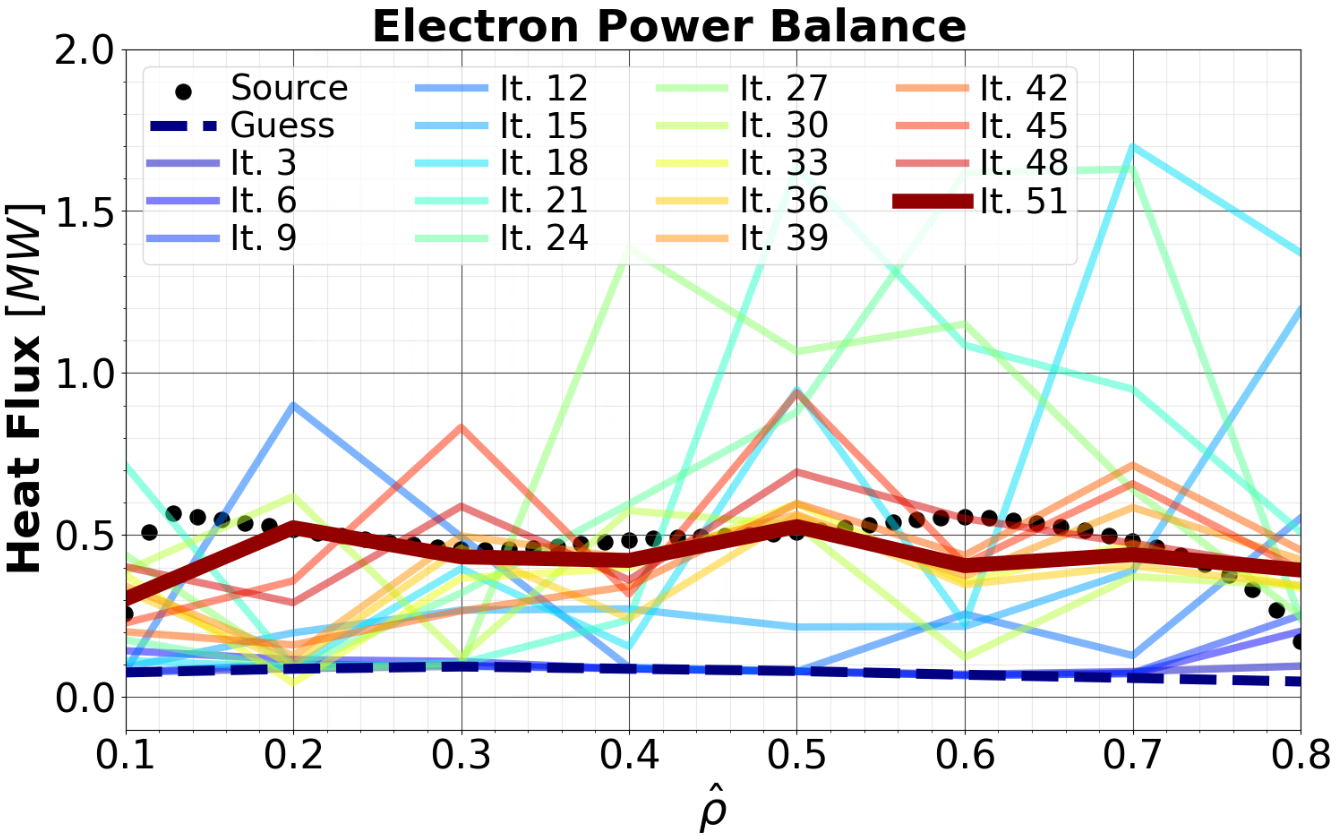}
        \caption{}
        \label{fig:flat_elec_bal}
    \end{subfigure}
    \hfill
    \begin{subfigure}{0.31\textwidth}
        \centering
        \includegraphics[width=\linewidth]{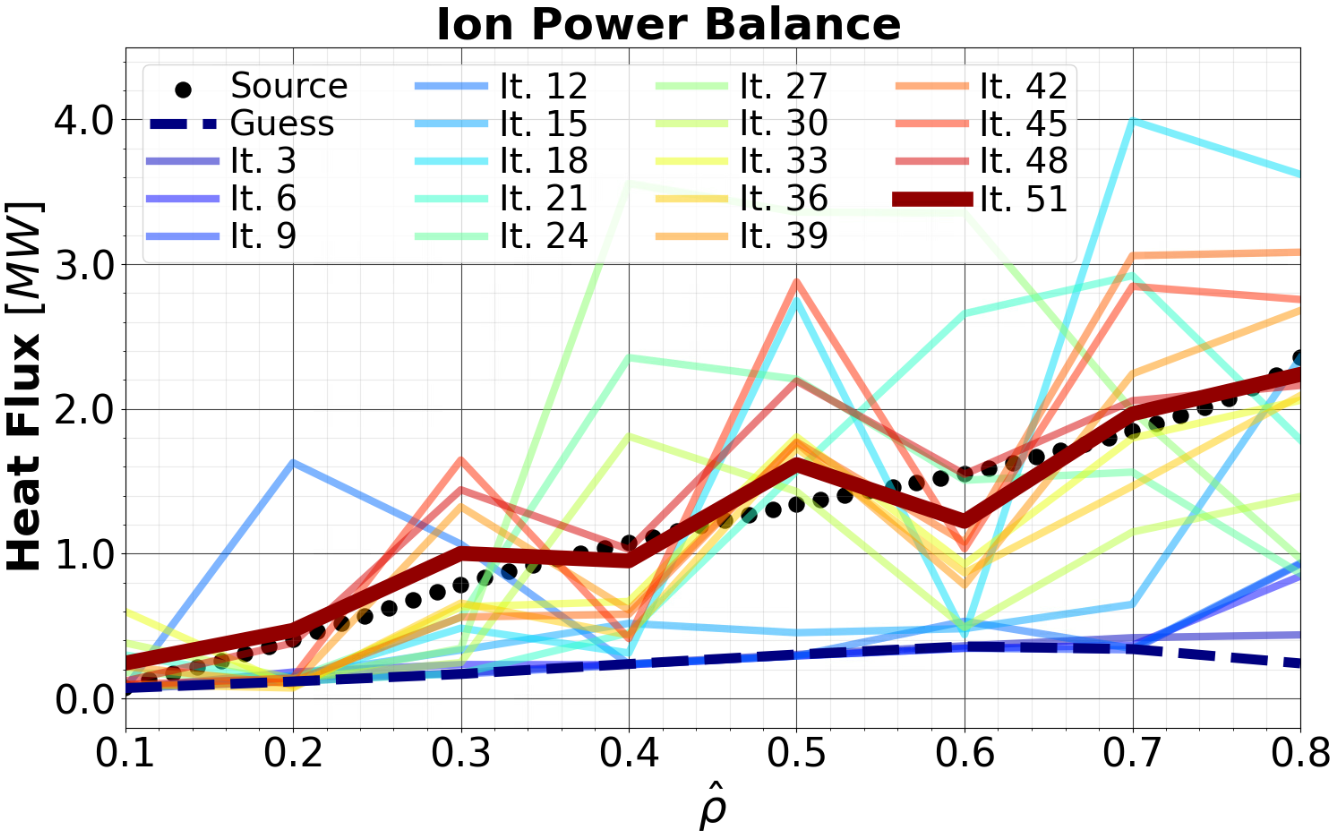}
        \caption{}
        \label{fig:flat_ion_bal}
    \end{subfigure}
    \hfill
    \begin{subfigure}{0.31\textwidth}
        \centering
        \includegraphics[width=\linewidth]{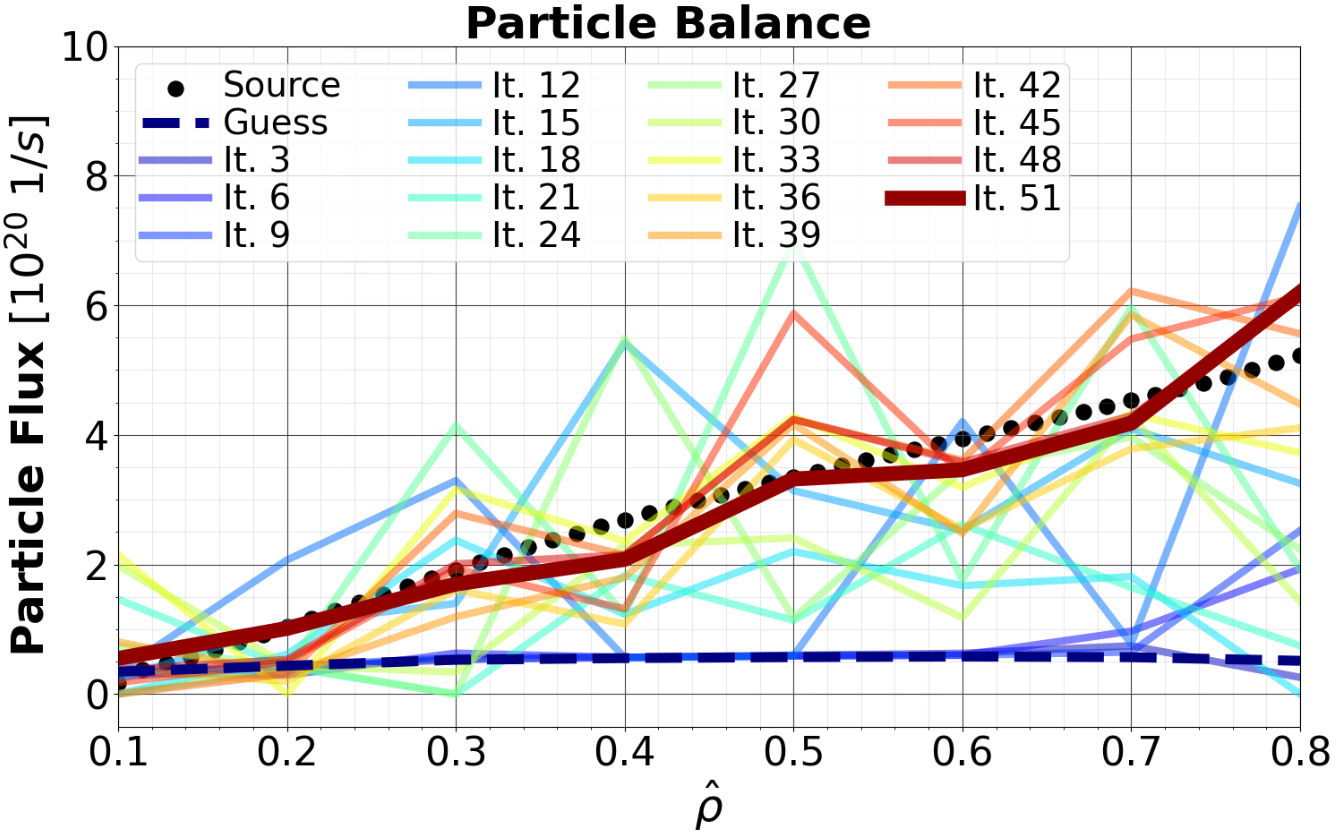}
        \caption{}
        \label{fig:flat_particle_bal}
    \end{subfigure}
    \begin{subfigure}{0.31\textwidth}
        \centering
        \includegraphics[width=\linewidth]{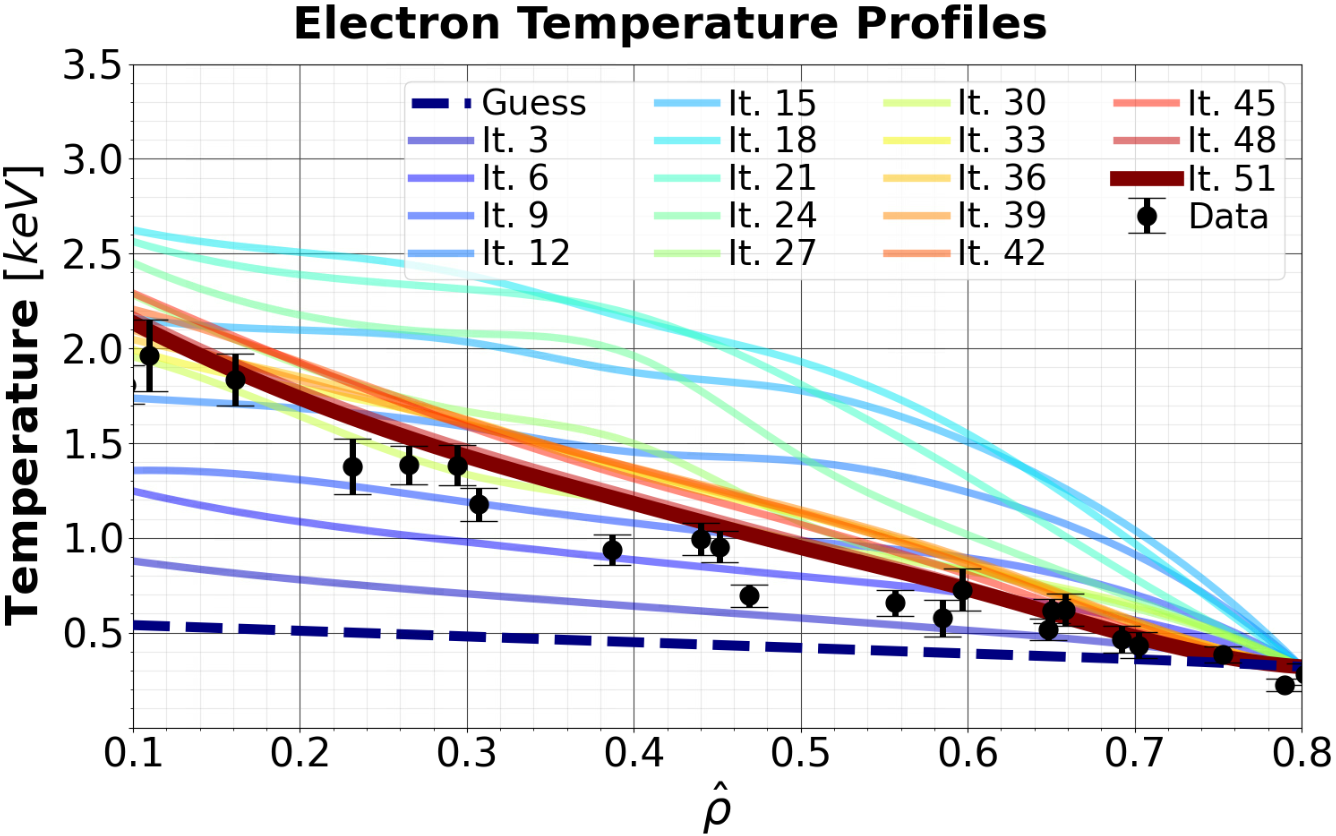}
        \caption{}
        \label{fig:flat_Te}
    \end{subfigure}
    \hfill
    \begin{subfigure}{0.31\textwidth}
        \centering
        \includegraphics[width=\linewidth]{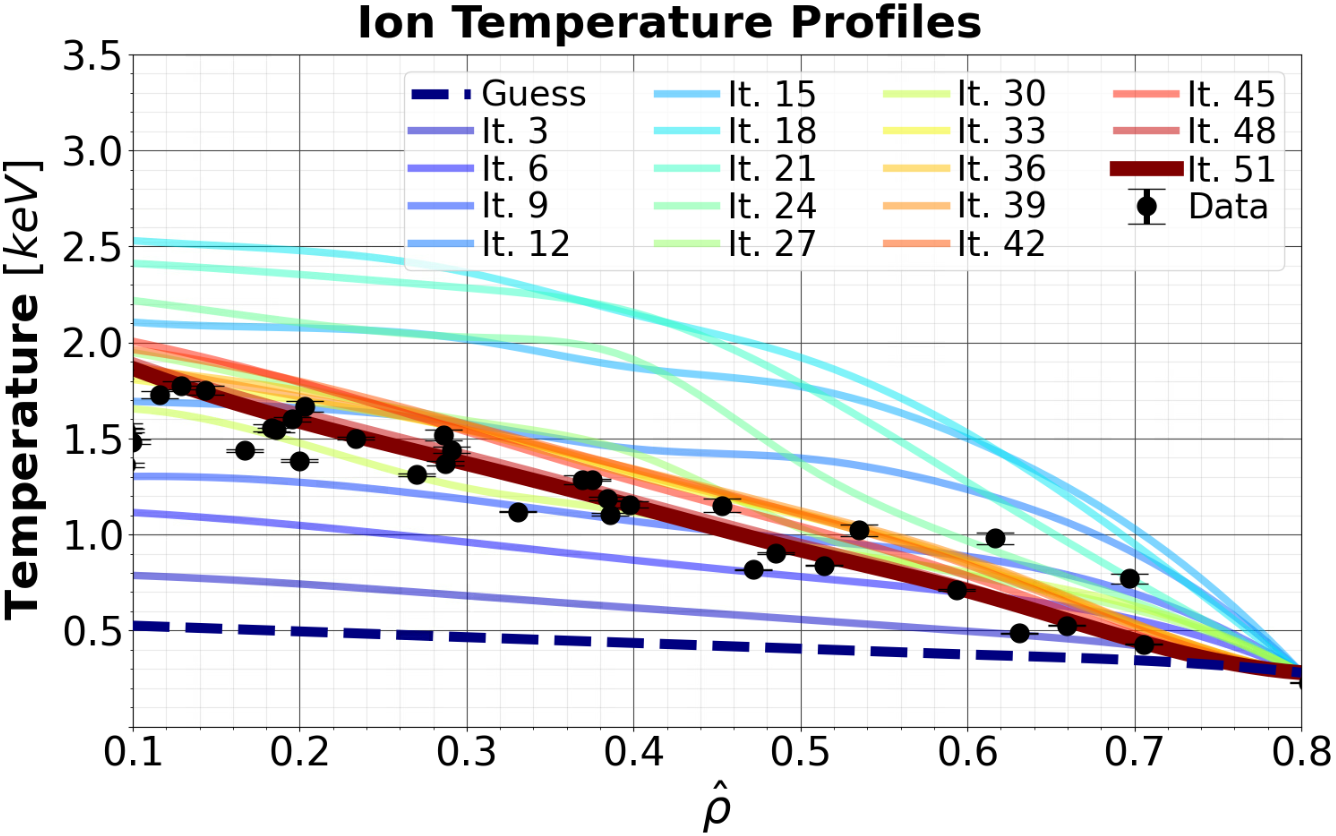}
        \caption{}
        \label{fig:flat_Ti}
    \end{subfigure}
    \hfill
    \begin{subfigure}{0.31\textwidth}
        \centering
        \includegraphics[width=\linewidth]{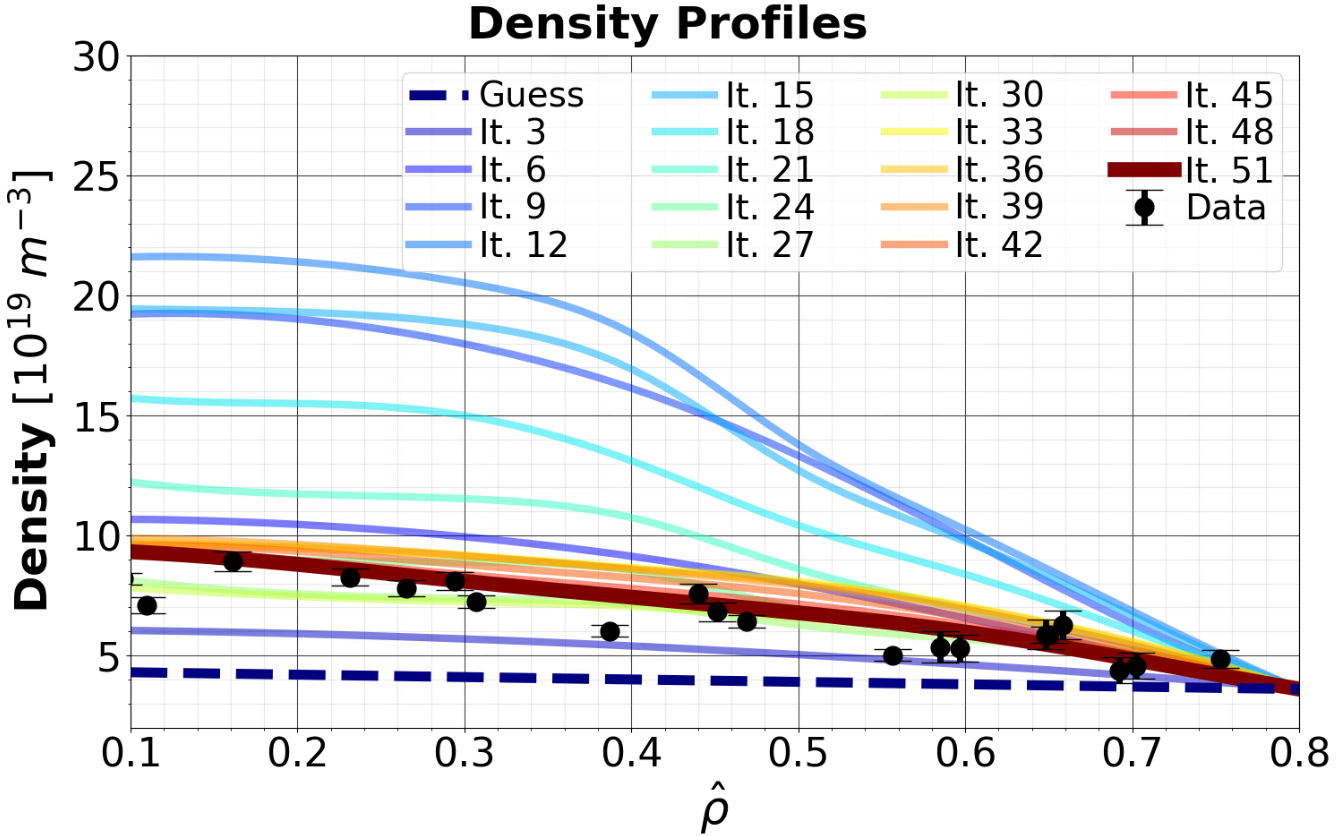}
        \caption{}
        \label{fig:flat_n}
    \end{subfigure}
    \caption{Demonstration of how the \texttt{GENE-KNOSOS-Tango} framework evolves the (\subref{fig:flat_Te}) electron temperature, (\subref{fig:flat_Ti}) ion temperature, and (\subref{fig:flat_n}) density profiles, which were all given initial guesses that are approximately flat. Convergence of the (\subref{fig:flat_elec_bal}) electron power, (\subref{fig:flat_ion_bal}) ion power, and (\subref{fig:flat_particle_bal}) particle balances was achieved in about 50 iterations.}
    \label{fig:flat_Tango}
\end{figure*}

A demonstration of how the \texttt{GENE-KNOSOS-Tango} framework evolves the plasma profiles is shown in Fig. \ref{fig:flat_Tango}. As a rule of thumb, fluxes that overshoot the sources at a given position will result in a local flattening of the relevant profile, while the profile is steepened at positions where the fluxes are under-predicted. In Fig. \ref{fig:flat_Tango}, approximately flat profiles were used as initial guesses for the plasma temperatures and density. In the context of the profiles, only the boundary values are absolutely important for \texttt{Tango}; the code will adjust every other point in the profile to satisfy the power and particle balances without the need for any external user prompt. However, while convergence can still be achieved, using poorly chosen initial profile guesses typically requires more iterations before balances are satisfactorily fulfilled. In this case, about 50 iterations were needed. \\

To make the validation study possible, the capabilities of \texttt{Tango} were expanded through the addition of a neoclassical $E_r \times B$ shear calculation \cite{Hammett2006} and a particle source model based on neutrals ionization. For the former, the radial derivative of the $E_r \times B$ velocity is linearized and evaluated at each radial position of interest. In Eq. \ref{eq:ExB_shear}, normalized dimensionless quantities are used, as signified by the circumflex accent:
\begin{align}
\label{eq:ExB_shear}
    \widehat{\gamma}_{Er \times B}(\hat{\rho}_0) &= - \left. \frac{d}{d\hat{\rho}} \left( \frac{\widehat{E}_r(\hat{\rho})}{\widehat{B}(\hat{\rho})} \right) \right|_{\hat{\rho} = \hat{\rho}_0} \nonumber
    \\&
    = - \left. \left[ \frac{d\widehat{E}_r}{d\hat{\rho}} - \frac{\widehat{E}_r}{\widehat{B}} (1 - \hat{s}) \right] \right|_{\hat{\rho} = \hat{\rho}_0}
\end{align}
For the latter, the neutrals density radial profile, $n_0({\hat{\rho}})$, is calculated using a short-mean-free-path one-dimensional transport model: \cite{Hazeltine1992, Thienpondt2023}
\begin{align}
\label{eq:neutrals_ioni}
    \frac{1}{r} \frac{d}{dr} \left[r \left(\frac{T_0}{m_0 \nu_{\text{CX}}} \right) \left( \frac{dn_0}{dr} + \frac{1}{T_0} \frac{dT_0}{dr} n_0 \right) \right] = \nu_{\text{ion}} n_0 
\end{align}
In Eq. \ref{eq:neutrals_ioni}, $T_0$ and $m_0$ are the neutral species temperature and mass, respectively. The ionization and charge exchange frequencies are given by $\nu_{ion}$ and $\nu_{CX}$, respectively. These are calculated by evaluating the analytical fits of the ionization and charge exchange cross-sections of hydrogen at the plasma parameters. \cite{Janev1993, Janev2003} That being said, it is assumed in the model that plasma ions and neutrals have the same temperature profile due to the high frequency of collisions and charge exchange reactions occurring between them. It is important to point out that the neutrals density at the outer boundary, $n_{0,edge}$, is a free parameter in Eq. \ref{eq:neutrals_ioni}. Reported measurements in W7-X electron-cyclotron-heated plasmas \cite{Beurskens2021,Romba2025} and particle confinement time calculations\cite{Kremeyer2022} were used as bases for $n_{0,edge}$ input values in \texttt{Tango}. \\

The ionization rate of the neutrals provides a particle source term for the density transport equation. In addition to this, for cases 3 and 4, the NBI particle fueling rates were obtained from \texttt{BEAMS3D} \cite{McMillan2014,Lazerson2020,Lazerson2021} simulations. Two fueling rates are provided by \texttt{BEAMS3D}, one for the electrons and another one for the thermal ions. The latter is slightly smaller than the former due to fast ion losses. Due to uncertainties with the degree of thermalization of the NBI ions, the electron fueling rate is used as the particle source for both plasma electrons and ions. This greatly simplified the simulations for this study; dedicated three-species simulations will be explored in a future study. On the other hand, for the pressure equation, heat sources and sinks consist of the ECRH and NBI power deposition profiles calculated using \texttt{TRAVIS} \cite{Marushchenko2014} and \texttt{BEAMS3D}, respectively, and the collisional thermalization \cite{NRL2019} between ions and electrons calculated in \texttt{Tango}.

\section{\label{sec:simulation_setup}Simulation Set-up and Model Hierarchy}

\begin{figure*}
    \centering
    \begin{subfigure}{\textwidth}
        \centering
        \includegraphics[width=\linewidth]{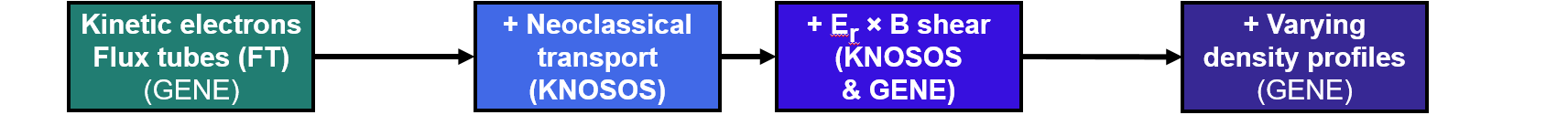}
    \end{subfigure}
    \begin{subfigure}{0.29\textwidth}
        \centering
        \includegraphics[width=\linewidth]{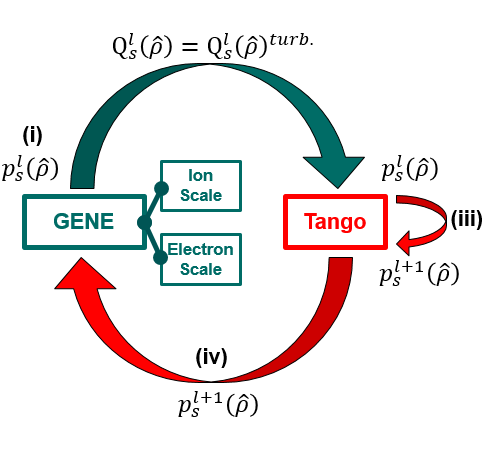}
        \caption{}
        \label{fig:framework_phase1}
    \end{subfigure}
    \hfill
    \begin{subfigure}{0.32\textwidth}
        \centering
        \includegraphics[width=\linewidth]{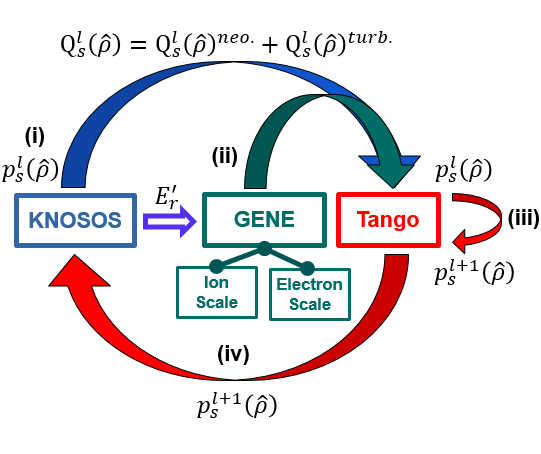}
        \caption{}
        \label{fig:framework_phase2&3}
    \end{subfigure}
    \hfill
    \begin{subfigure}{0.32\textwidth}
        \centering
        \includegraphics[width=\linewidth]{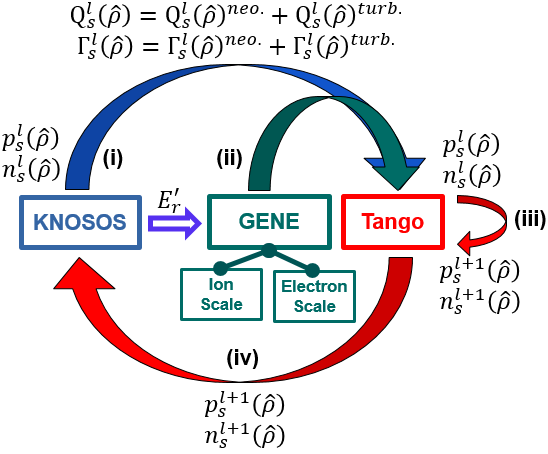}
        \caption{}
        \label{fig:framework_phase4}
    \end{subfigure}
    \caption{Model hierarchy of the \texttt{GENE-KNOSOS-Tango} framework: (a) Fixed-density simulations without neoclassical transport, (b) Inclusion of \texttt{KNOSOS} in the simulation loop for calculating neoclassical heat fluxes, (c) Calculation of the neoclassical $E_r \times B$ shear for \texttt{GENE} simulations, and (d) full framework with varying density profiles.}
    \label{fig:framework_hierarchy}
\end{figure*}

Instead of applying the full framework immediately, a model hierarchy was adapted in which physics is incrementally added to the simulations. This way, it is easier to understand how each of these effects can impact the converged plasma state. The workflow has four phases, as seen in Fig. \ref{fig:framework_hierarchy}, and the converged plasma profiles of one phase are used as initial inputs for the next one. \\

For phase 1, only \texttt{GENE} and \texttt{Tango} were included in the simulation loop, and the density profiles were kept fixed. As such, only the temperature profiles were varied based on how well the power balances had been met. To decrease the number of iterations necessary to reach convergence, the statistical fits of the experimental profiles were used as the initial guesses for the plasma profiles. The temperature and density values at $\hat{\rho} = 0.8$ were fixed as boundary conditions for \texttt{Tango}. The converged phase 1 plasma profiles were used as initial guesses for phase 2, where neoclassical heat fluxes were computed by \texttt{KNOSOS}. The neoclassical radial electric field $E_r$ from \texttt{KNOSOS} was passed to \texttt{Tango} in phase 3 to determine the neoclassical $E_r \times B$ shear profile. This was used as an additional input to \texttt{GENE}. Finally, in phase 4, the constraint on keeping the density profile fixed was lifted. With this, the particle fluxes, both turbulent from \texttt{GENE} and neoclassical from \texttt{KNOSOS}, were taken into account by \texttt{Tango}, which now also solved the transport equation for density. \\

For the \texttt{GENE} simulations, eight radial positions spanning $\hat{\rho}$ of 0.1 to 0.8 with a spacing of 0.1 were selected. The dimensionless radial coordinate $\hat{\rho}$ is defined by the magnetic toroidal flux passing through a given magnetic surface $\Psi$:
\begin{align}
\label{eq:rho_toroidal}
    \hat{\rho} = \frac{1}{a} \sqrt{\frac{\Psi}{\Psi_{LCFS}}}
\end{align}
In Eq. \ref{eq:rho_toroidal}, $a$ is the effective minor radius of the device and LCFS refers to the last closed flux surface, which is located at $\hat{\rho}$ = 1. At each $\hat{\rho}$ and each iteration, ion-scale kinetic-electron and electron-scale adiabatic-ion flux-tube simulations were performed. \\

The choice for this radial domain of the simulations was motivated by the task of validating the framework over a large portion of the stellarator core while keeping the computational costs of the study tractable. First, adding an additional point close to the magnetic axis ($\hat{\rho} \approx$ 0.0) was deemed unnecessary since the heat and particle fluxes in this region are generally negligible. While on-axis heating can be significant in some cases, its impact is tempered by the small plasma volume near the axis. Moreover, the quantities in this region are not measured well experimentally, providing less motivation to compare simulated results for $\hat{\rho}$ < 0.1 with experimental data. Next, there were some considerations for limiting the simulations to $\hat{\rho}$ = 0.8. Simulations in the vicinity of the LCFS at $\hat{\rho}$ = 1.0 will require a higher resolution in at least the radial and parallel directions due to the larger magnetic shear and stronger plasma shaping close to the plasma edge, respectively. More importantly, the kinetic simulations are only two-species, with hydrogen ions and electrons forming the plasma. As one moves closer to the LCFS, the effects of impurities and neutrals on the turbulence become increasingly important. In this case, they could have a large effect on the plasma and should no longer be disregarded. For instance, high impurity concentrations lead to higher radiation losses, which will subsequently reduce plasma temperatures. \\

Before starting any of the phases, convergence tests were first performed to determine the nominal resolution parameters that would be used throughout the study. After convergence had been achieved with respect to both power and particle balances in phase 4, several high-resolution runs were carried out to ensure that heat and particle fluxes stayed within $\pm20\%$ of the values obtained using the nominal resolution. This allowance is justified by the stiffness of the fluxes, as a moderate change in fluxes only slightly alters the gradients and, consequently, the profiles. Table \ref{tab:resolution} shows the numerical parameters for nominal and high resolution, where $l_i$ and $n_i$ are the extension of the simulation box and the number of grid points in the $i$ direction, respectively. The minimum value of the binormal wavenumber is given by $k_{y_{min}}$. The resolution parameters in the y-direction for the electron-scale simulation remain unchanged between the nominal- and high-resolution cases since these were constrained by the separation of ion- and electron-scales. Eq. \ref{eq:ion_electron_ky} must hold true to ensure that no overlap between the $k_y$ ranges of the two simulations occurred. In this equation, $\rho_s$ is the reference gyroradius for species $s$.
\begin{align}
\label{eq:ion_electron_ky}
    k_{y_{min, e}} \rho_e = \sqrt{\frac{m_i}{m_e}} k_{y_{min, e}} \rho_i > k_{y_{max, i}} \rho_i
\end{align}

\begin{table}
    \centering
    \begin{ruledtabular}
    \begin{tabular}{ccccc}
    \multirow{2}{*}{Parameter} & \multicolumn{2}{c}{Ion scale} & \multicolumn{2}{c}{Electron scale} \\
    & Nominal & High & Nominal & High \\
    \midrule[0.15em]
    $l_x$         & 128  & 256   & 128   & 256  \\
    $n_{k_x}$     & 64   & 256   & 64    & 256  \\
    $k_{y_{min}}$ & 0.10 & 0.05  & 0.075 & 0.075 \\
    $n_{k_y}$     & 32   & 64    & 48    & 48   \\
    $n_z$         & 96   & 128   & 96    & 128  \\
    $l_{v_{||}}$  & 2.5  & 3     & 2.5   & 3    \\
    $n_{v_{||}}$  & 16   & 48    & 16    & 48   \\
    $l_{\mu}$     & 9    & 9     & 9     & 9   \\
    $n_{\mu}$     & 12   & 12    & 12    & 12   \\
    \end{tabular}
    \end{ruledtabular}
    \caption{Numerical parameters for the resolution of the \texttt{GENE} simulations}
    \label{tab:resolution}
\end{table}

Since $k_{y_{min}}$ was not modified, $n_{k_y}$ was maintained at the nominal value as well so that the high-resolution simulation covers the same $k_y$ range. Lastly, it should be noted that the actual value of radial box extent in units of gyroradii, $l_x$, used by \texttt{GENE} is not always equal to the input due to its quantization based on $k_{y_{min}}$, the magnetic shear $\hat{s}$, the safety factor $q$, and other parameters. In practice, $l_x$ became as low as 80 in some cases, giving a radial resolution close to one gyroradius.

\begin{figure*}[t]
    \centering
    \begin{subfigure}{0.45\textwidth}  
        \centering
        \includegraphics[width=\textwidth]{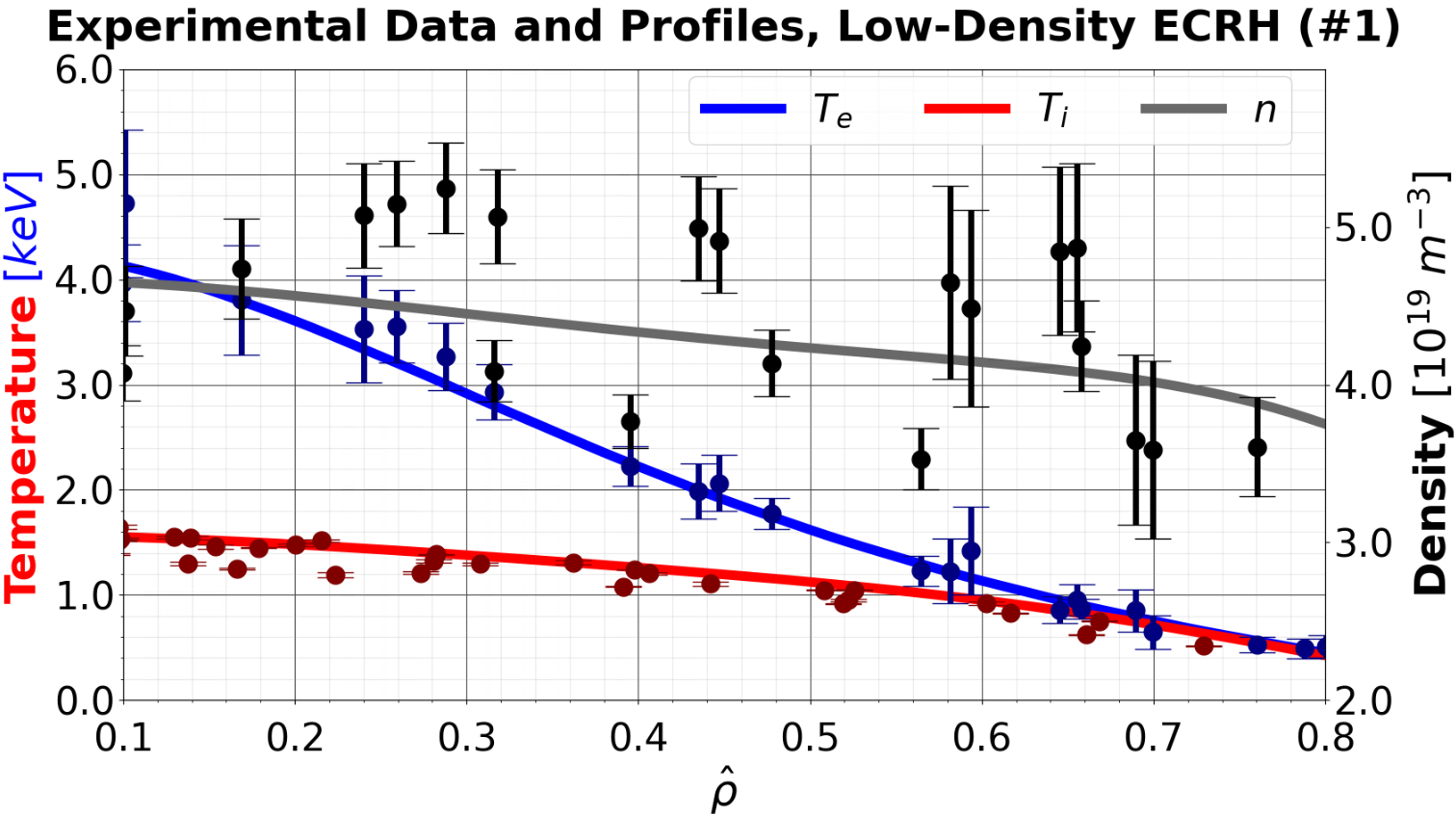} 
        \caption{}
        \label{fig:Elow_Tn_expt}
    \end{subfigure}
    \hspace{0.5cm}
    \begin{subfigure}{0.45\textwidth}  
        \centering
        \includegraphics[width=\textwidth]{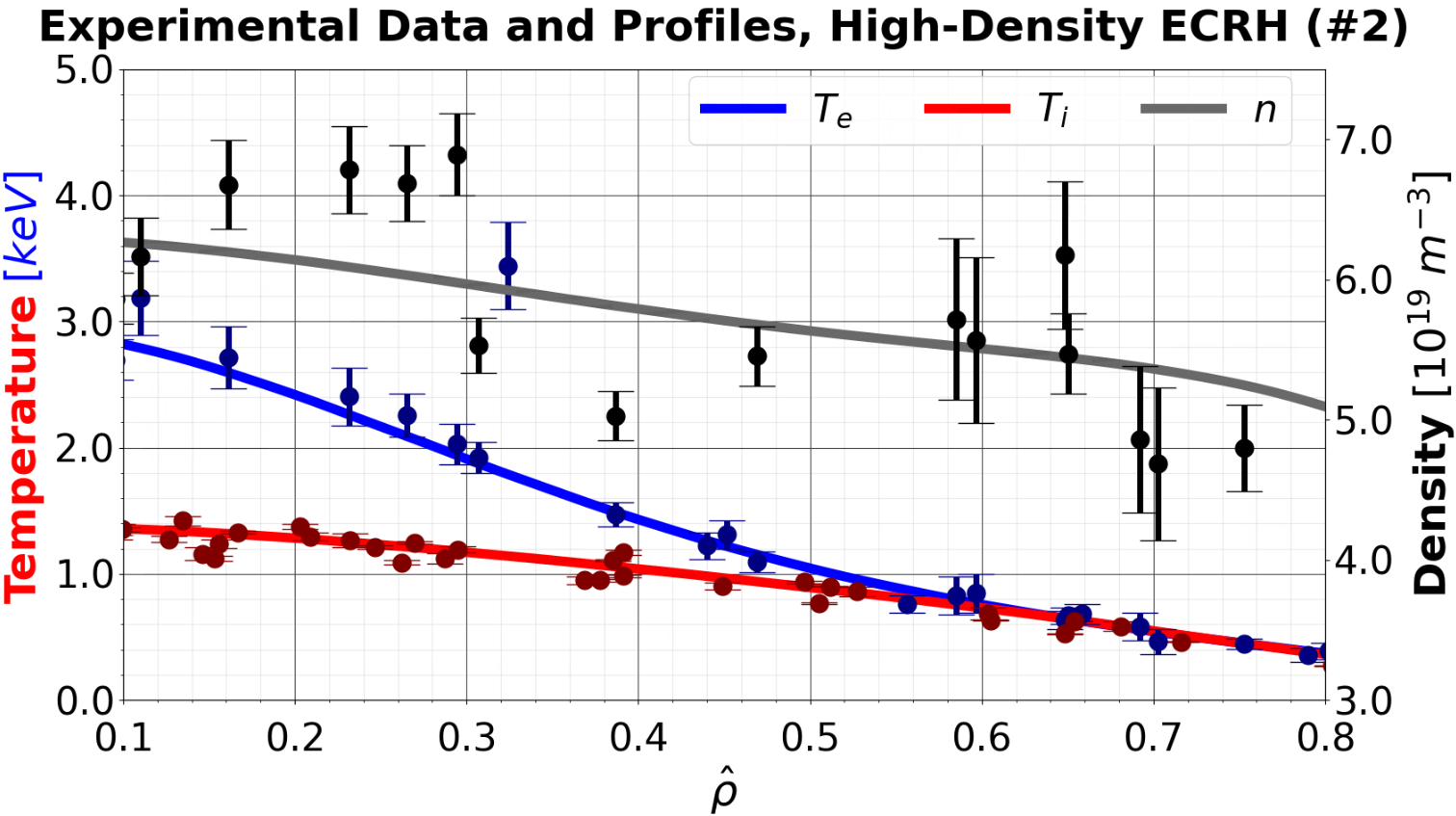}  
        \caption{}
        \label{fig:Ehigh_Tn_expt}
    \end{subfigure}
    \vspace{0.5cm}
    \begin{subfigure}{0.45\textwidth}
        \centering
        \includegraphics[width=\textwidth]{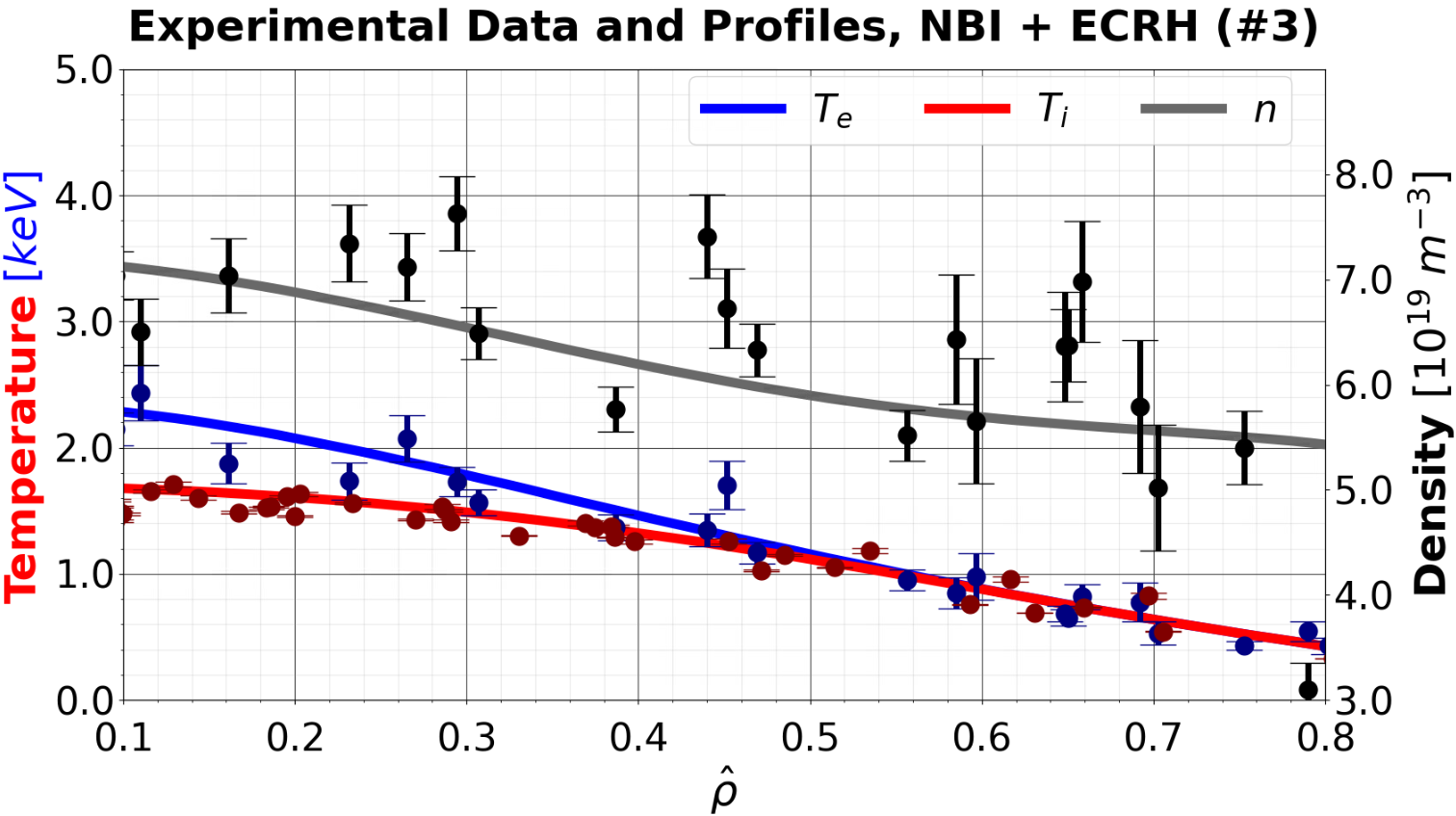}
        \caption{}
        \label{fig:NII_Tn_expt}
    \end{subfigure}
    \hspace{0.5cm}
    \begin{subfigure}{0.45\textwidth}
        \centering
        \includegraphics[width=\textwidth]{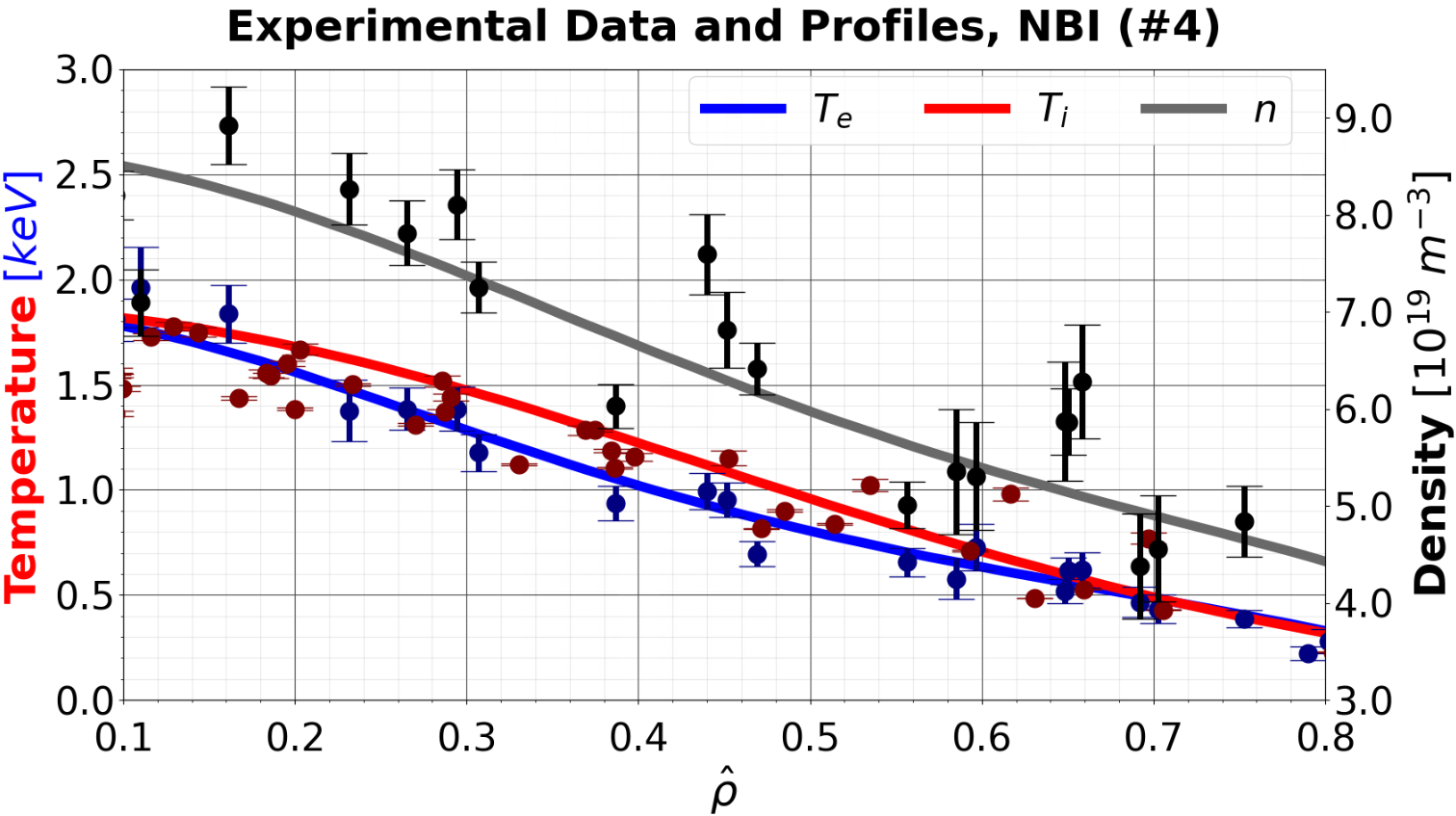}
        \caption{}
        \label{fig:NIII_Tn_expt}
    \end{subfigure}
    \caption{The experimental data points and the statistical fits, which were used as the initial profiles for the simulation framework, for the (\subref{fig:Elow_Tn_expt}) low-density ECRH, (\subref{fig:Ehigh_Tn_expt}) high-density ECRH, (\subref{fig:NII_Tn_expt}) NBI + ECRH, and (\subref{fig:NIII_Tn_expt}) NBI scenarios}
    \label{fig:Tn_data}
\end{figure*}

The kinetic-electron ion-scale high-resolution simulations proved to be extremely computationally expensive. As such, only two positions for the four scenarios were checked. Since the edge region significantly influences the plasma profiles, $\hat{\rho} = 0.7$ was chosen as the first point. Next, to introduce variation to this test, $\hat{\rho} = 0.4$ was selected as the second point. This position lies sufficiently distant from both the edge region, which is already represented by the $\hat{\rho} = 0.7$ simulation, and the magnetic axis, which is generally less interesting from a simulation viewpoint as mentioned earlier. For both positions, good agreement between the nominal- and high-resolution fluxes, both ion-scale and electron-scale, is observed for all four scenarios. This result confirms that the nominal resolution parameters were indeed acceptable. 

\section{\label{sec:results}Simulation Results}

The experimental data points and the statistical fits are shown in Fig. \ref{fig:Tn_data}. In the next sections, the statistical fit for the temperature profiles will no longer be shown to simplify the plots. The $n_e$ profiles were obtained through an interferometer \cite{Brunner2018} and a Thomson scattering system\cite{Bozhenkov2017}. The Thomson scattering system was also used to obtain the $T_e$ profiles. Lastly, the $T_i$ profiles were measured using an X-ray imaging crystal spectrometer (XICS) \cite{Kring2018,Pablant2021} and a charge exchange recombination spectroscopy system (CXRS)\cite{Ford2020}. A systematic bias was found for the former, but applying a correction for this improved the agreement between the XICS and CXRS measurements. \cite{Carralero2021} The error bars of the data provides leeway for the simulated profiles to be adjusted by \texttt{Tango} while still allowing the experimental data to be adequately matched. For the first three phases of the workflow, the $T_e$, $T_i$, and $n_e$ values at $\hat{\rho}$ = 0.8 were used and kept fixed. In the fourth phase, where the density profiles were allowed to evolve, the impact of adjustments to the boundary conditions was explored. \\

Several questions were also explored and addressed in each phase. In phase 1, what is relative contribution of ion- and electron-scale turbulence to the heat fluxes for each scenario? Moreover, how different are the heat fluxes for different flux tubes located on the same surface? In phase 2, how large are the neoclassical heat fluxes? Moreover, how do the neoclassical heat fluxes vary from one position to another? In phase 3, how significant is the reduction in the heat fluxes due to the neoclassical $E_r \times B$ shear? Lastly, in phase 4, what are the parameters that affect the particle flux the most? \\

The next sub-sections will focus exclusively on the low-density ECRH scenario. The trends from other cases will be presented only when significant differences arise among the four. In phase 4, which is the culmination of the validation study simulations, the simulation results will be shown for all four scenarios. Following this, selected turbulent characteristics and trends extracted from the simulations will be compared with those obtained from the experimental data.

\subsection{\label{sec:results_phase1}Phase 1: Fixed Density Profile, without \texttt{KNOSOS}}

\begin{figure}
    \centering
    \includegraphics[width=\linewidth]{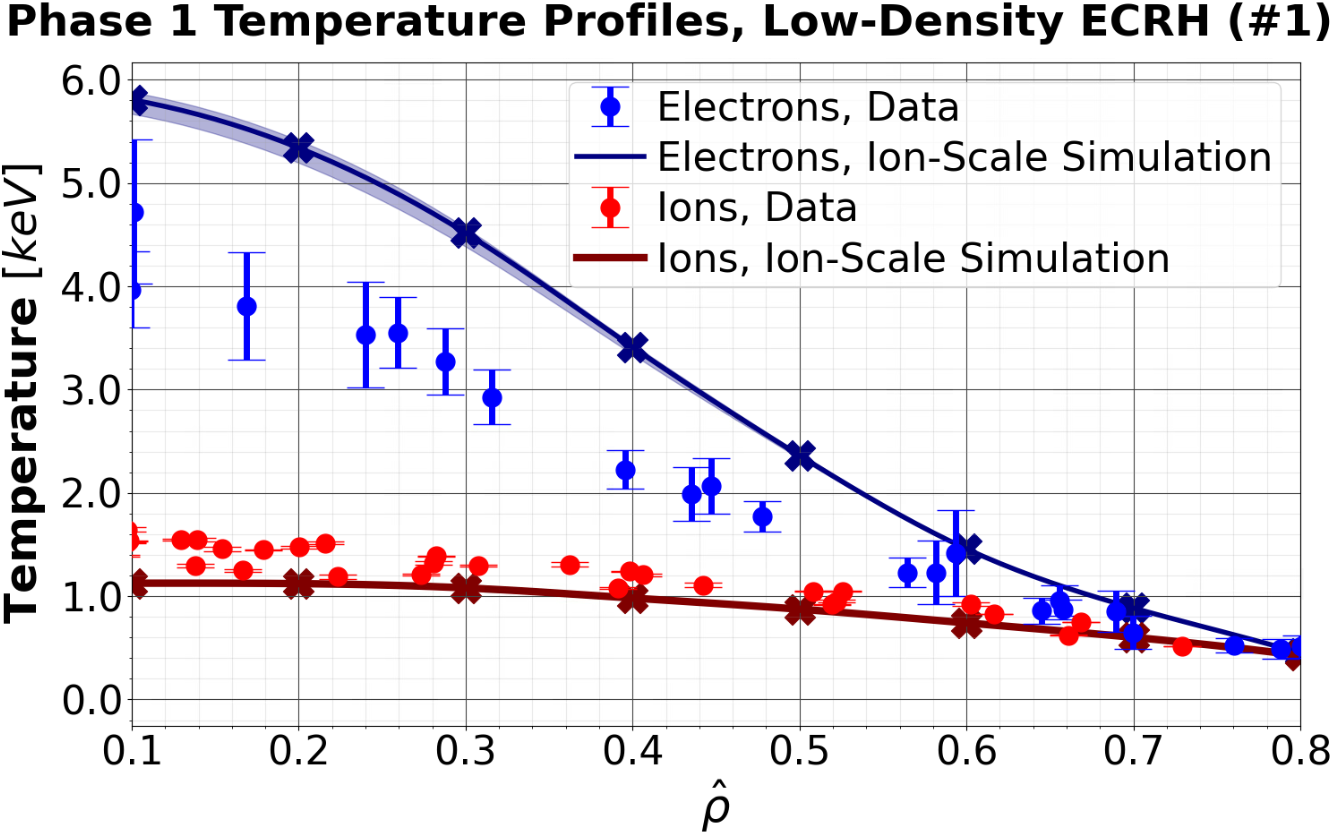}
    \caption{Temperature profiles for the low-density ECRH case without the electron-scale simulation. The mean, minimum, and maximum from the last several iterations are shown.}
    \label{fig:ECRH_low_noETG_temp}
\end{figure}

\begin{figure}[h!]
    \centering
    \begin{subfigure}[b]{0.235\textwidth}
        \includegraphics[width=\textwidth]{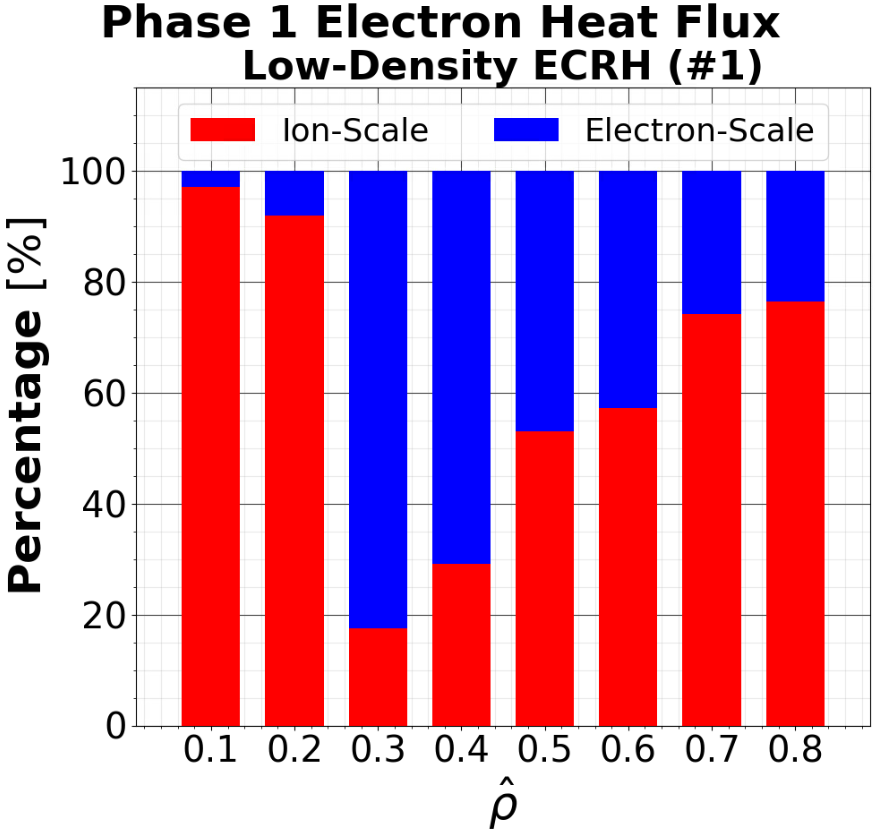}
        \caption{}
        \label{fig:Q_breakdown_Elow}
    \end{subfigure}
    \hfill
    \begin{subfigure}[b]{0.235\textwidth}
        \includegraphics[width=\textwidth]{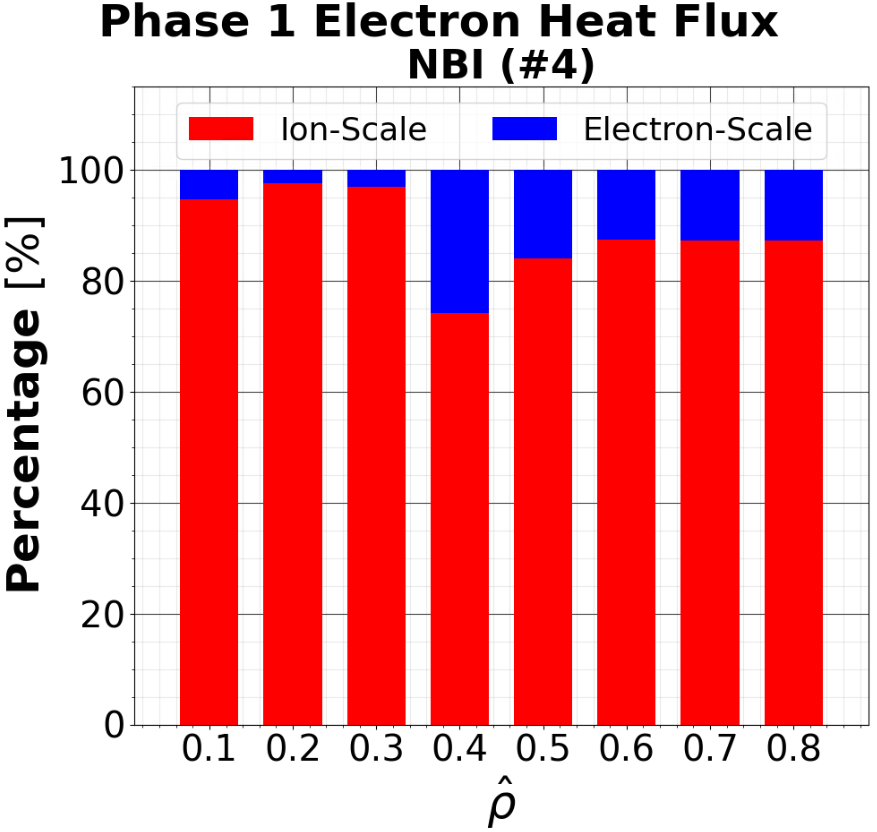}
        \caption{}
        \label{fig:Q_breakdown_NIII}
    \end{subfigure}
    \caption{Breakdown of electron heat fluxes for (\subref{fig:Q_breakdown_Elow}) low-density ECRH and (\subref{fig:Q_breakdown_NIII}) NBI scenarios.}
    \label{fig:Q_breakdown_phase1}
\end{figure}

In the beginning of this phase, the electron-scale simulation was first removed from the iteration loop. This was done for the low-density ECRH case to check the effect of electron-scale turbulence on the converged plasma profiles. The power balance was satisfied within 30 iterations. However, the $T_e$ profile shown in Fig. \ref{fig:ECRH_low_noETG_temp} deviated significantly from the experimental data, especially for $\hat{\rho} < 0.4$. This was not surprising, given that the electron-scale heat fluxes could be significant for $\hat{\rho} \ge 0.4$. For this case, the contribution to the total electron heat flux $Q_e$ in this range varied between 20 – 80\%, as shown in Fig. \ref{fig:Q_breakdown_Elow}. This is equivalent to 20 - 40\% of the total heat flux. The sharp increase in the electron-scale flux contribution between $\hat{\rho= 0.2}$ and 0.3 is attributed to the significant increase in the normalized electron temperature gradient $\omega_{Te}$ by about 33\%. In the absence of this heat loss channel, $T_e$ would indeed increase. This result already highlights the importance of including electron temperature gradient (ETG) turbulence in modeling experiments, especially for electron-heated plasmas. The high-density ECRH and NBI + ECRH cases exhibited similar electron-scale heat flux contributions of about 60\% for $\hat{\rho} \ge 0.4$. The least contribution was observed for the NBI case, as depicted in Fig. \ref{fig:Q_breakdown_NIII}, where the electron-scale heat flux amounted to $10 - 25$\% of $Q_e$. \\

\begin{figure}
    \centering
    \includegraphics[width=\linewidth]{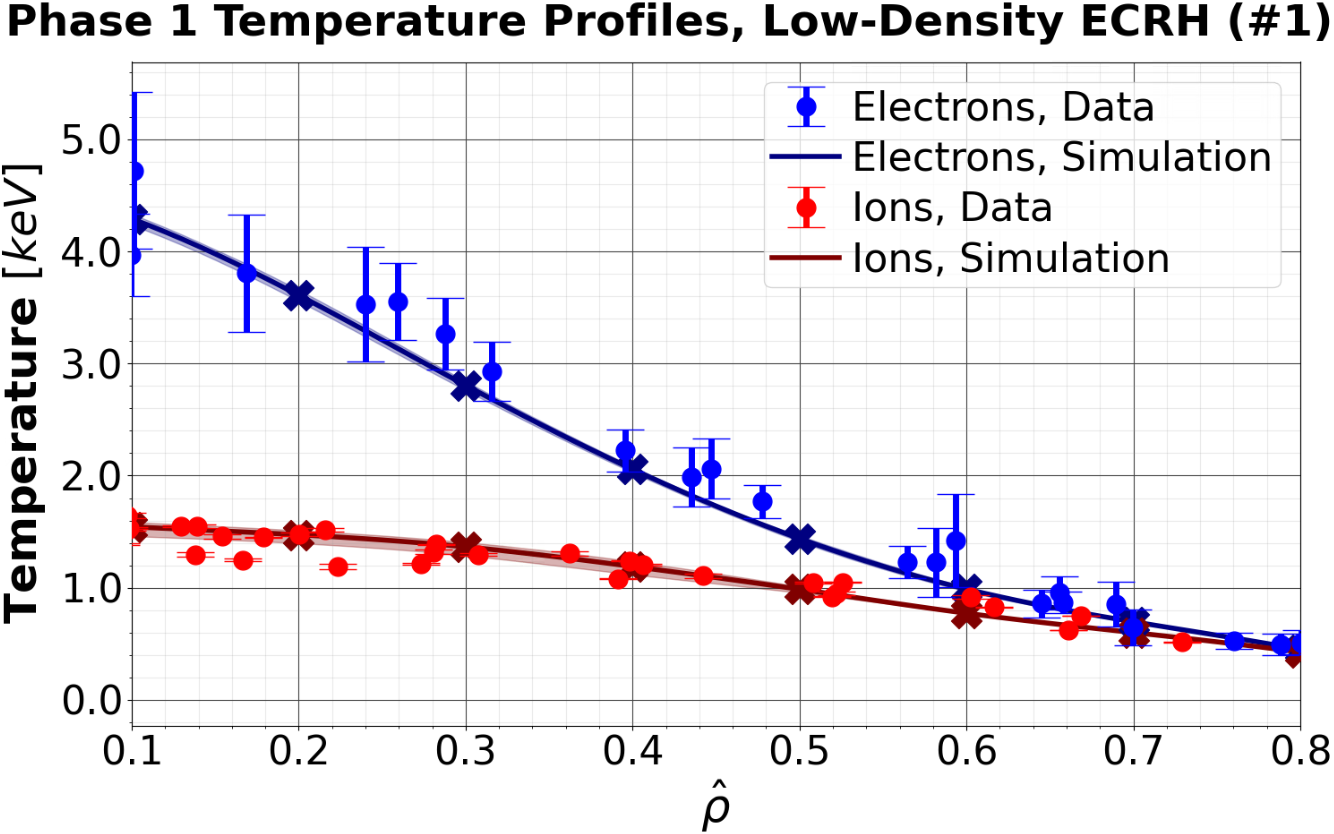}
    \caption{Phase 1 temperature profiles for the low-density ECRH case. Both ion- and electron-scale simulations are present in the iteration loop.}
    \label{fig:phase1}
\end{figure}

With the inclusion of electron-scale turbulence, good agreement with the simulated and experimental profiles was achieved, as seen in Fig. \ref{fig:phase1}. The increase in electron heat flux lowered the $T_e$ profile and augmented the ion heating through collisional heat transfer. It may be tempting to stop at this point and immediately conclude that the \texttt{GENE-KNOSOS-Tango} framework does indeed work, but the simulations were still lacking several physics and, more importantly, the density profile was still fixed. \\

\begin{figure}
    \centering
    \includegraphics[width=0.85\linewidth]{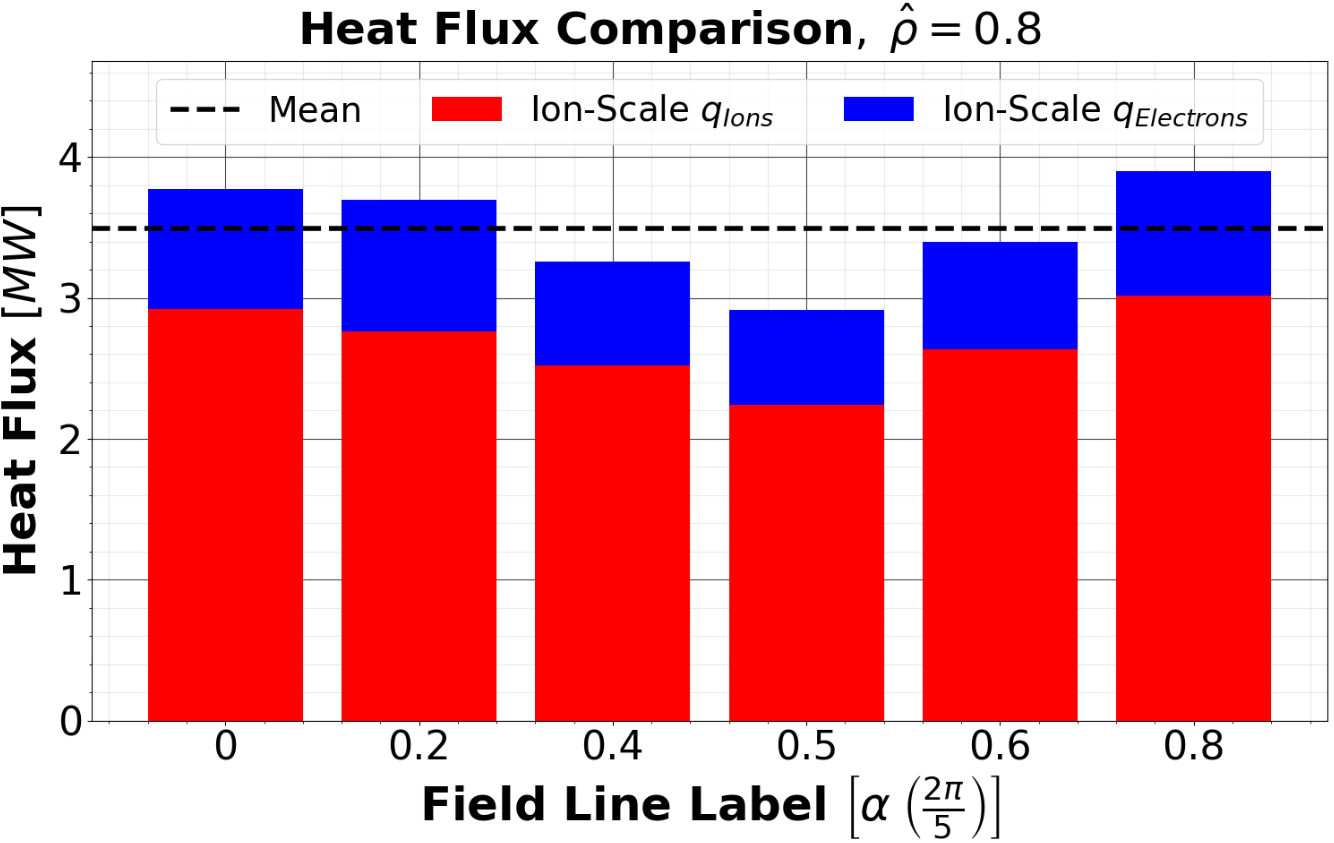}
    \caption{Total turbulent heat fluxes for different flux tubes located on the same flux surface. The variation of the total heat flux among the flux tubes was moderate.}
    \label{fig:different_alpha}
\end{figure}

To check the validity of the results thus far, simulations for different flux tubes on a single surface were also performed. This was done to gauge the variation of the heat flux magnitude with respect to the $\alpha = 0$ flux tube, which was where the simulations have all been carried out. For reference, this is the flux tube that passes through the outboard mid-plane of the bean-shaped cross-section of W7-X. The position of $\hat{\rho}$ = 0.8 was selected and flux-tube simulations were carried out for five additional field lines as denoted by the field line label $\alpha$ in Fig. \ref{fig:different_alpha}. The field line label $\alpha$ covers [0, $\frac{2\pi}{5}$] due to the 5-fold symmetry of W7-X. The maximum error of the $\alpha = 0$ heat flux with respect to that of the other field lines was about 23\%, while the difference to the mean was only 6\%. Due to the stiff dependency of the fluxes on their respective gradient drive, choosing a different flux tube for the simulations would most likely still yield the same temperature profiles. 

\subsection{\label{sec:results_phase2}Phase 2: Fixed Density Profile, with Neoclassical Heat Transport}

\begin{figure}
    \centering
    \includegraphics[width=\linewidth]{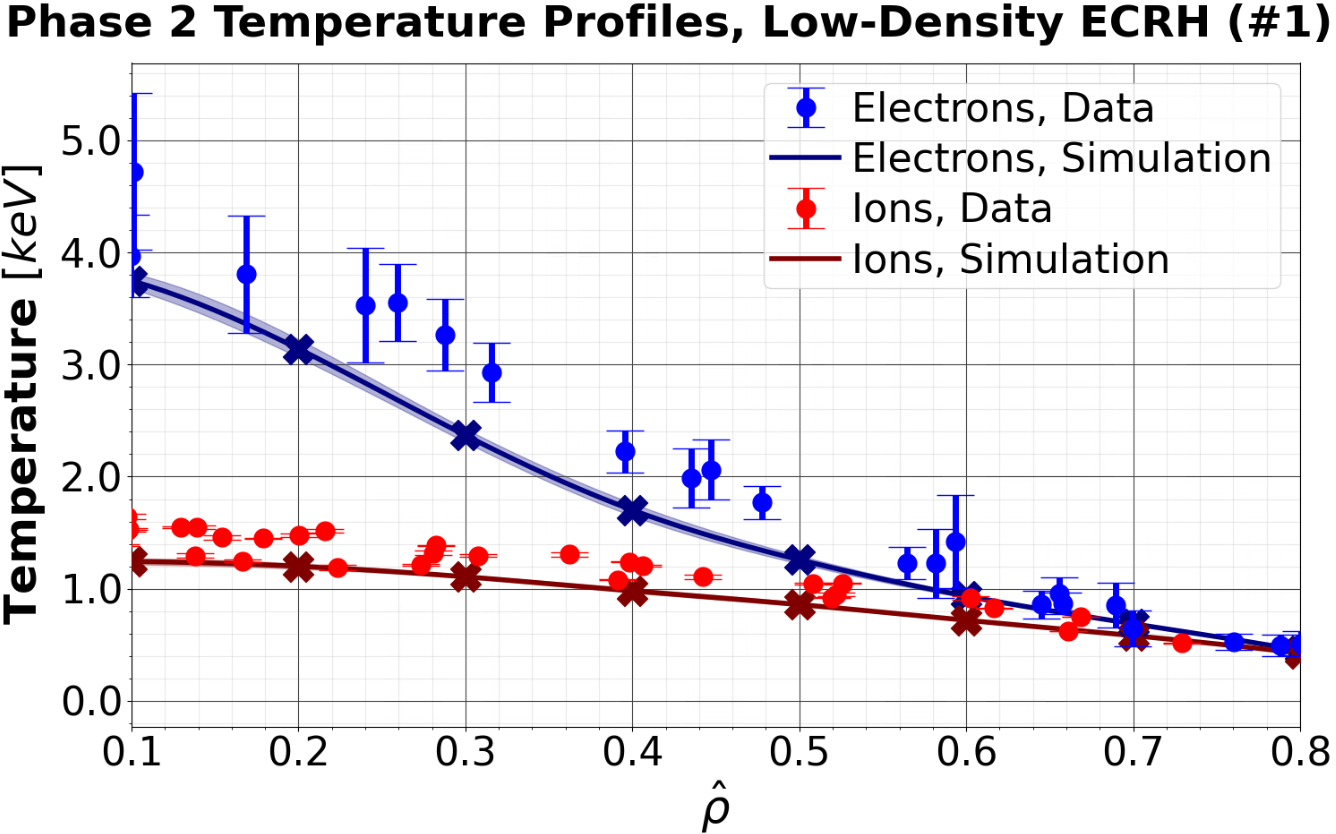}
    \caption{Phase 2 temperature profiles for the low-density ECRH case. The addition of neoclassical heat fluxes caused a small reduction in the plasma temperatures.}
    \label{fig:phase2}
\end{figure}

\begin{figure}
    \centering
    \begin{subfigure}[b]{0.235\textwidth}
        \includegraphics[width=\textwidth]{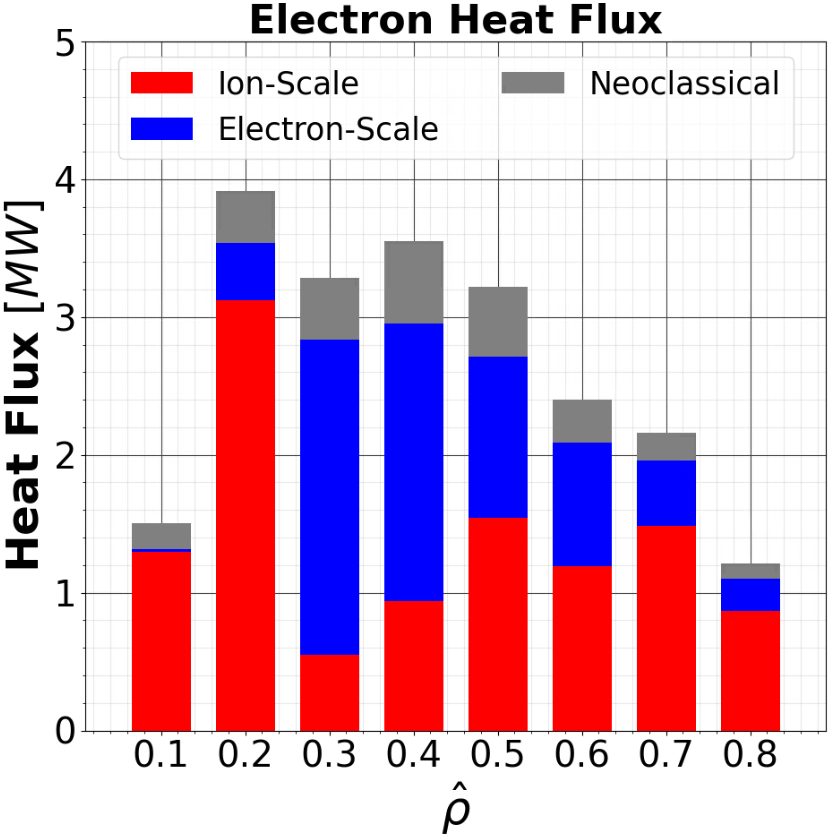}
        \caption{}
        \label{fig:phase2_absoluteQ_electrons}
    \end{subfigure}
    \hfill
    \begin{subfigure}[b]{0.235\textwidth}
        \includegraphics[width=\textwidth]{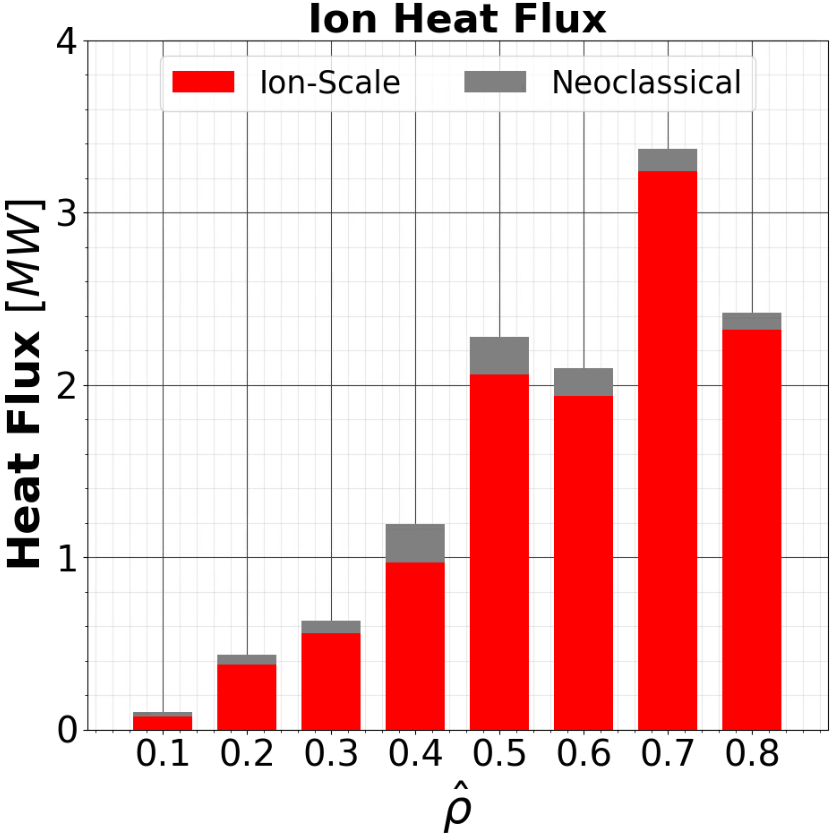}
        \caption{}
        \label{fig:phase2_absoluteQ_ions}
    \end{subfigure}
    \caption{Phase 2 flux breakdown for the (\subref{fig:phase2_absoluteQ_electrons}) electron and (\subref{fig:phase2_absoluteQ_ions}) ions in the low-density ECRH case. The contribution of neoclassical heat flux is larger for the electrons.}
    \label{fig:phase2_absoluteQ_breakdown}
\end{figure}

With the addition of neoclassical heat transport, the temperature profiles decreased as seen in Fig. \ref{fig:phase2}. The agreement worsened for the inner radial positions, especially for the electrons. On the other hand, the ions were less affected and exhibited a smaller temperature reduction. The breakdown of total electron and ion heat fluxes could explain these observations. In Fig. \ref{fig:phase2_absoluteQ_breakdown}, it can be seen that the neoclassical electron fluxes of about 0.5 MW are relatively larger than those of the ions, with 0.1 MW on average. With a larger additional energy loss channel, the phase 1 electron temperature profile was flattened more by \texttt{Tango} to reduce the turbulent heat flux and maintain power balance. \\

The neoclassical heat fluxes $Q_{neo}$ show good agreement with the \texttt{KNOSOS}-corrected calculations made using \texttt{NEOTRANSP} \cite{Smith2022} and \texttt{DKES} \cite{vanRij1989} mono-energetic coefficients. \cite{Carralero2022} The maximum was generally found in $0.3 \le \hat{\rho} \le 0.6$, with $Q_{neo}$ gradually decreasing as one moved away from this range in either direction. The simulated neoclassical flux profiles were generally more skewed toward $\hat{\rho} = 0.3$, which are in good agreement with the results shown in Ref. \citenum{Carralero2022}.

\subsection{\label{sec:results_phase3}Phase 3: Fixed Density Profile, with Neoclassical Heat Transport and Neoclassical $\mathrm{E_r \times B}$ shear}

\begin{figure}
    \centering
    \begin{subfigure}[b]{0.235\textwidth}
        \includegraphics[width=\textwidth]{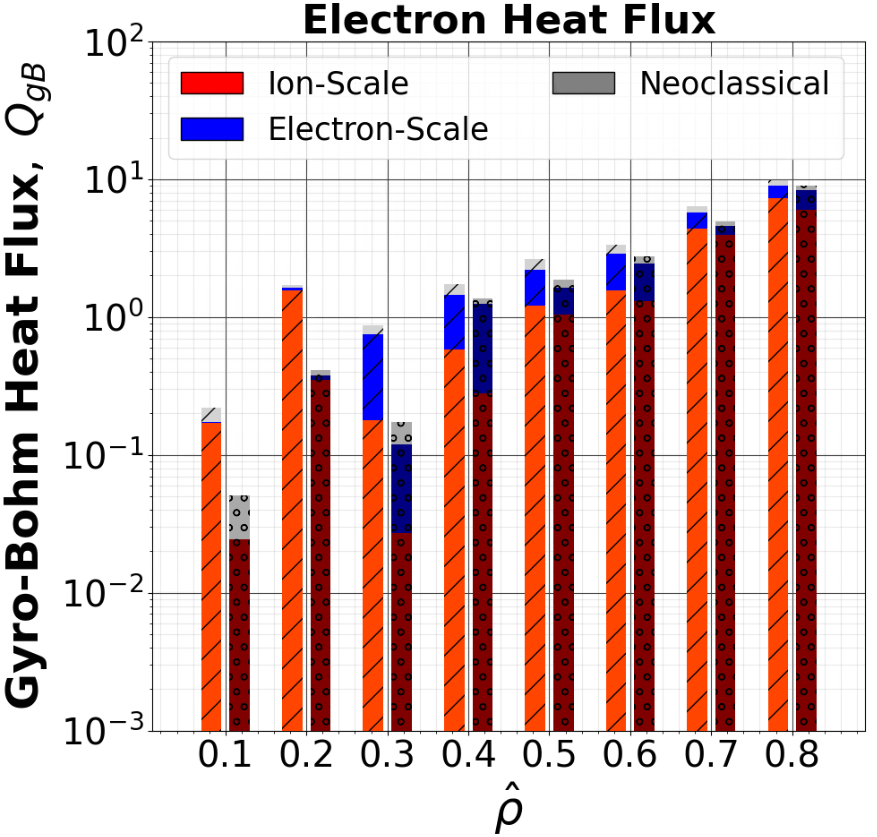}
        \caption{}
        \label{fig:phase_2_3_elec_GB}
    \end{subfigure}
    \hfill
    \begin{subfigure}[b]{0.235\textwidth}
        \includegraphics[width=\textwidth]{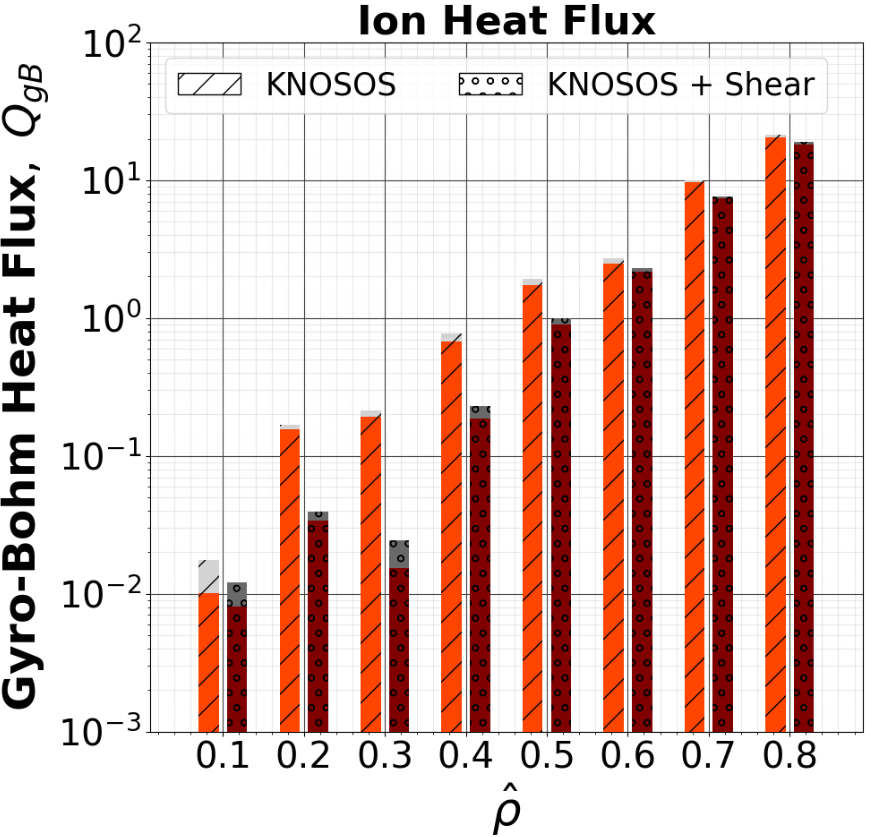}
        \caption{}
        \label{fig:phase_2_3_ion_GB}
    \end{subfigure}
    \caption{Comparison of the gyro-Bohm (\subref{fig:phase_2_3_elec_GB}) electron and (\subref{fig:phase_2_3_ion_GB}) ion heat flux breakdown between phases 2 and 3 for the low-density ECRH case. The inclusion of the shear reduced the heat fluxes, especially the ion-scale contribution.}
    \label{fig:phase_2_3_GB}
\end{figure}

In Fig. \ref{fig:phase_2_3_GB}, the heat flux breakdown in gyro-Bohm units at each $\hat{\rho}$ is shown for phases 2, which included neoclassical transport but not the neoclassical $E_r \times B$ shear yet, and 3, where the shear was added to the \texttt{GENE} simulations. The side-by-side comparison shows that the heat fluxes were reduced due to this effect. This reduction was more significant in the inner and intermediate positions, specifically for $\hat{\rho} \le 0.6$. \\

\begin{figure}
    \centering
    \includegraphics[width=\linewidth]{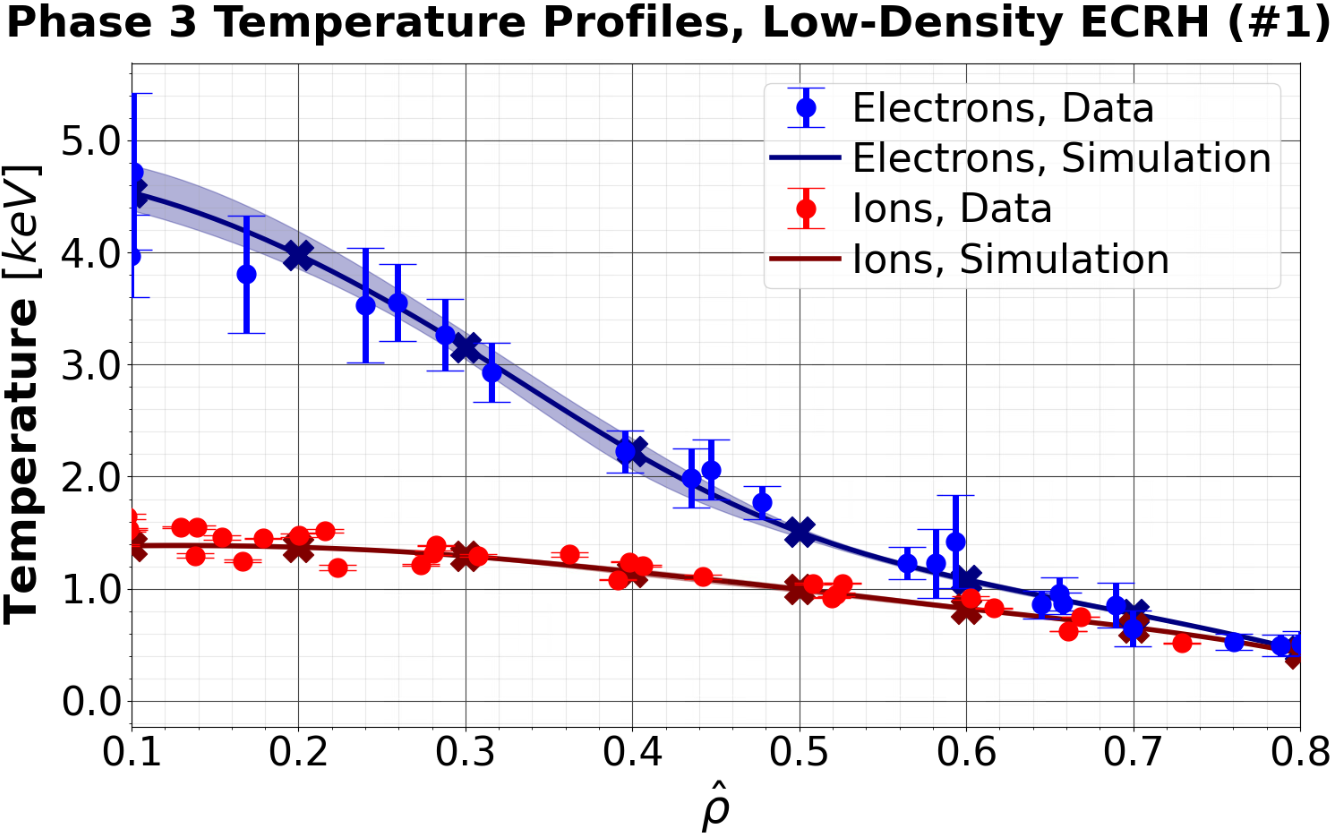}
    \caption{Phase 3 temperature profiles for the low-density ECRH case. Good agreement with the experimental temperature data was recovered relative to phase 2.}
    \label{fig:phase3}
\end{figure}

More importantly, the ion-scale heat fluxes were more impacted by the addition of neoclassical $E_r \times B$ shear. Since shear diminishes the size of turbulent structures and eddies in the system, ion-scale structures were subjected to a larger reduction than those on the electron-scale, which are already relatively smaller to begin with. Moreover, an electron-ion root transition\cite{Yokoyama2007,Klinger2016} occurred in $0.3 \le \hat{\rho} \le 0.4$, causing the largest reduction in ion-scale heat fluxes. Due to this turbulence-suppressing effect, the temperature profiles increased with the inclusion of the shear. The agreement with experimental temperature data improved with respect to phase 2. Both experimental temperature profiles were recovered well by the simulations, as seen in Fig. \ref{fig:phase3}. However, in some cases, $T_e$ can be slightly higher in the inner radial region than the experimental profile, such as in the NBI scenario.

\subsection{\label{sec:results_phase4}Phase 4: Varying Density Profile, with Neoclassical Heat Transport and Neoclassical $\mathrm{E_r \times B}$ shear}

Similar to phase 1, the experimental fit of the density data is used as the initial guess for the \texttt{GENE} and \texttt{KNOSOS} simulations. The boundary values of the temperature and density profiles from the statistical fit were initially used and kept fixed but adjustments were eventually made, as discussed later in this sub-section.

\begin{figure*}[t]
    \centering
    \begin{subfigure}{0.23\textwidth}  
        \centering
        \includegraphics[width=\textwidth]{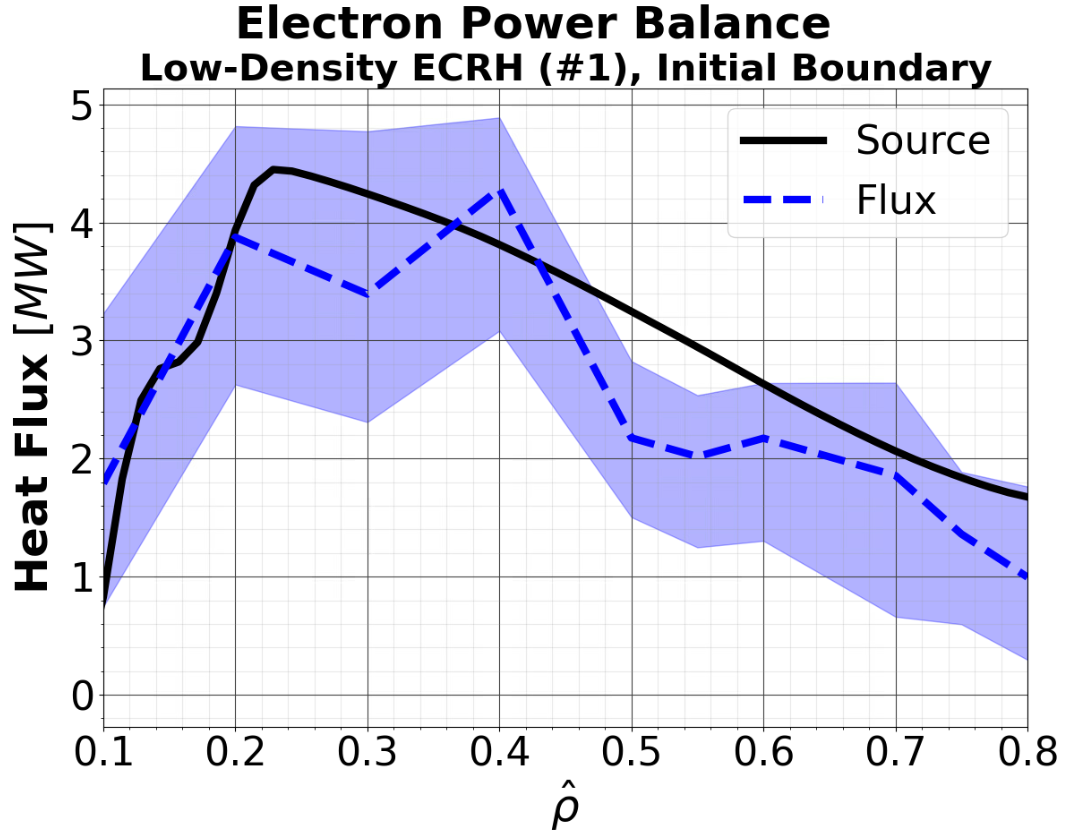} 
        \caption{}
        \label{fig:phase4_base_power_elec_Elow}
    \end{subfigure}
    \hspace{0.3cm}
    \begin{subfigure}{0.23\textwidth}  
        \centering
        \includegraphics[width=\textwidth]{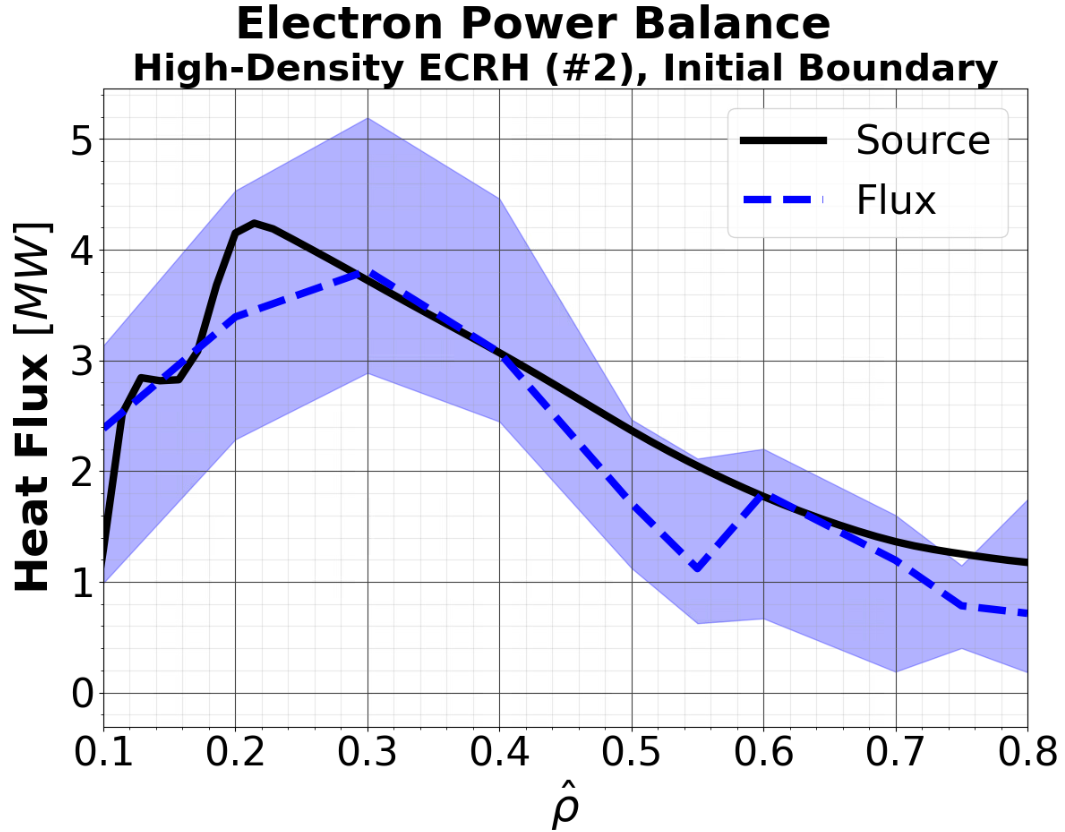}  
        \caption{}
        \label{fig:phase4_base_power_elec_Ehigh}
    \end{subfigure}
    \hspace{0.3cm}
    \begin{subfigure}{0.23\textwidth}  
        \centering
        \includegraphics[width=\textwidth]{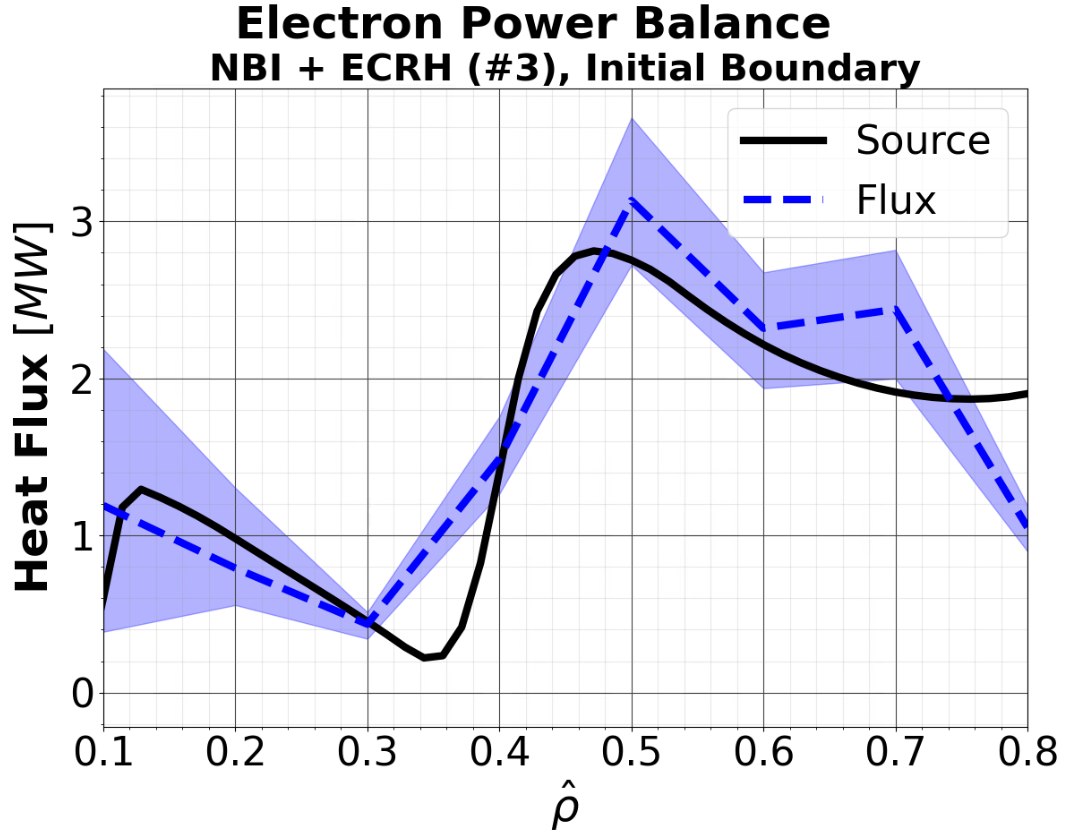}  
        \caption{}
        \label{fig:phase4_base_power_elec_NII}
    \end{subfigure}
    \hspace{0.3cm}
    \begin{subfigure}{0.23\textwidth}  
        \centering
        \includegraphics[width=\textwidth]{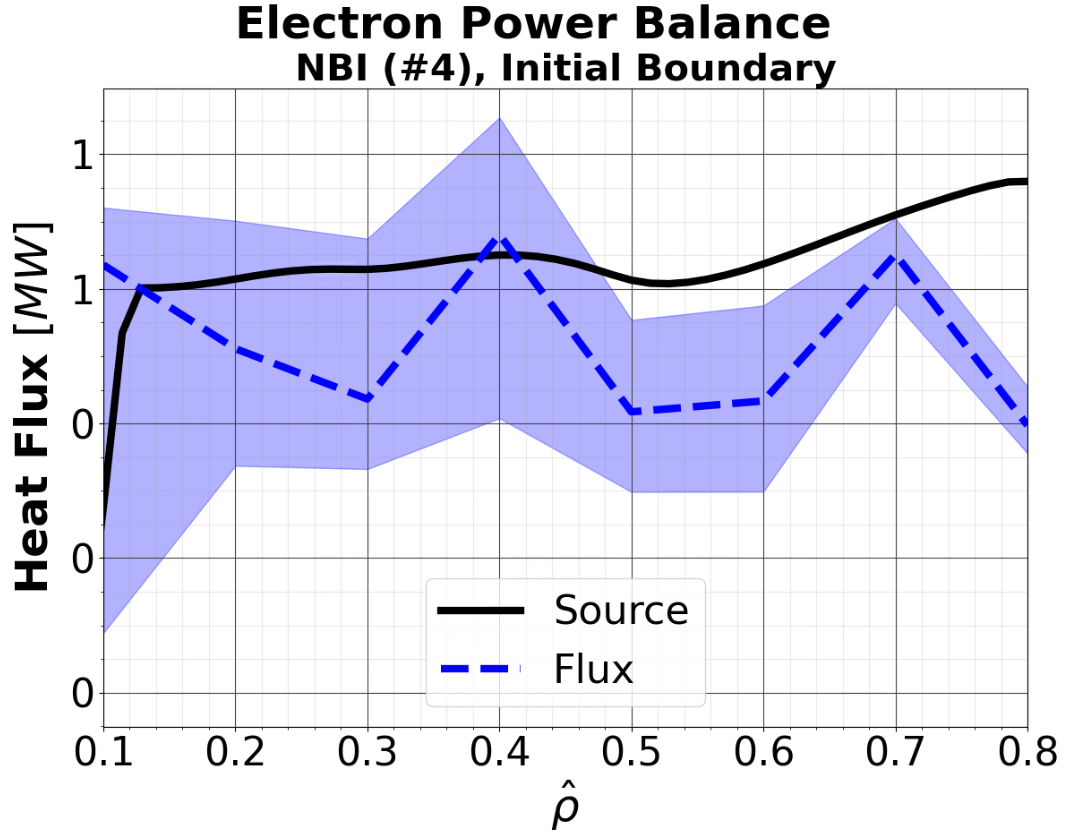}  
        \caption{}
        \label{fig:phase4_base_power_elec_NIII}
    \end{subfigure}
    
    \vspace{0.5cm}

    \begin{subfigure}{0.23\textwidth}  
        \centering
        \includegraphics[width=\textwidth]{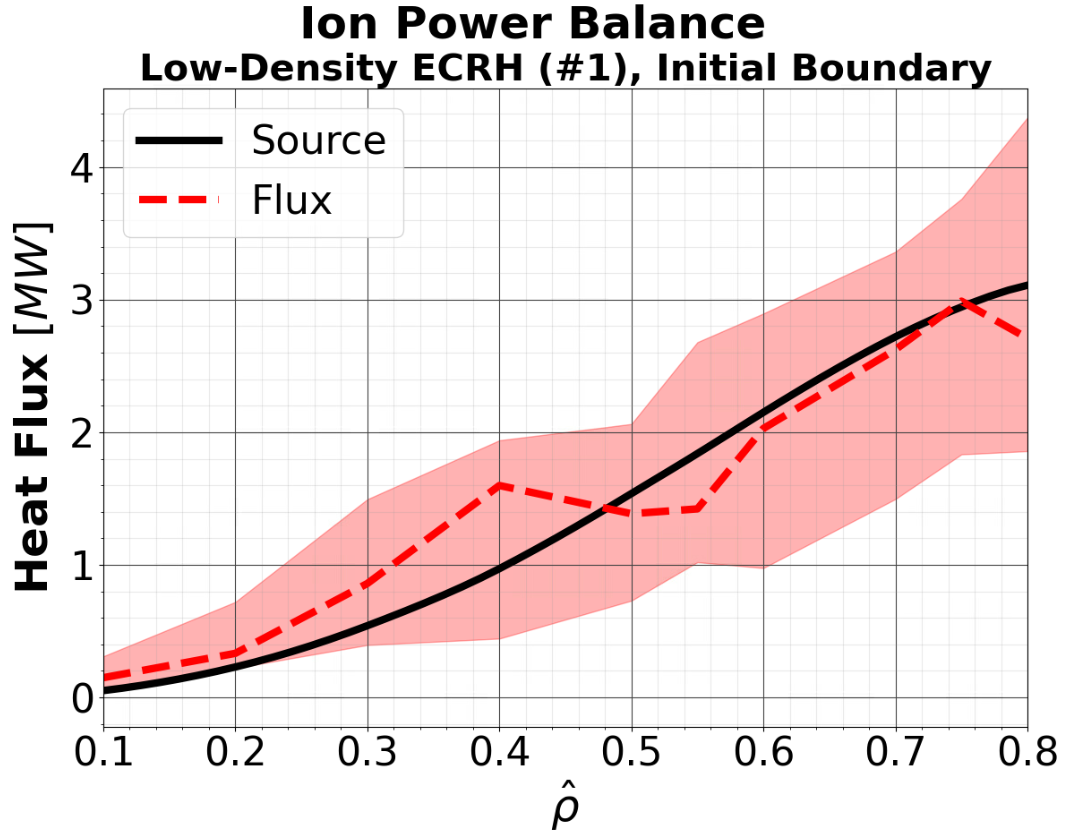} 
        \caption{}
        \label{fig:phase4_base_power_ion_Elow}
    \end{subfigure}
    \hspace{0.3cm}
    \begin{subfigure}{0.23\textwidth}  
        \centering
        \includegraphics[width=\textwidth]{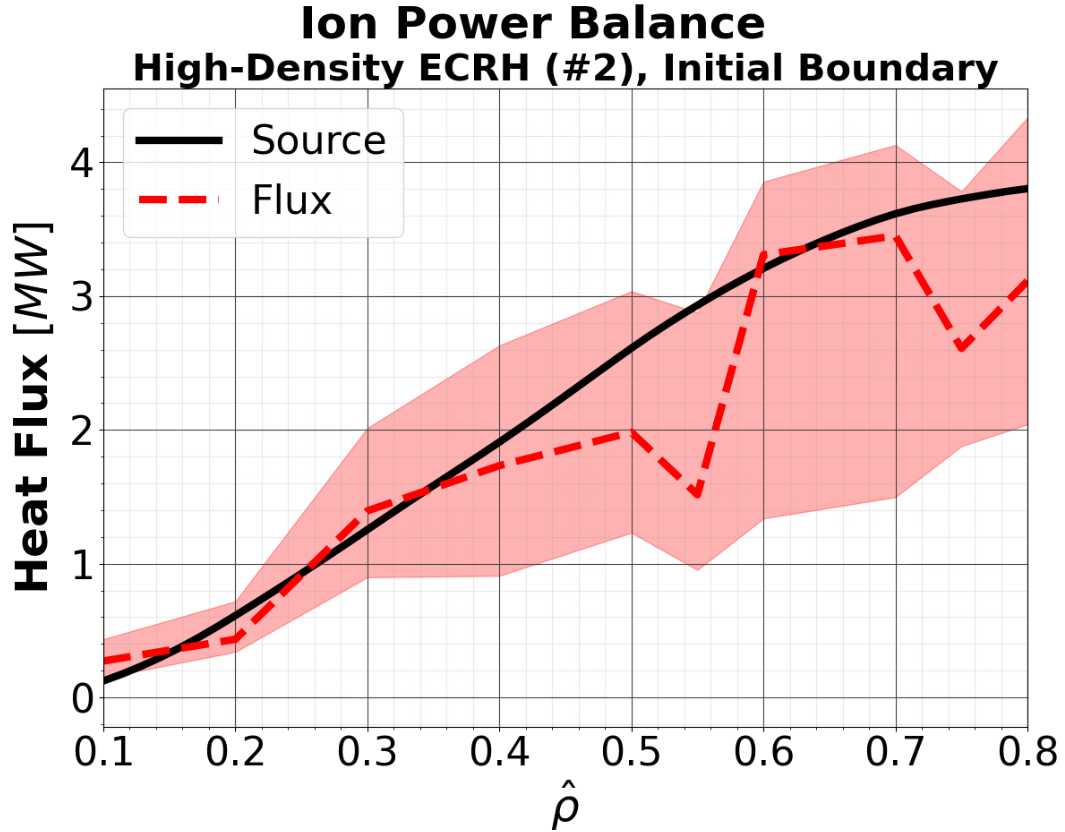}  
        \caption{}
        \label{fig:phase4_base_power_ion_Ehigh}
    \end{subfigure}
    \hspace{0.3cm}
    \begin{subfigure}{0.23\textwidth}  
        \centering
        \includegraphics[width=\textwidth]{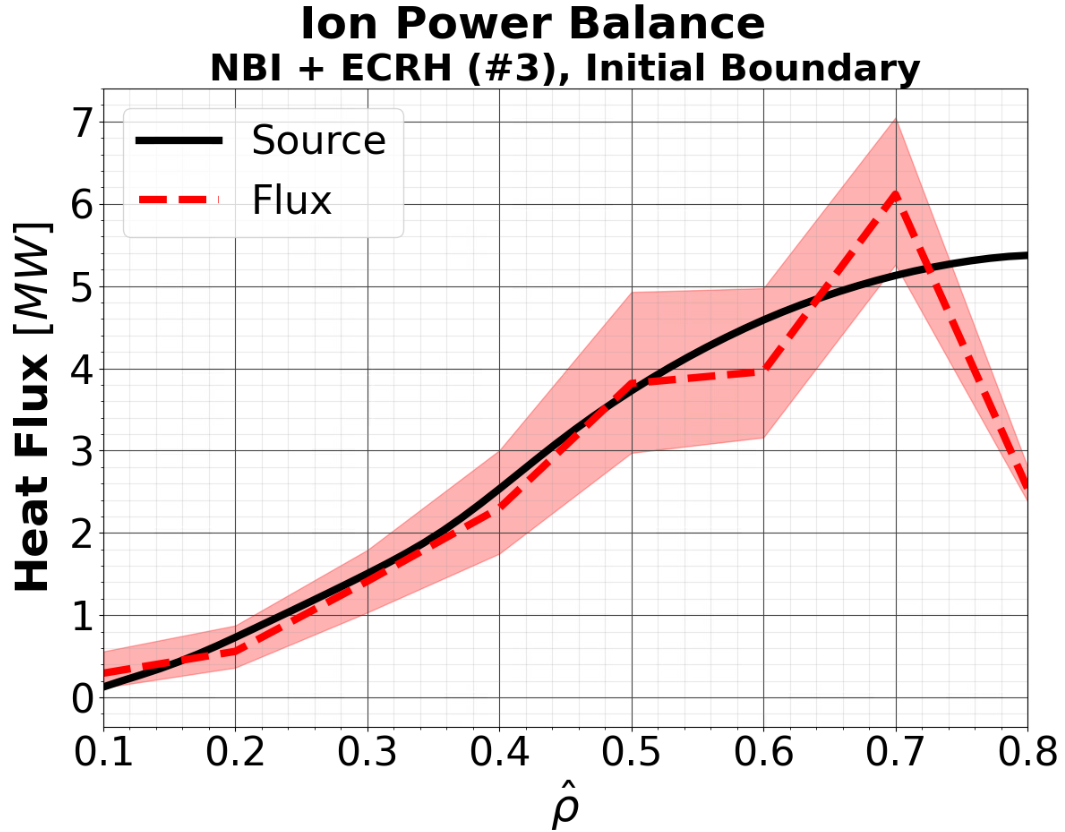}  
        \caption{}
        \label{fig:phase4_base_power_ion_NII}
    \end{subfigure}
    \hspace{0.3cm}
    \begin{subfigure}{0.23\textwidth}  
        \centering
        \includegraphics[width=\textwidth]{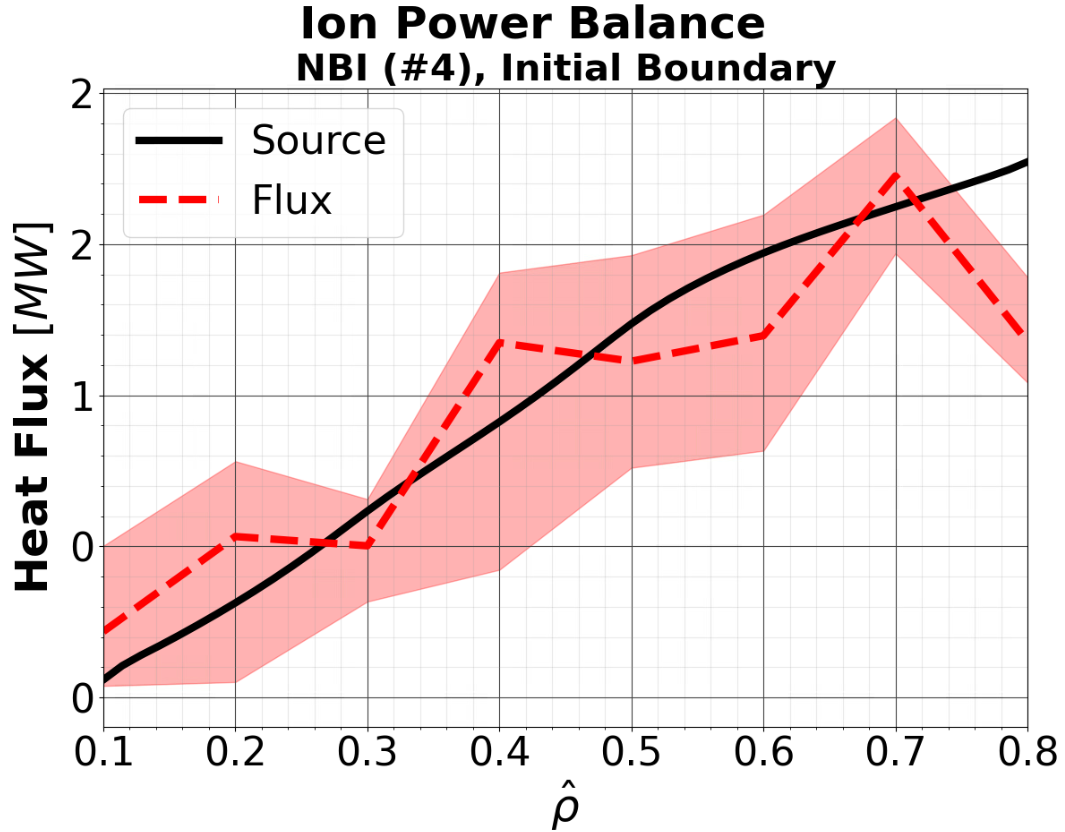}  
        \caption{}
        \label{fig:phase4_base_power_ion_NIII}
    \end{subfigure}
    
    \vspace{0.5cm}
    
    \begin{subfigure}{0.23\textwidth}  
        \centering
        \includegraphics[width=\textwidth]{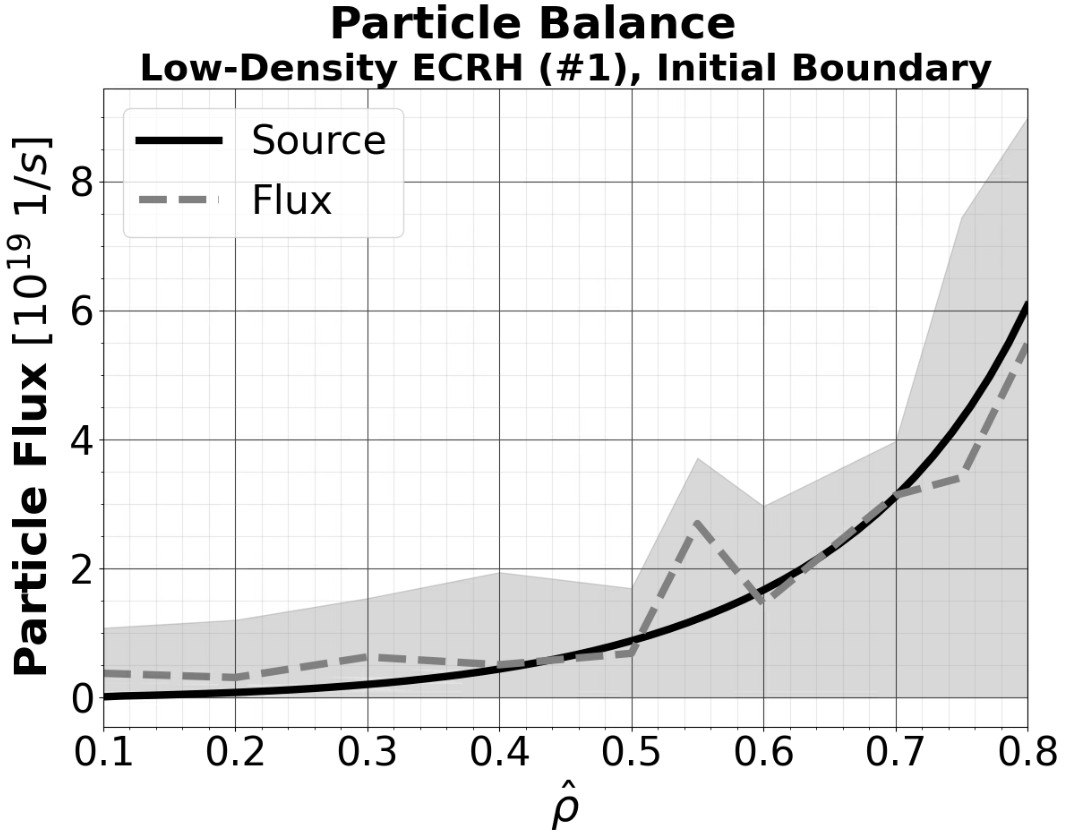} 
        \caption{}
        \label{fig:phase4_base_particle_Elow}
    \end{subfigure}
    \hspace{0.3cm}
    \begin{subfigure}{0.23\textwidth}  
        \centering
        \includegraphics[width=\textwidth]{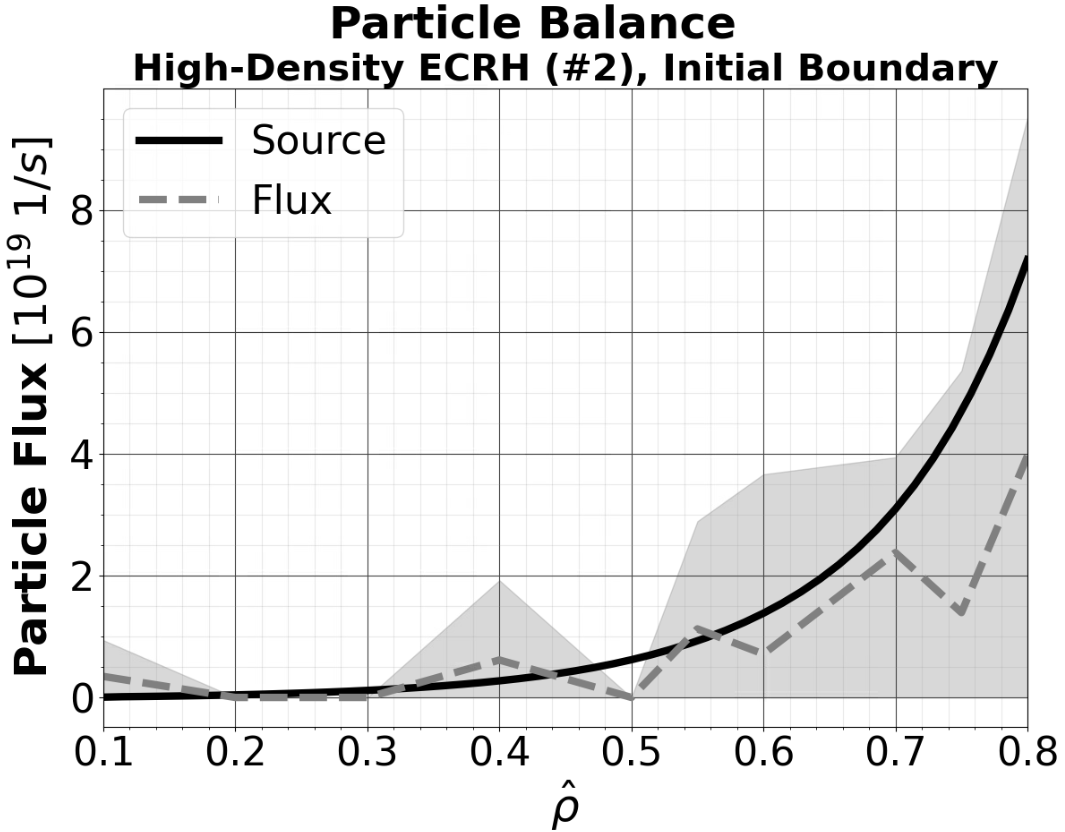}  
        \caption{}
        \label{fig:phase4_base_particle_Ehigh}
    \end{subfigure}
    \hspace{0.3cm}
    \begin{subfigure}{0.23\textwidth}
        \centering
        \includegraphics[width=\textwidth]{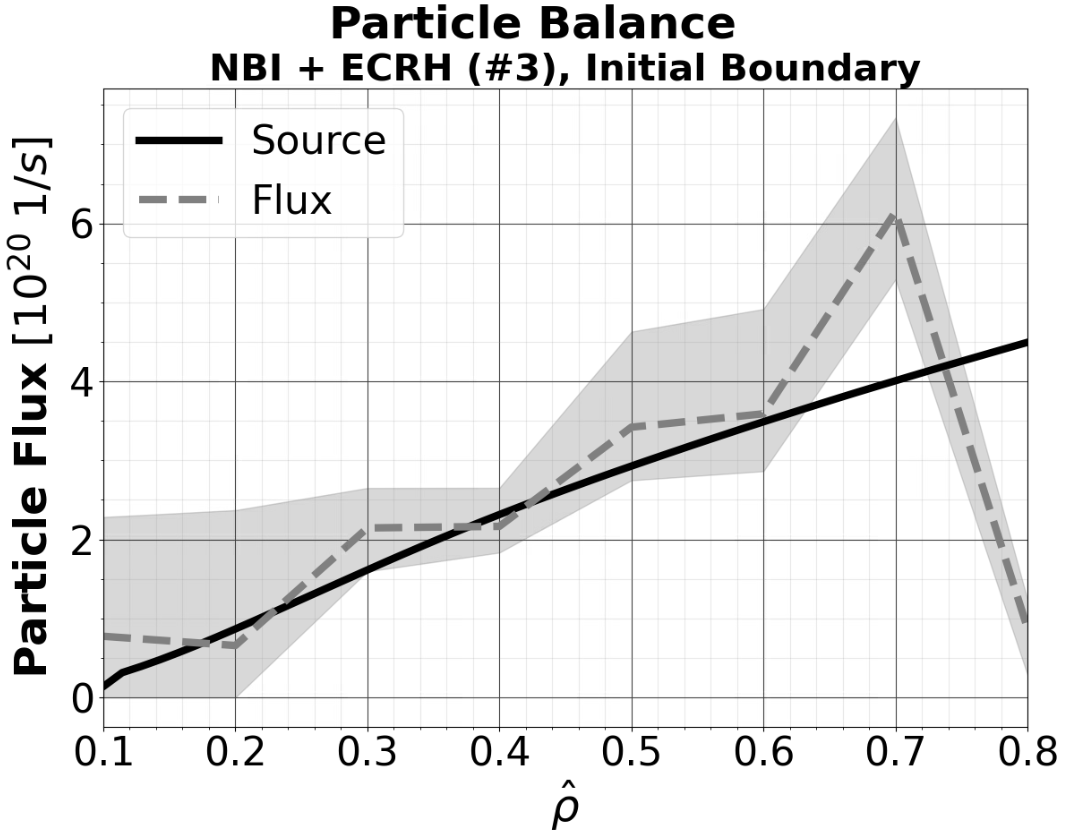}
        \caption{}
        \label{fig:phase4_base_particle_NII}
    \end{subfigure}
    \hspace{0.3cm}
    \begin{subfigure}{0.23\textwidth}
        \centering
        \includegraphics[width=\textwidth]{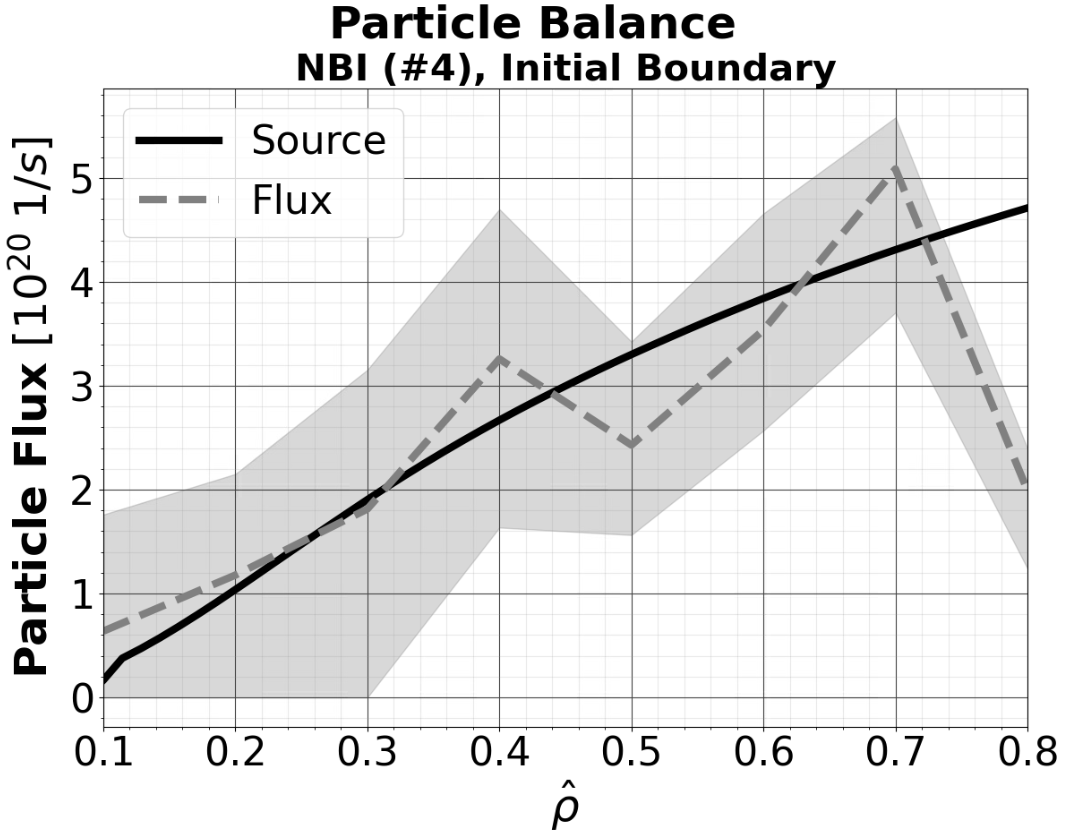}
        \caption{}
        \label{fig:phase4_base_particle_NIII}
    \end{subfigure}
    
    \caption{Power and particle balances of the low-density ECRH (\subref{fig:phase4_base_power_elec_Elow},  \subref{fig:phase4_base_power_ion_Elow}, and \subref{fig:phase4_base_particle_Elow}), high-density ECRH (\subref{fig:phase4_base_power_elec_Ehigh}, \subref{fig:phase4_base_power_ion_Ehigh}, and \subref{fig:phase4_base_particle_Ehigh}), NBI + ECRH (\subref{fig:phase4_base_power_elec_NII}, \subref{fig:phase4_base_power_ion_NII}, and \subref{fig:phase4_base_particle_NII}), and NBI (\subref{fig:phase4_base_power_elec_NIII}, \subref{fig:phase4_base_power_ion_NIII}, and \subref{fig:phase4_base_particle_NIII}) scenarios using the boundaries of the profile data fits. Two points, specifically $\hat{\rho} = 0.55$ and $0.75$, were added for the low- and high-density ECRH cases due to convergence problems in the particle balance.}
    \label{fig:phase4_base_balances}
\end{figure*}

\subsubsection{\label{sec:results_phase4_base}Fixed Profile Boundaries with 
$\bm{n_{0,\text{edge}}\;=\;10^{13}}$~$\bm{m^{-3}}$}
\begin{figure*}
    \centering
    \begin{subfigure}{0.43\textwidth}  
        \centering
        \includegraphics[width=\textwidth]{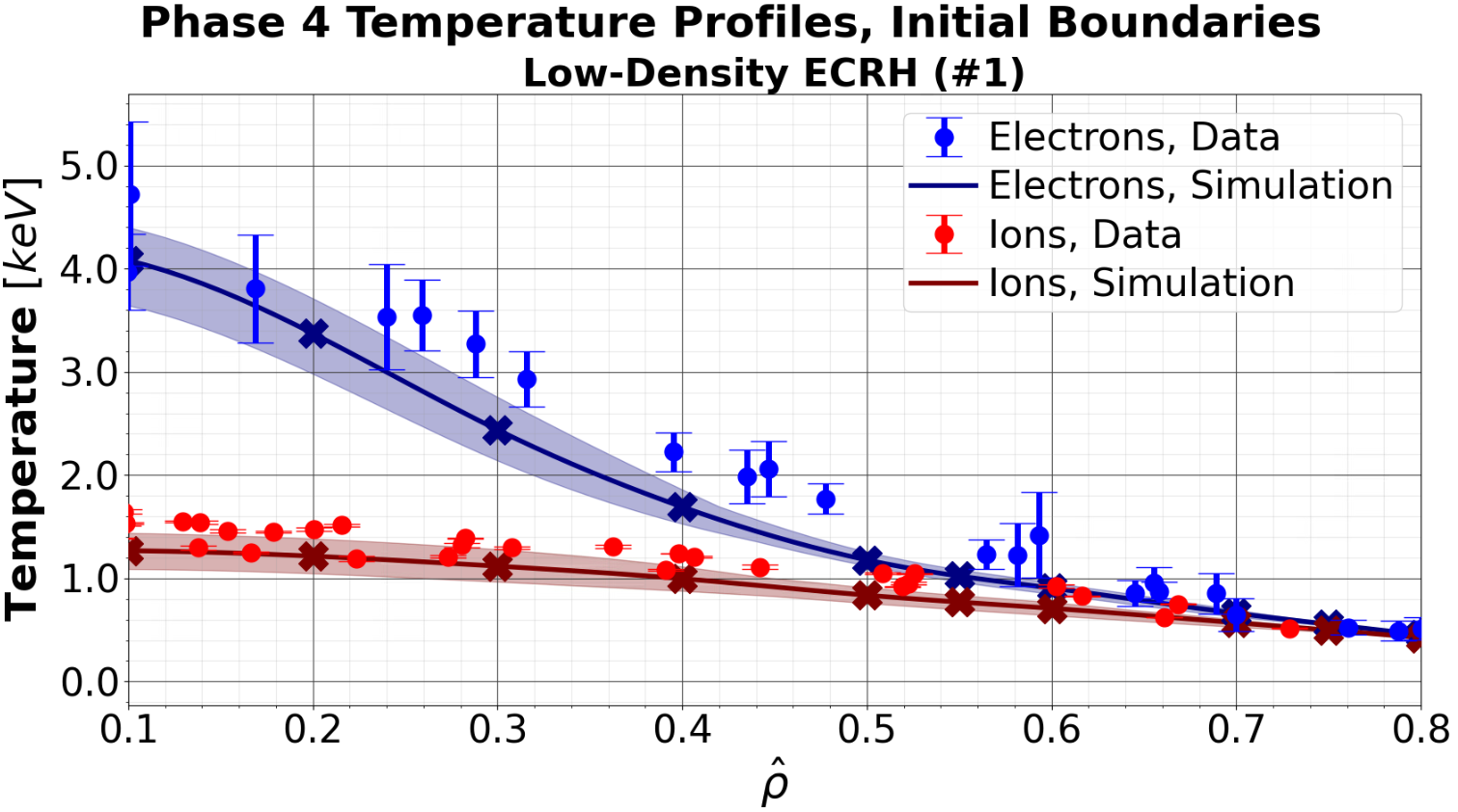} 
        \caption{}
        \label{fig:phase4_base_T_Elow}
    \end{subfigure}
    \hspace{0.5cm}
    \begin{subfigure}{0.43\textwidth}  
        \centering
        \includegraphics[width=\textwidth]{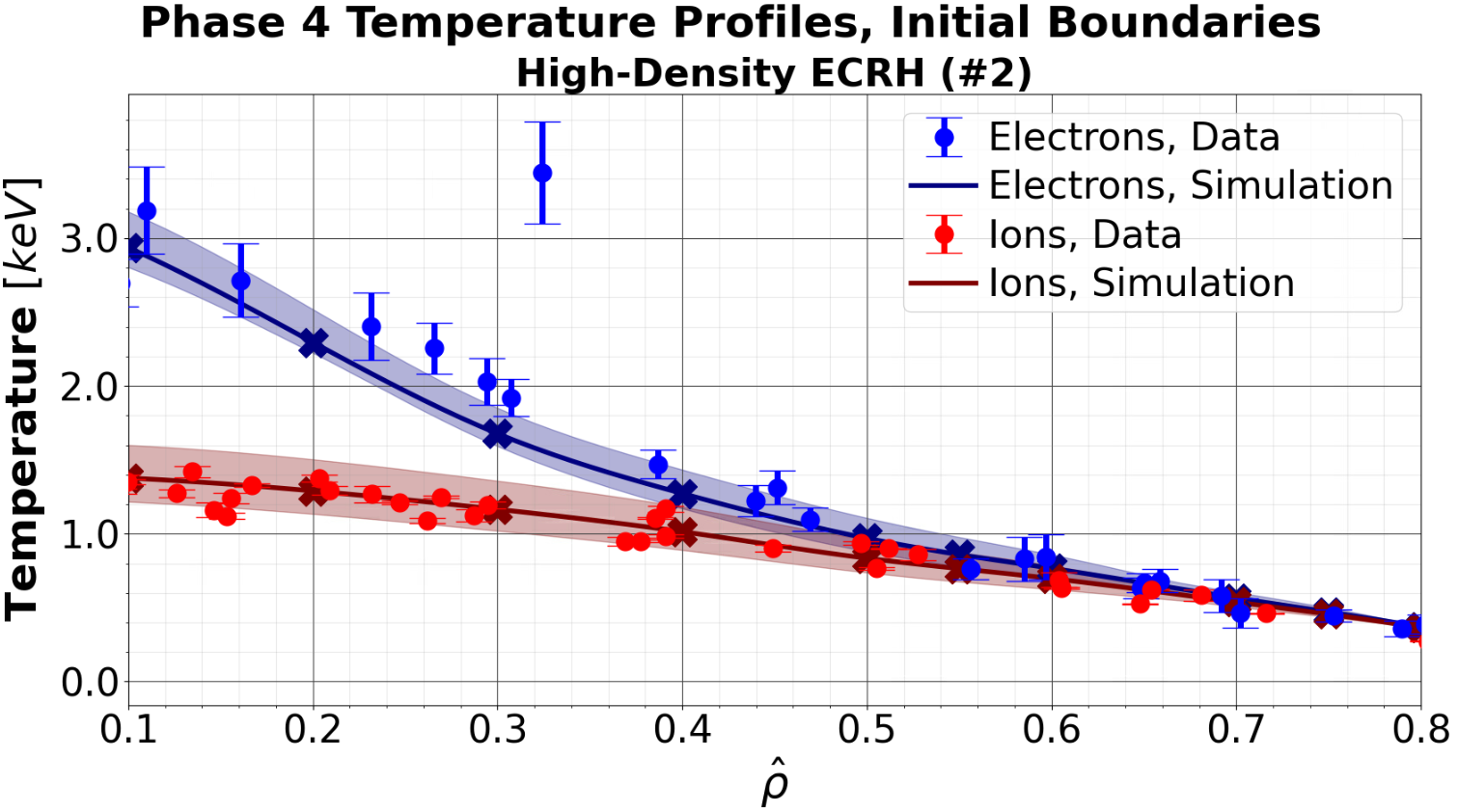}  
        \caption{}
        \label{fig:phase4_base_T_Ehigh}
    \end{subfigure}
    \vspace{0.5cm}  
    \begin{subfigure}{0.43\textwidth}
        \centering
        \includegraphics[width=\textwidth]{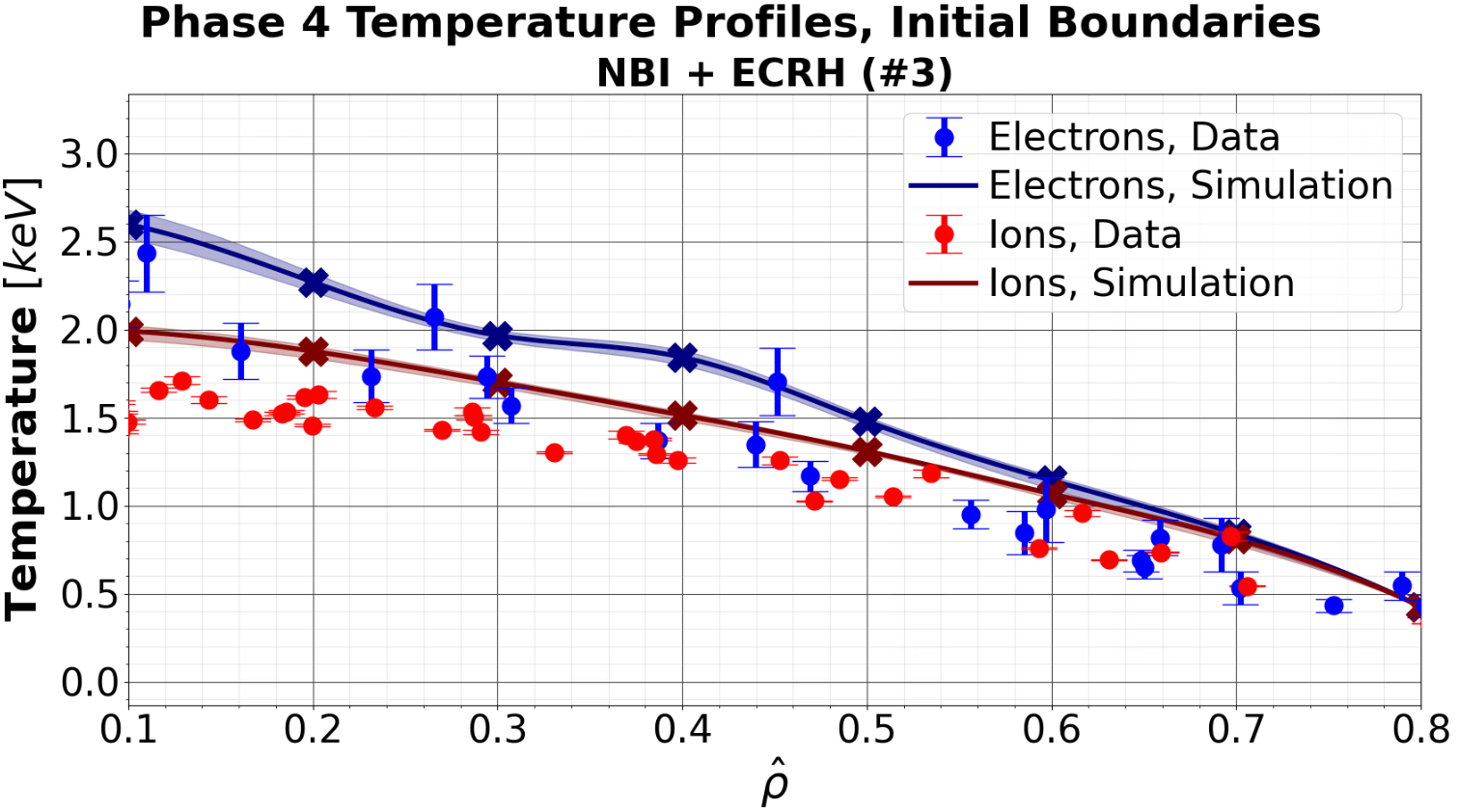}
        \caption{}
        \label{fig:phase4_base_T_NII}
    \end{subfigure}
    \hspace{0.5cm}
    \begin{subfigure}{0.43\textwidth}
        \centering
        \includegraphics[width=\textwidth]{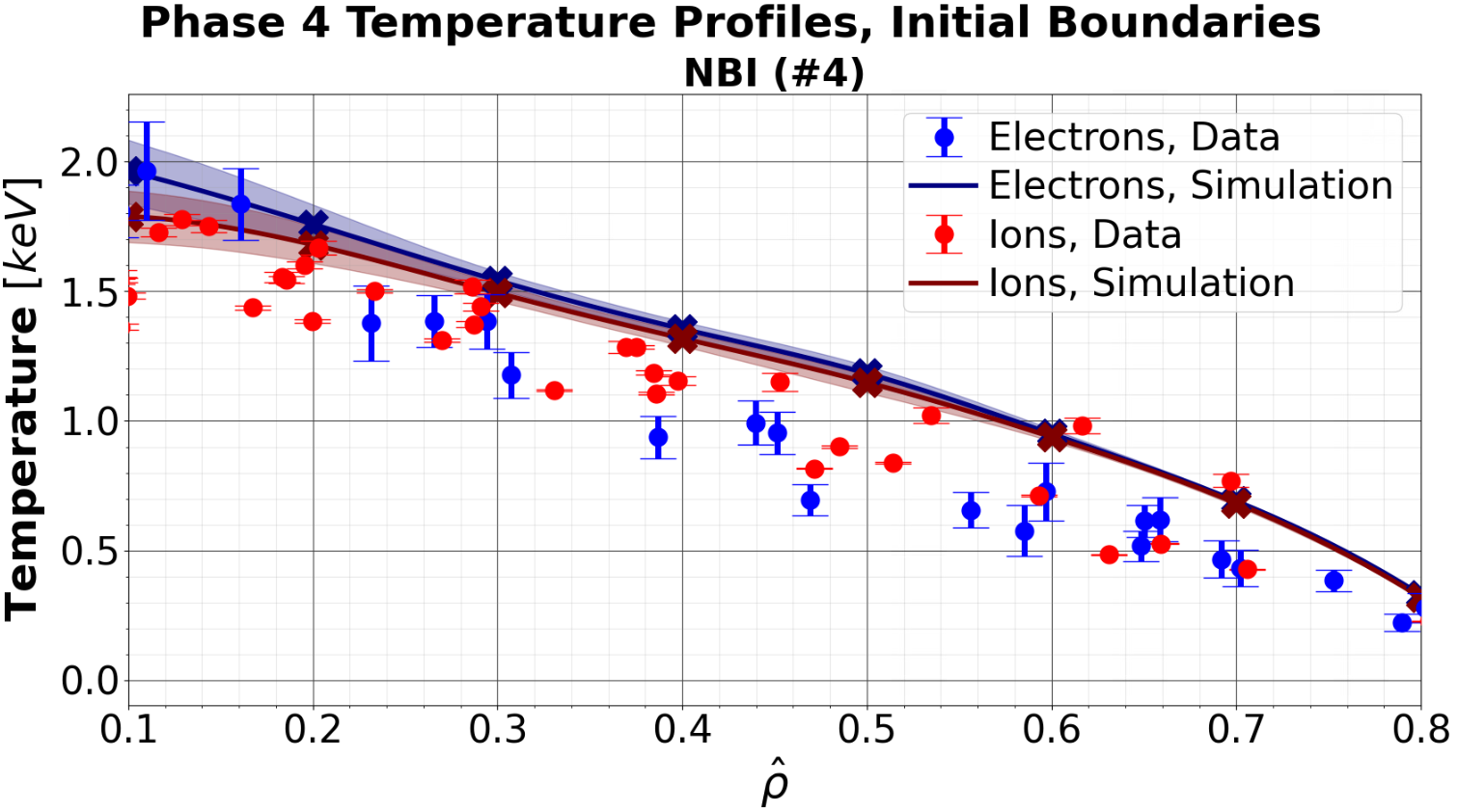}
        \caption{}
        \label{fig:phase4_base_T_NIII}
    \end{subfigure}
    \vspace{0.5cm}
    \begin{subfigure}{0.43\textwidth}  
        \centering
        \includegraphics[width=\textwidth]{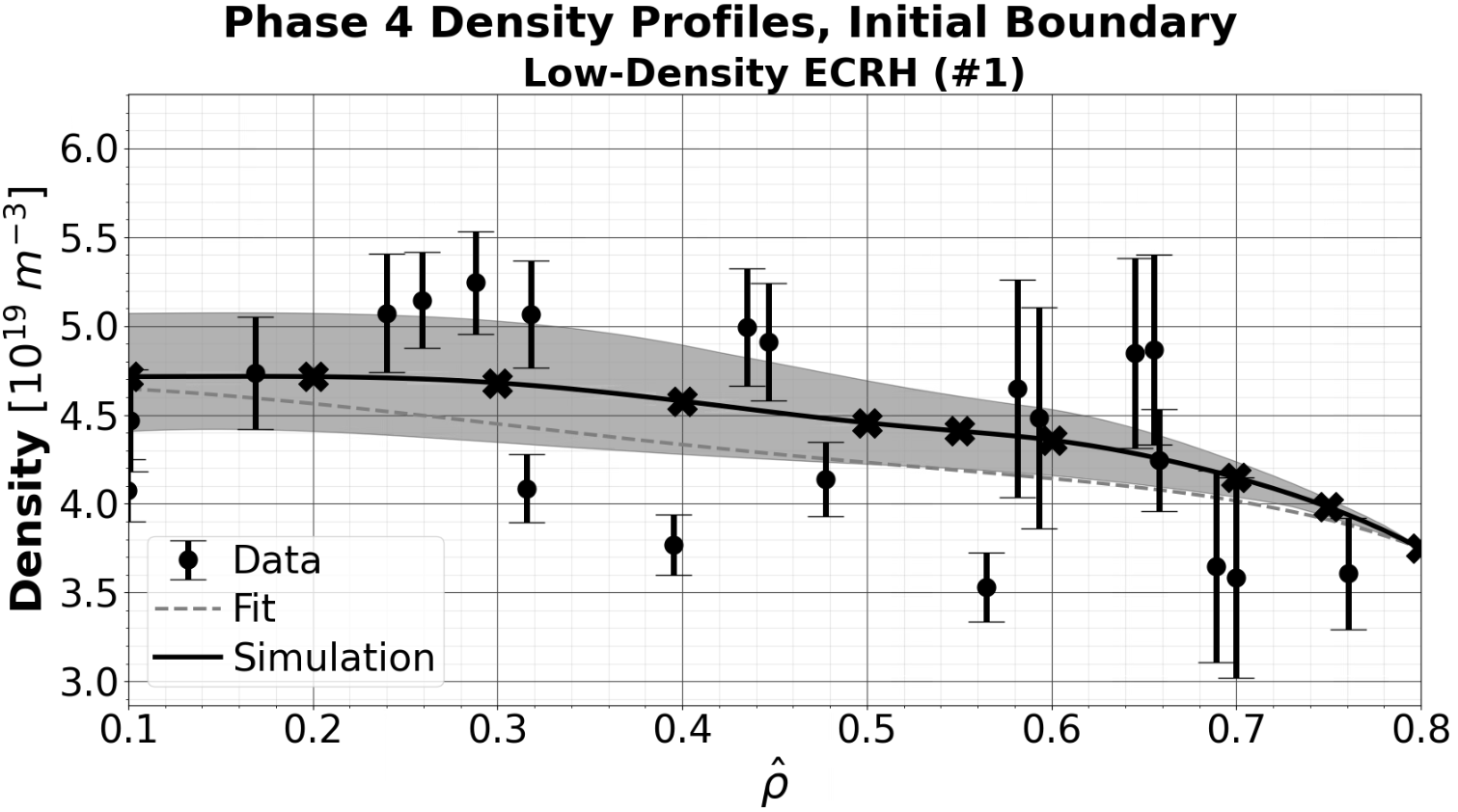} 
        \caption{}
        \label{fig:phase4_base_n_Elow}
    \end{subfigure}
    \hspace{0.5cm}
    \begin{subfigure}{0.43\textwidth}  
        \centering
        \includegraphics[width=\textwidth]{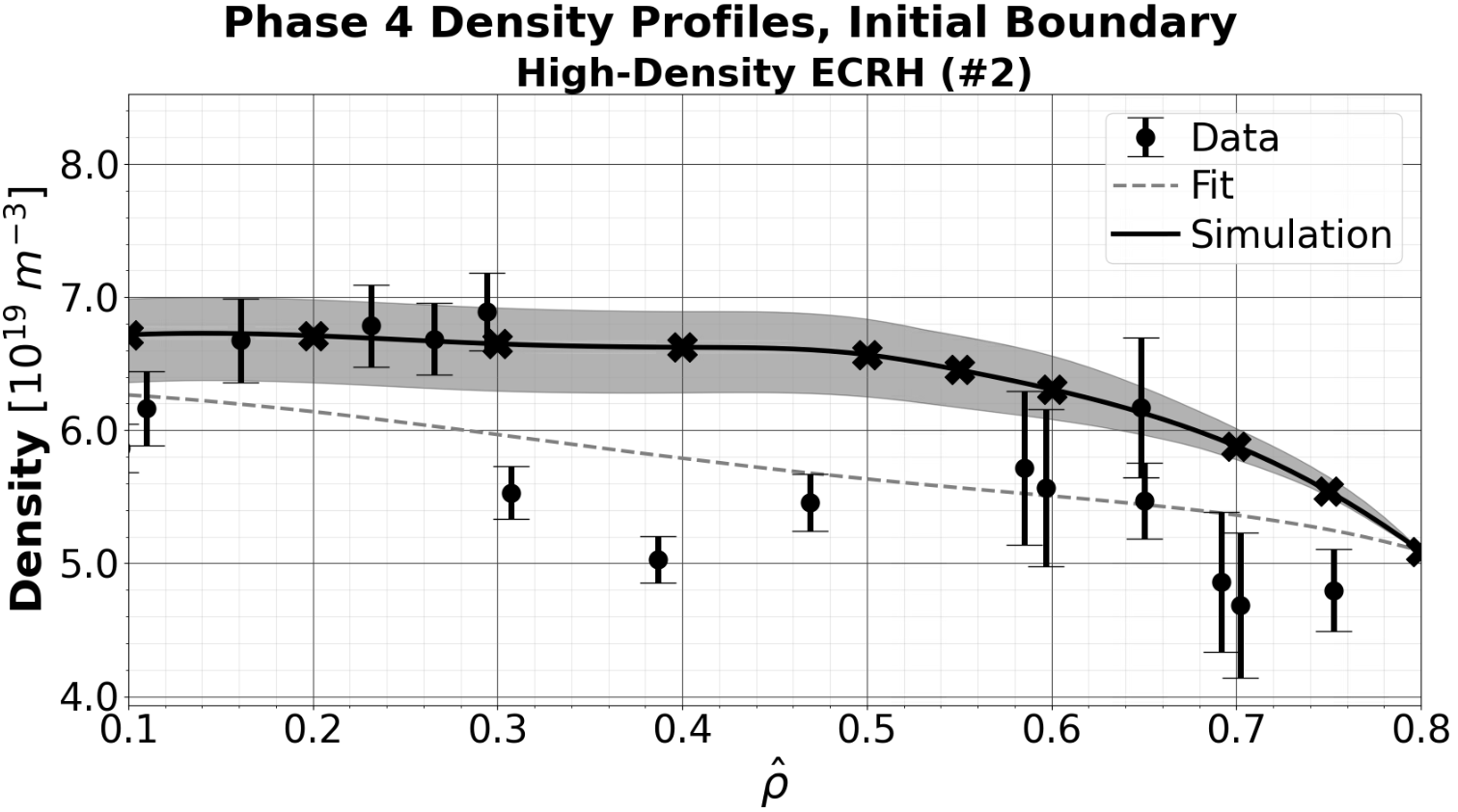}  
        \caption{}
        \label{fig:phase4_base_n_Ehigh}
    \end{subfigure}
    \hspace{0.5cm}
    \begin{subfigure}{0.43\textwidth}
        \centering
        \includegraphics[width=\textwidth]{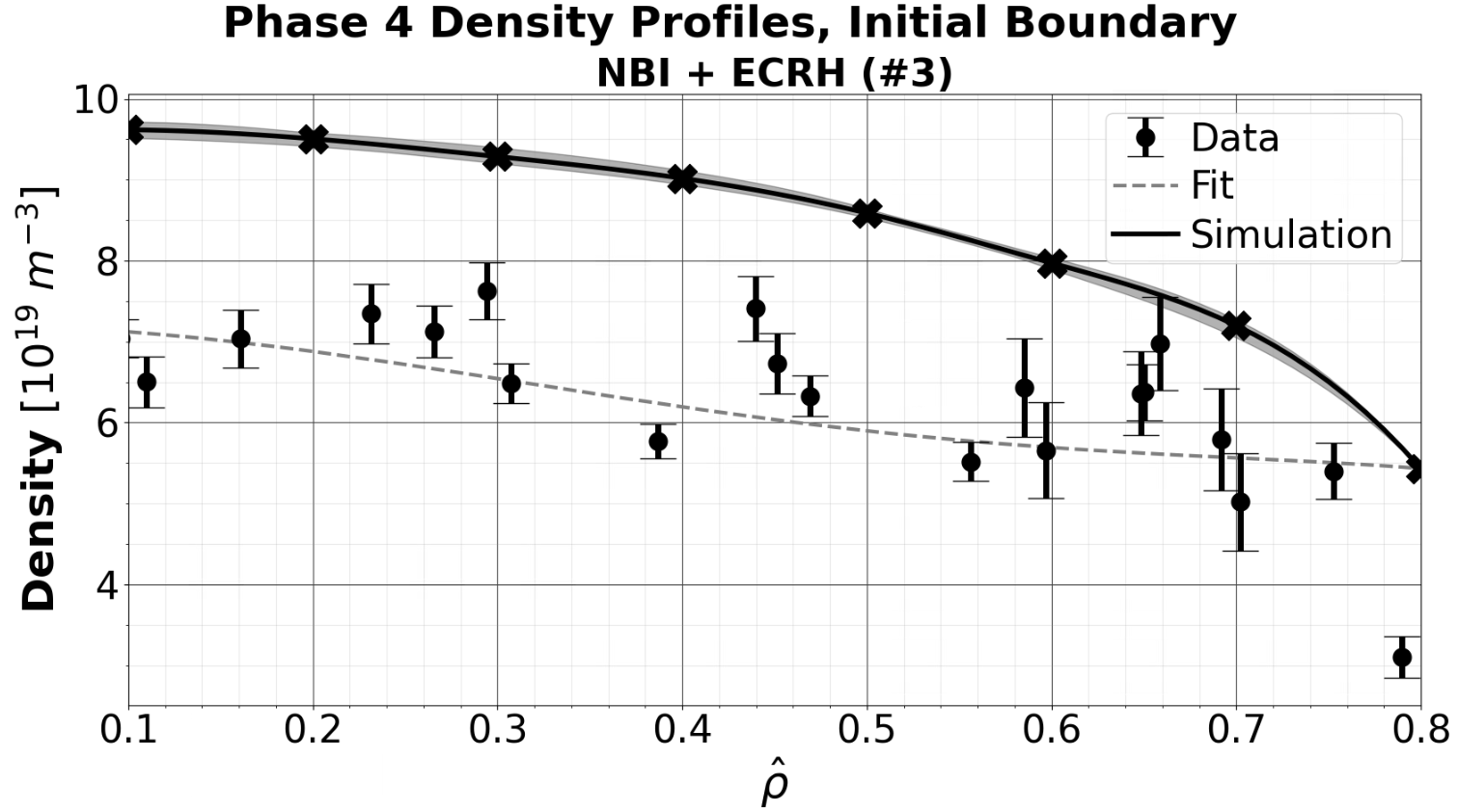}
        \caption{}
        \label{fig:phase4_base_n_NII}
    \end{subfigure}
    \hspace{0.5cm}
    \begin{subfigure}{0.43\textwidth}
        \centering
        \includegraphics[width=\textwidth]{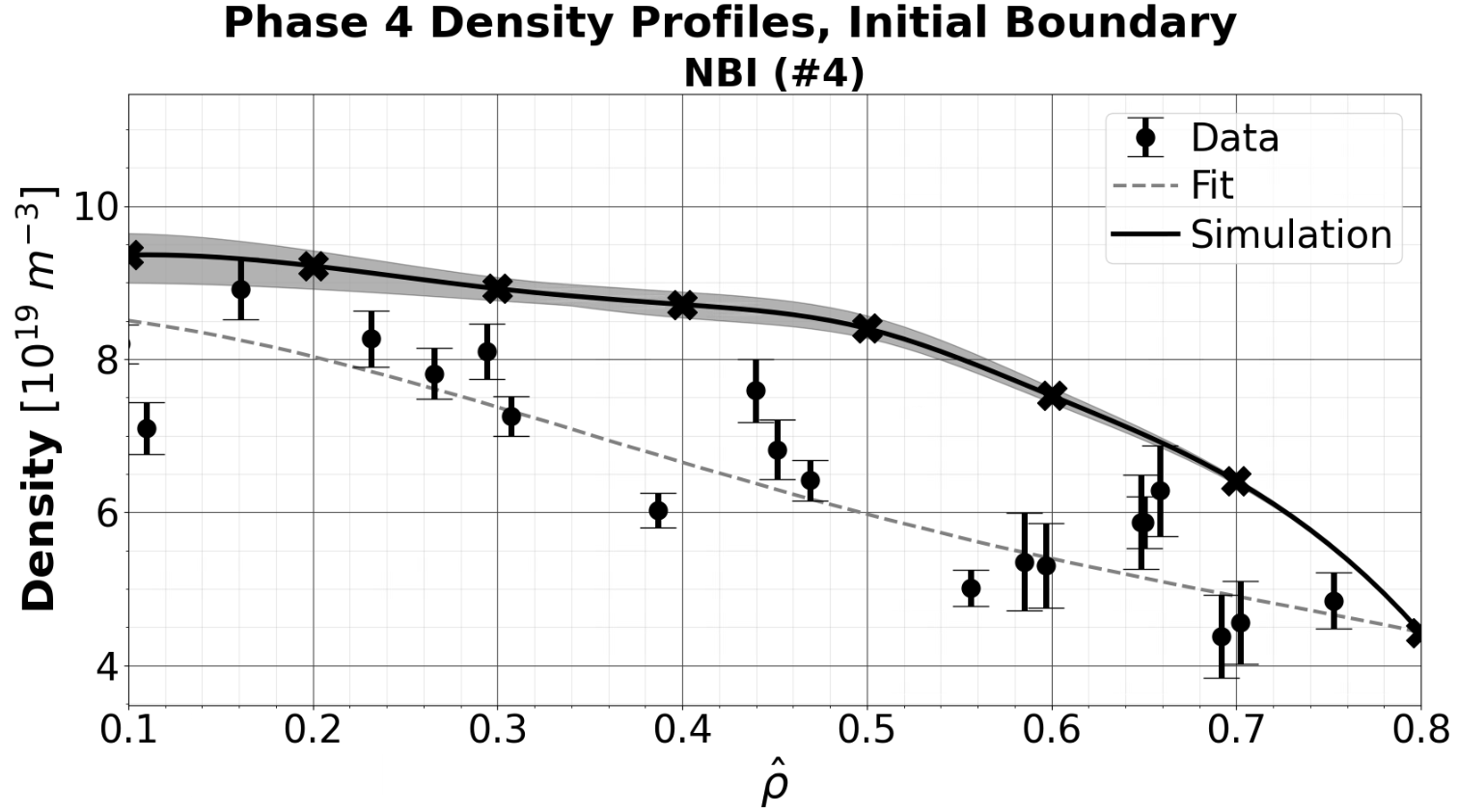}
        \caption{}
        \label{fig:phase4_base_n_NIII}
    \end{subfigure}
    
    \caption{Converged temperature and density profiles of the low-density ECRH (\subref{fig:phase4_base_T_Elow} and \subref{fig:phase4_base_n_Elow}), high-density ECRH (\subref{fig:phase4_base_T_Ehigh} and \subref{fig:phase4_base_n_Ehigh}), NBI + ECRH (\subref{fig:phase4_base_T_NII} and \subref{fig:phase4_base_n_NII}), and NBI (\subref{fig:phase4_base_T_NIII} and \subref{fig:phase4_base_n_NIII}) scenarios using the boundaries of the profile data fits. Agreement with the experimental data could be improved, especially for the two NBI scenarios as seen in (\subref{fig:phase4_base_T_NII}), (\subref{fig:phase4_base_T_NIII}), (\subref{fig:phase4_base_n_NII}), and (\subref{fig:phase4_base_n_NIII}).}
    \label{fig:phase4_base_Tn}
\end{figure*}

Several complications were encountered in this phase. First, the low- and high-density ECRH cases could not match the particle flux $\Gamma$ at all positions consistently. Small changes in the plasma profiles led to large swings in the magnitude and sign of $\Gamma$, while some positions persistently had a negative total $\Gamma$. To remedy this convergence problem in the particle balance, two points, specifically $\hat{\rho} = 0.55$ and $0.75$, were added for the low- and high-density ECRH cases. The two ECRH particle balances consequently improved, and the converged results for all four scenarios are shown in Fig. \ref{fig:phase4_base_balances}. \\

However, disagreements are present between the experimental data and simulated plasma profiles. Although the two ECRH cases showed satisfactory agreement with some of the profile data points, the same could not be said for the two NBI cases. The $T_e$, $T_i$, and $n_e$ profiles shown in Figs. \ref{fig:phase4_base_T_NII}, \ref{fig:phase4_base_n_NII}, \ref{fig:phase4_base_T_NIII}, and \ref{fig:phase4_base_n_NIII} diverged from the experimental data. This was more pronounced for the density profiles, especially in the inner positions. \texttt{Tango} determined that a higher density gradient was necessary toward the edge of the radial domain, at $0.5 \le \hat{\rho} \le 0.8$. This resulted from encountering negative or inward $\Gamma$ at these positions. More specifically, for the first iteration of phase 4 or the standalone simulation, the total $\Gamma$ at $\hat{\rho} \ge 0.5$ were immediately negative for all four scenarios. This result was not compatible with the positive particle sources at these positions coming from neutrals ionization and/or NBI fueling. To recover total positive $\Gamma$ from the \texttt{KNOSOS} and \texttt{GENE} simulations, \texttt{Tango} increased the density gradient at $\hat{\rho} \ge 0.5$. On the other hand, to match the negligible $\Gamma$ in the inner radial positions, it was flattened for $\hat{\rho} \le 0.3$. This is consistent with the requirement of having large positive turbulent $\Gamma$ for $\hat{\rho} \ge 0.5$ to compensate for the smaller neoclassical $\Gamma$ in this region.\cite{Thienpondt2023}

\begin{figure*}[t]
    \centering
    \begin{subfigure}{0.23\textwidth}  
        \centering
        \includegraphics[width=\textwidth]{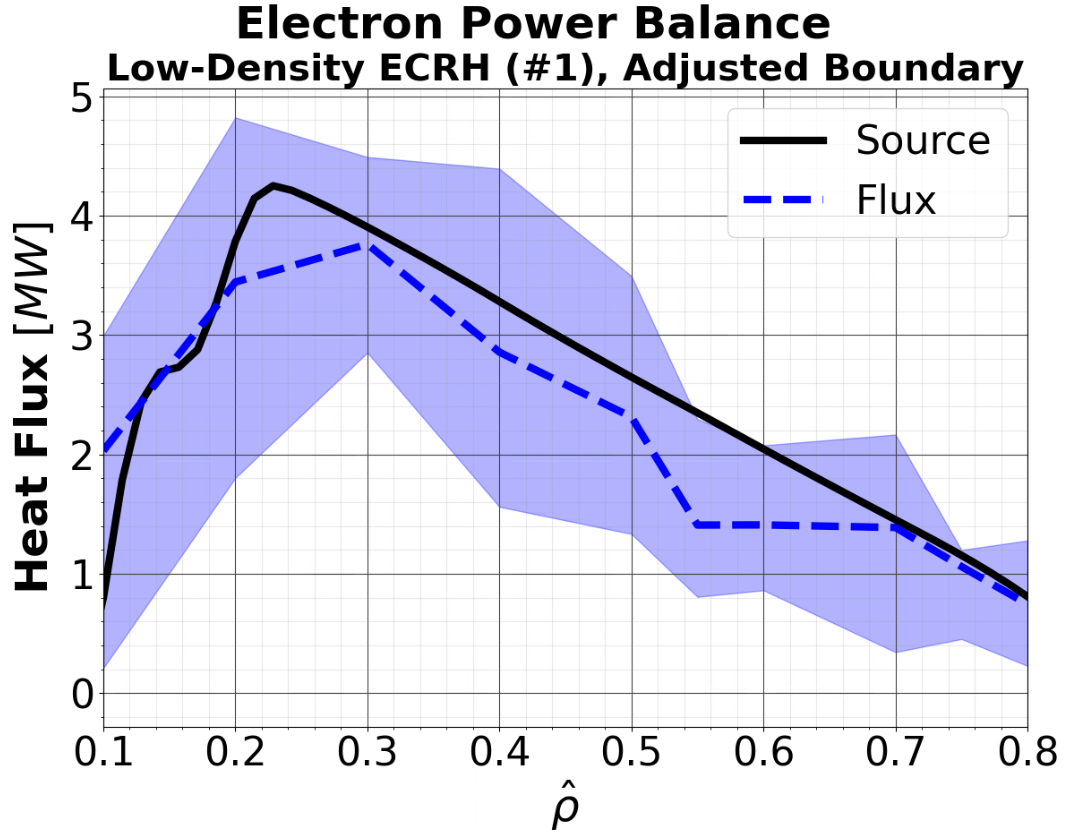} 
        \caption{}
        \label{fig:phase4_mod_power_elec_Elow}
    \end{subfigure}
    \hspace{0.3cm}
    \begin{subfigure}{0.23\textwidth}  
        \centering
        \includegraphics[width=\textwidth]{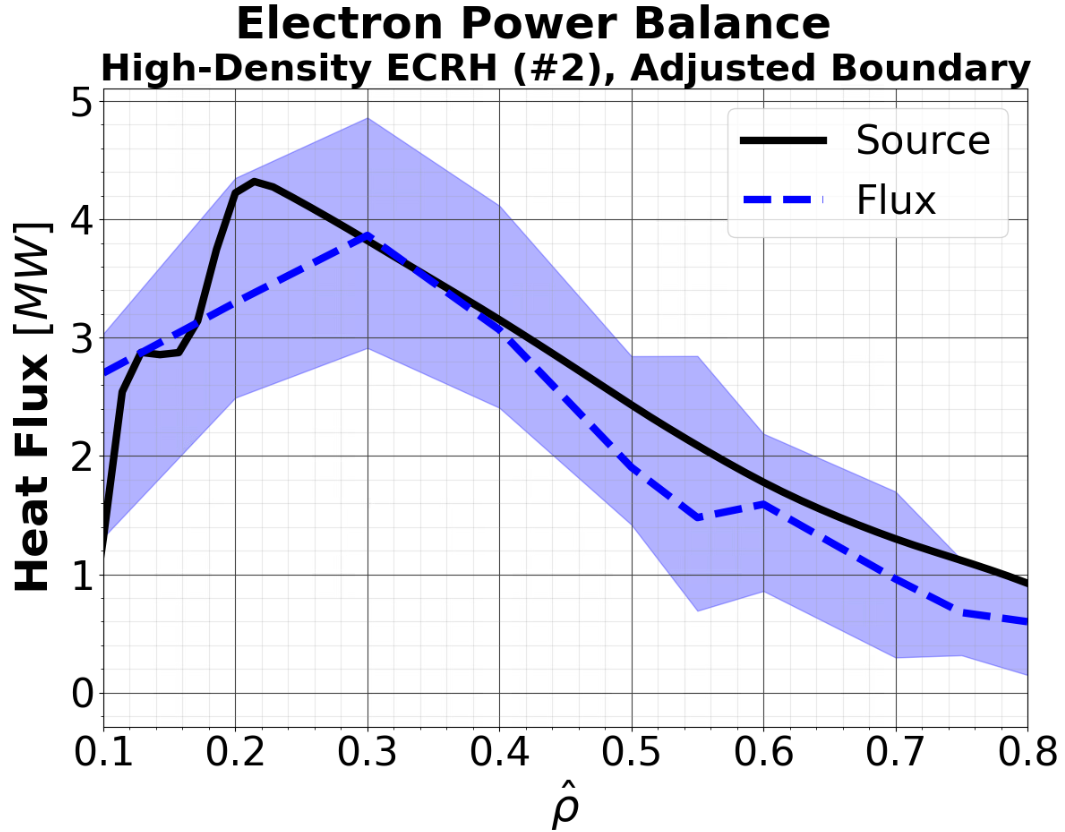}  
        \caption{}
        \label{fig:phase4_mod_power_elec_Ehigh}
    \end{subfigure}
    \hspace{0.3cm}
    \begin{subfigure}{0.23\textwidth}  
        \centering
        \includegraphics[width=\textwidth]{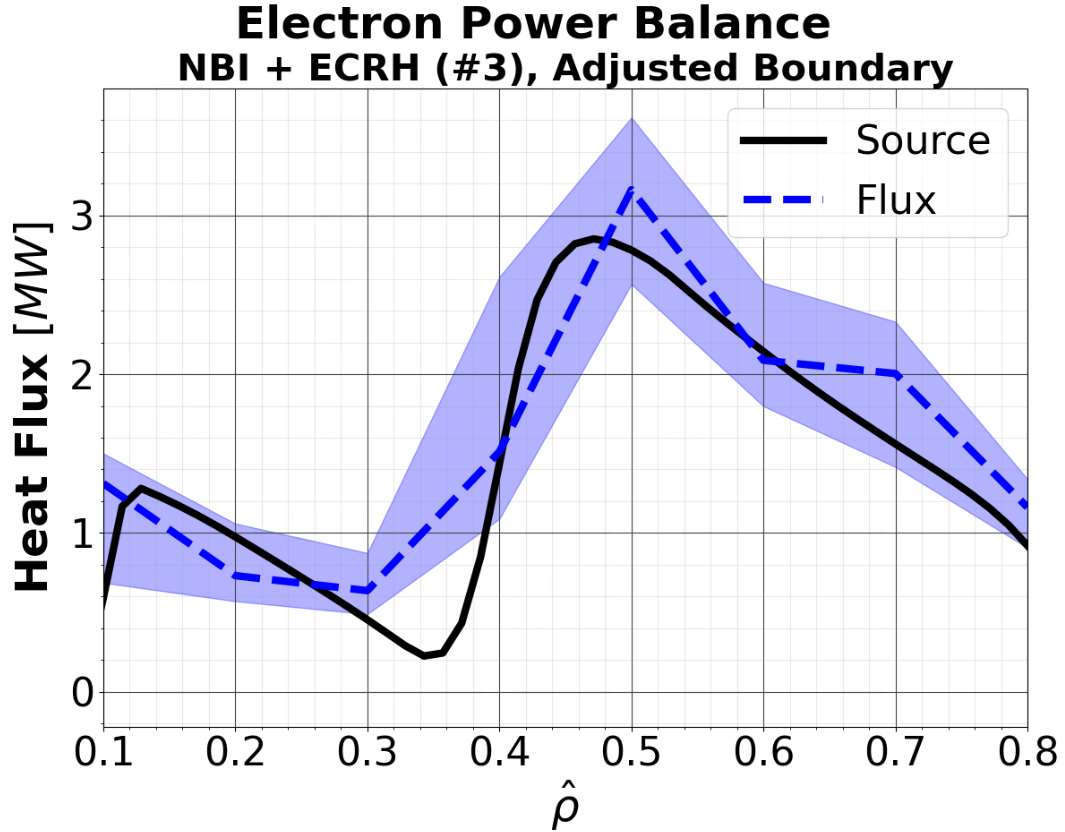}  
        \caption{}
        \label{fig:phase4_mod_power_elec_NII}
    \end{subfigure}
    \hspace{0.3cm}
    \begin{subfigure}{0.23\textwidth}  
        \centering
        \includegraphics[width=\textwidth]{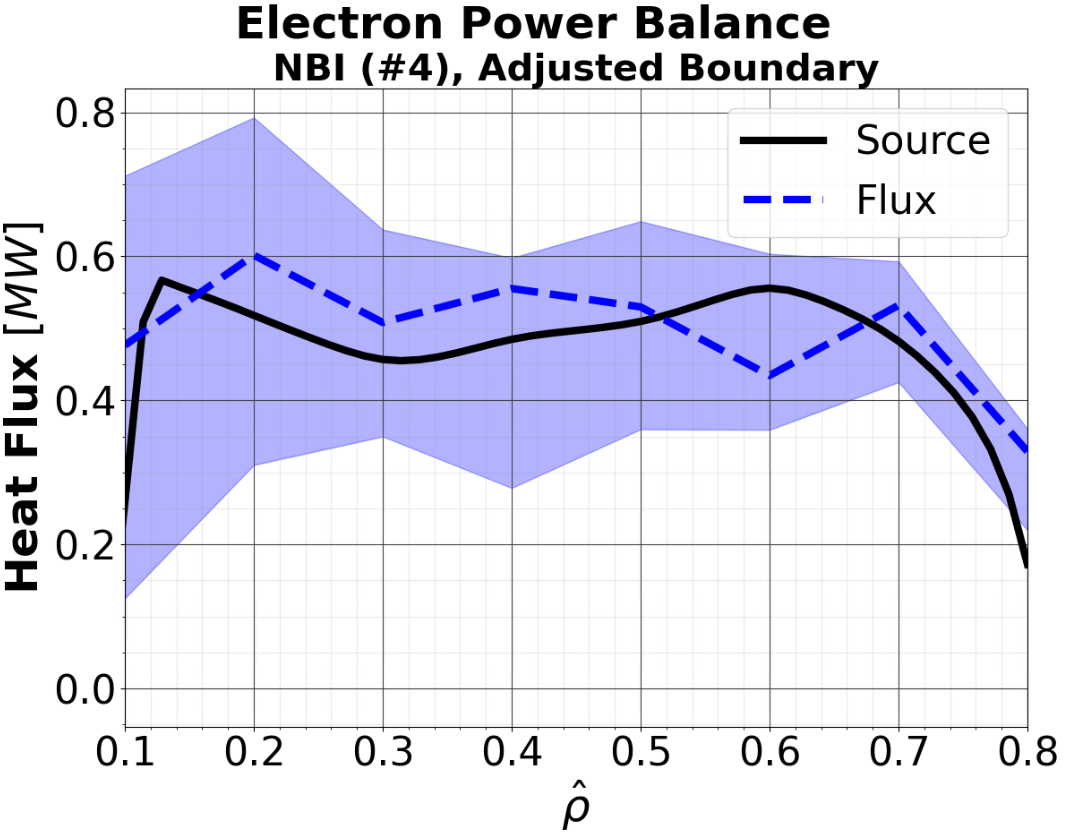}  
        \caption{}
        \label{fig:phase4_mod_power_elec_NIII}
    \end{subfigure}
    
    \vspace{0.5cm}

        \begin{subfigure}{0.23\textwidth}  
        \centering
        \includegraphics[width=\textwidth]{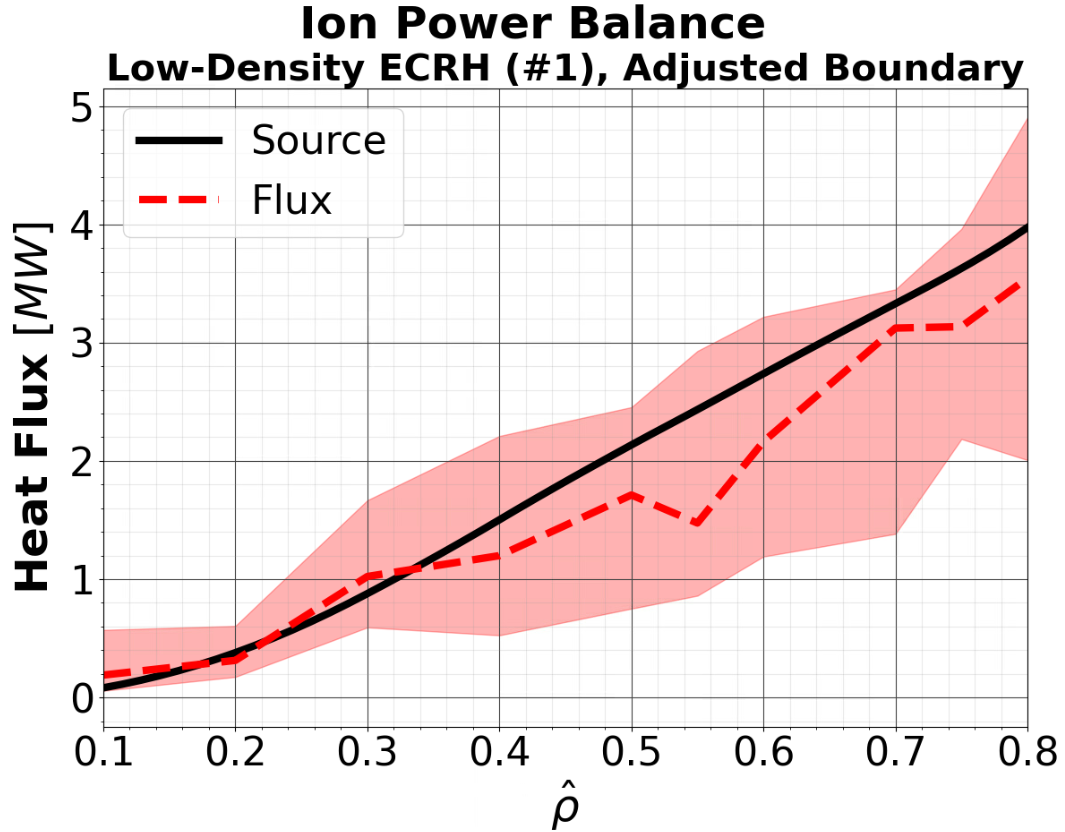} 
        \caption{}
        \label{fig:phase4_mod_power_ion_Elow}
    \end{subfigure}
    \hspace{0.3cm}
    \begin{subfigure}{0.23\textwidth}  
        \centering
        \includegraphics[width=\textwidth]{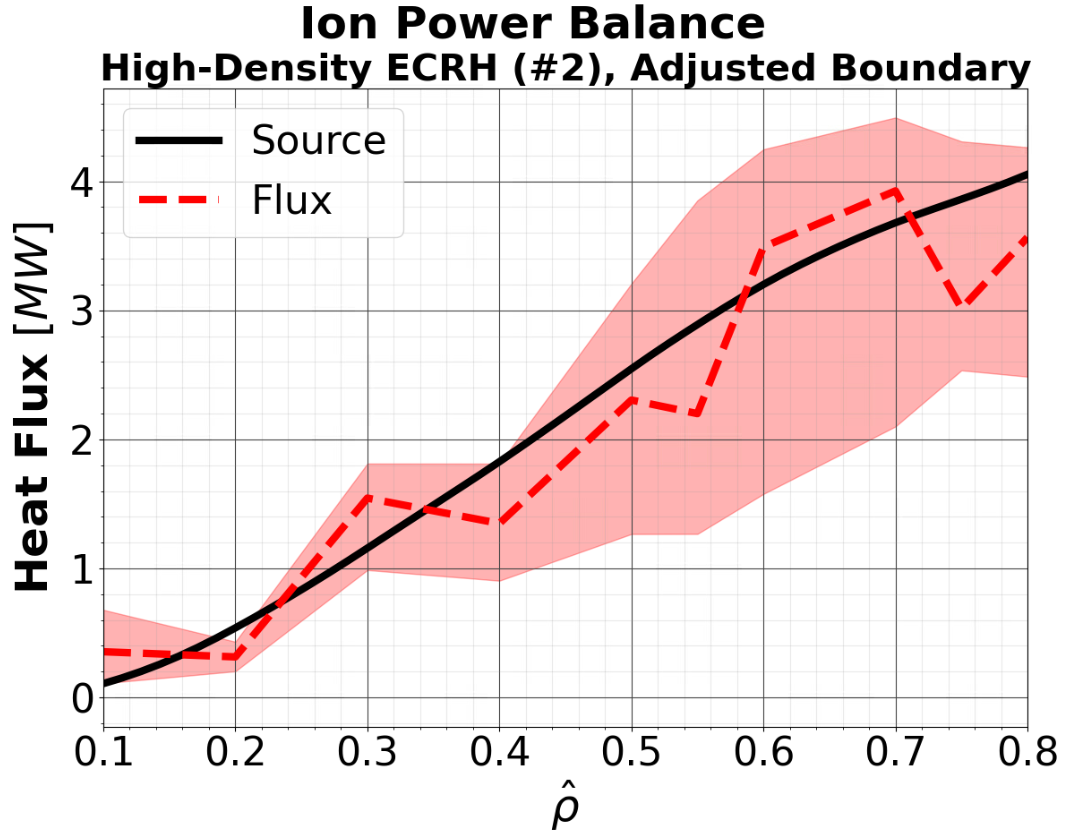}  
        \caption{}
        \label{fig:phase4_mod_power_ion_Ehigh}
    \end{subfigure}
    \hspace{0.3cm}
    \begin{subfigure}{0.23\textwidth}  
        \centering
        \includegraphics[width=\textwidth]{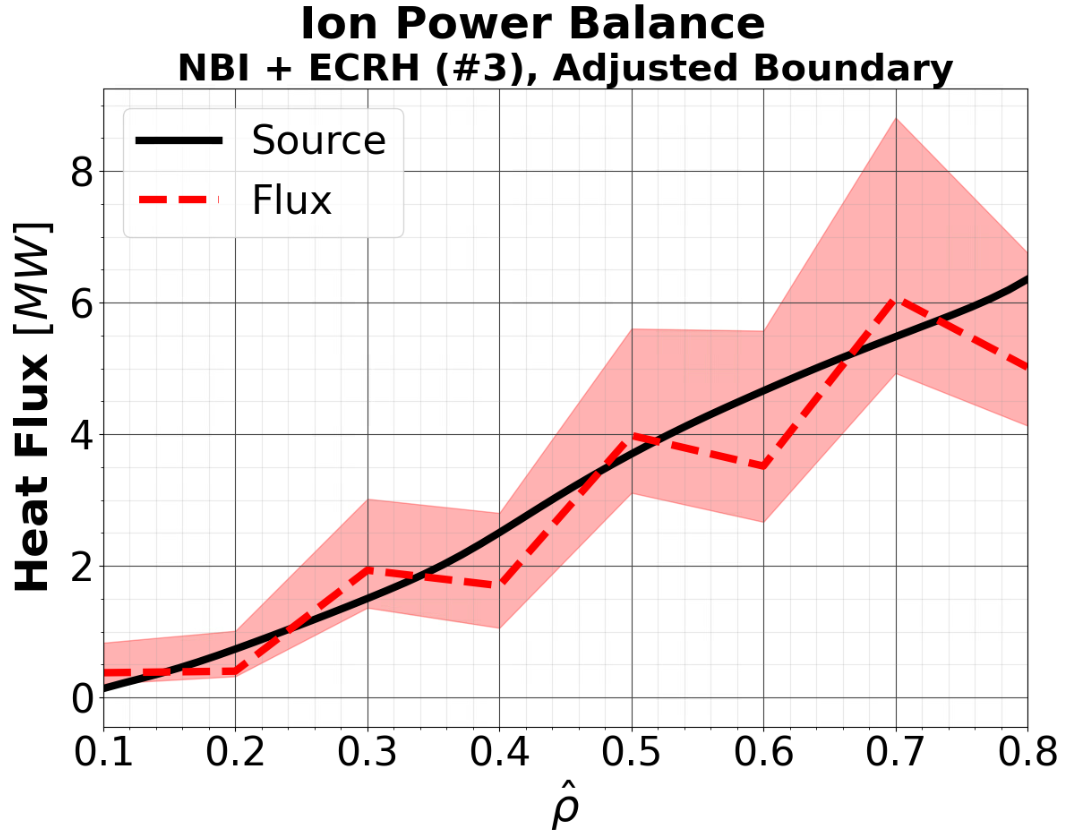}  
        \caption{}
        \label{fig:phase4_mod_power_ion_NII}
    \end{subfigure}
    \hspace{0.3cm}
    \begin{subfigure}{0.23\textwidth}  
        \centering
        \includegraphics[width=\textwidth]{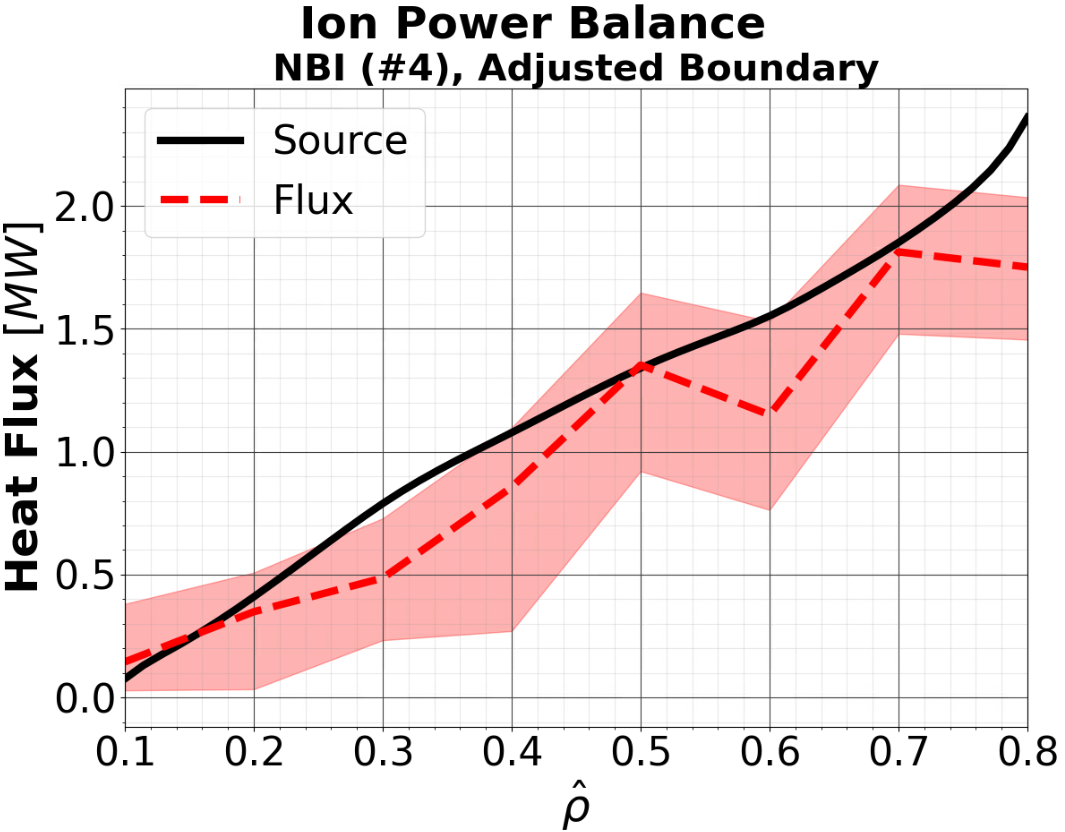}  
        \caption{}
        \label{fig:phase4_mod_power_ion_NIII}
    \end{subfigure}
    
    \vspace{0.5cm}
    
    \begin{subfigure}{0.23\textwidth}  
        \centering
        \includegraphics[width=\textwidth]{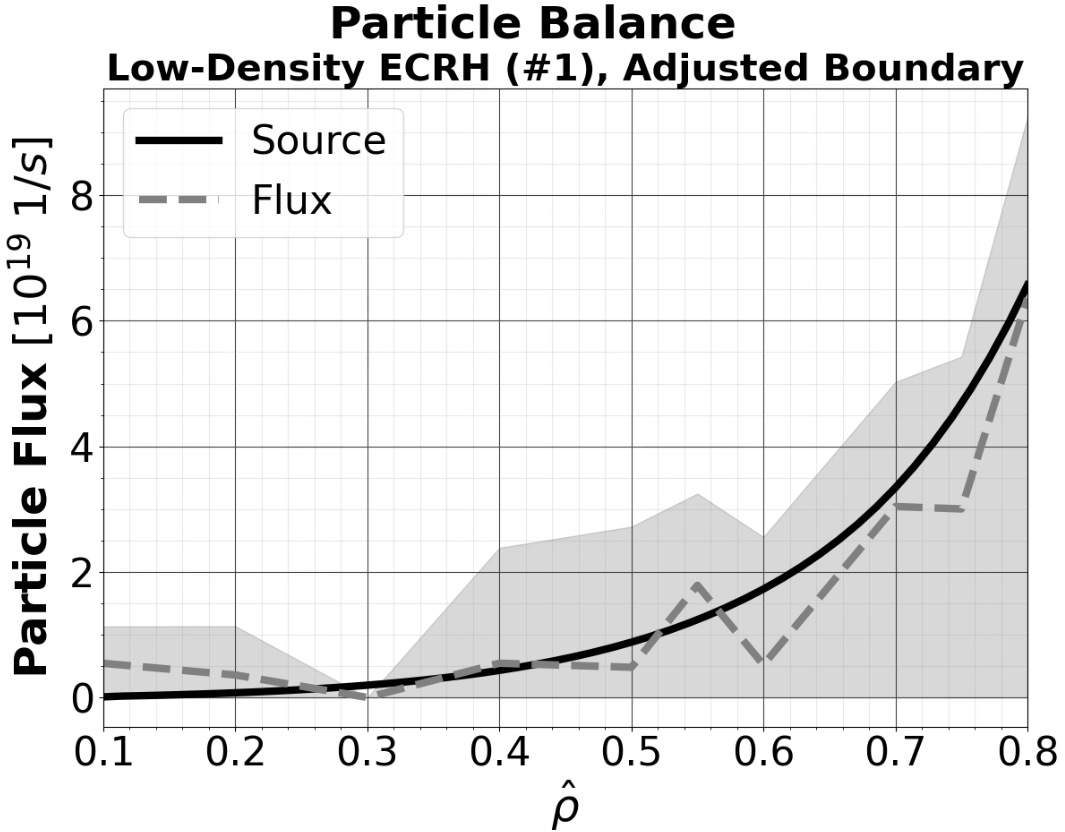} 
        \caption{}
        \label{fig:phase4_mod_particle_Elow}
    \end{subfigure}
    \hspace{0.3cm}
    \begin{subfigure}{0.23\textwidth}  
        \centering
        \includegraphics[width=\textwidth]{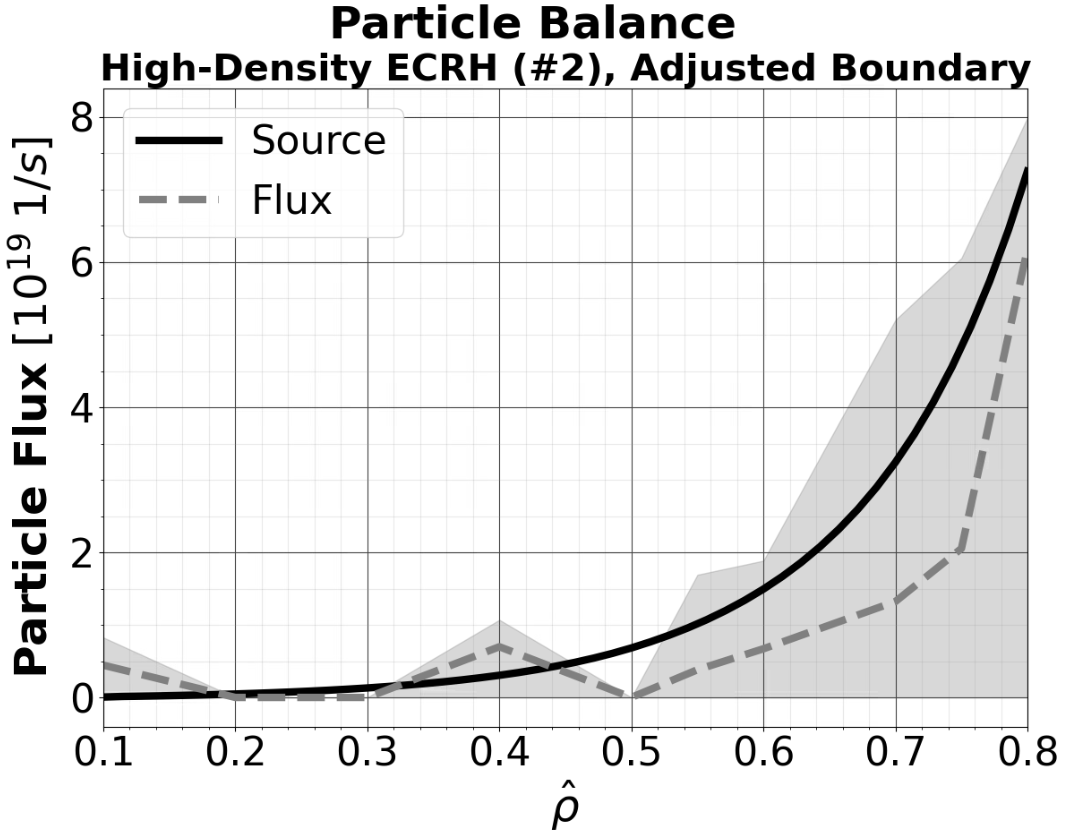}  
        \caption{}
        \label{fig:phase4_mod_particle_Ehigh}
    \end{subfigure}
    \hspace{0.3cm}
    \begin{subfigure}{0.23\textwidth}
        \centering
        \includegraphics[width=\textwidth]{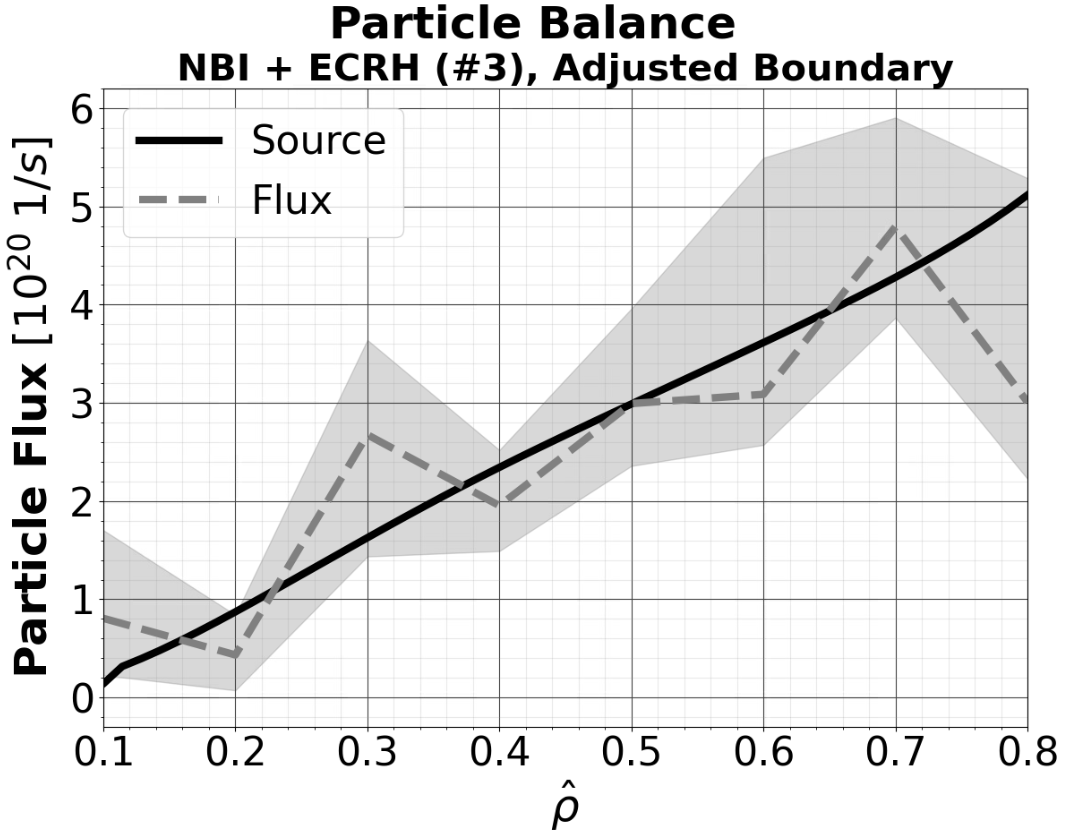}
        \caption{}
        \label{fig:phase4_mod_particle_NII}
    \end{subfigure}
    \hspace{0.3cm}
    \begin{subfigure}{0.23\textwidth}
        \centering
        \includegraphics[width=\textwidth]{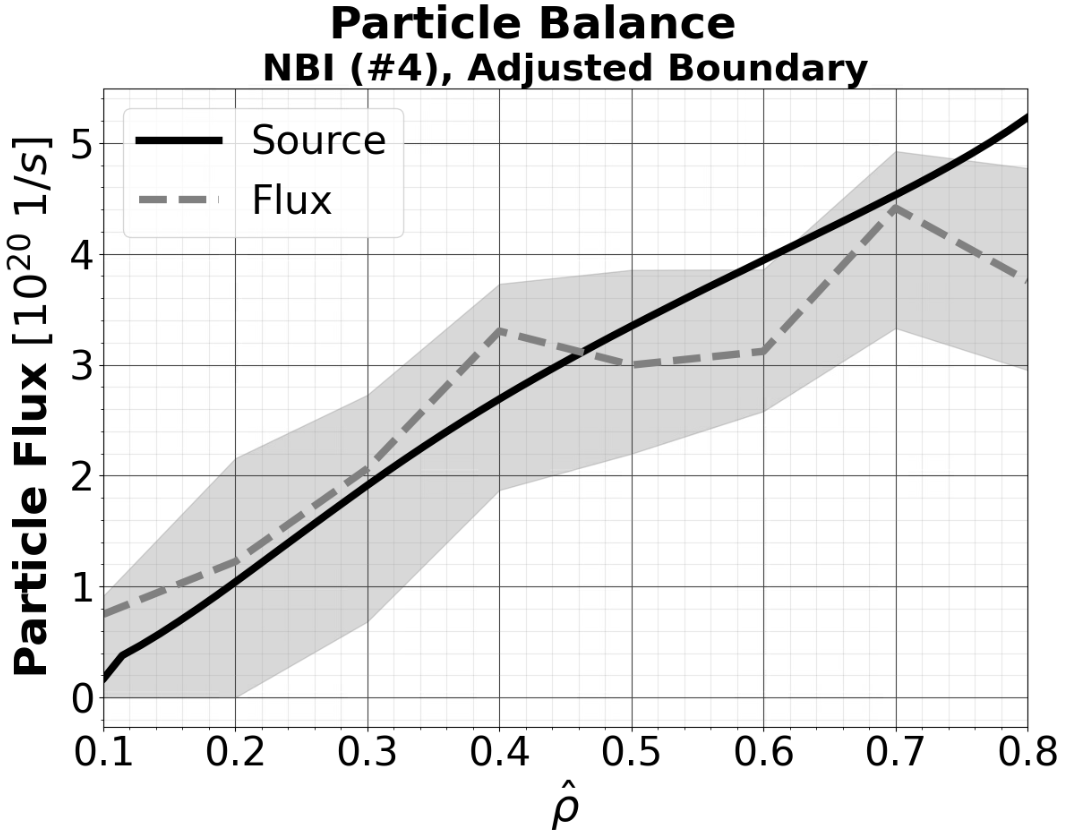}
        \caption{}
        \label{fig:phase4_mod_particle_NIII}
    \end{subfigure}
    
    \caption{Power and particle balances of the low-density ECRH (\subref{fig:phase4_mod_power_elec_Elow}, \subref{fig:phase4_mod_power_ion_Elow}, and \subref{fig:phase4_mod_particle_Elow}), high-density ECRH (\subref{fig:phase4_mod_power_elec_Ehigh}, \subref{fig:phase4_mod_power_ion_Ehigh}, and \subref{fig:phase4_mod_particle_Ehigh}), NBI + ECRH (\subref{fig:phase4_mod_power_elec_NII}, \subref{fig:phase4_mod_power_ion_NII}, and \subref{fig:phase4_mod_particle_NII}), and NBI (\subref{fig:phase4_mod_power_elec_NIII}, \subref{fig:phase4_mod_power_ion_NIII}, and \subref{fig:phase4_mod_particle_NIII}) scenarios with adjusted boundaries for the plasma profiles and $n_{0,edge} = 10^{13}\;m^{-3}$. Fulfillment of the balances slightly improved, but the particle balances of the two ECRH cases (\subref{fig:phase4_mod_particle_Elow} and \subref{fig:phase4_mod_particle_Ehigh}) still vary significantly between successive iterations.}
    \label{fig:phase4_mod_balances}
\end{figure*}

\subsubsection{\label{sec:results_phase4_1e13}Adjusted Profile Boundaries with $\bm{n_{0,\text{edge}}\;=\;10^{13}}$~$\bm{m^{-3}}$}

Several options are available to improve the agreement of the simulated plasma profiles with the experimental data points. A separate study has been started due to the aforementioned prevalence of negative $\Gamma$ in the outer half of the simulated radial domain. In addition to the density gradient, it was found in this other study that the other primary factor dictating the sign of $\Gamma$ is the collision frequency $\nu_c$. More details will be provided in a different publication but one of the key takeaways is that $\Gamma$ becomes positive for sufficiently low values of $\nu_c$. Given the linear relationship between $\nu_c$ and $n_e$, one possible adjustment to the density profile was to lower the boundary $n_e$. This was expected to reduce the collision frequency for all flux-tube simulations since the edge conditions dictate the core profile. \cite{BanonNavarro2024} In other words, to the zeroth order, moving the boundary density by a certain fraction will cause the rest of the profile to move by a similar magnitude. This would then give \texttt{Tango} a larger leeway in adjusting the density while simultaneously allowing the profile to stay close to the data points. Another possible adjustment is to increase the boundary $T_e$. Stemming from the inverse quadratic relationship between $T_e$ and $\nu_c$, smaller adjustments in $T_e$ are enough to reproduce the reduction in $\nu_c$ by decreasing the boundary $n_e$. \\

These adjustments were the focus of the second batch of phase 4 simulations. The simulations were once again initialized using the converged plasma profiles from phase 3. Instructions were then provided to \texttt{Tango} to gradually adjust the temperature and density boundaries. The adjustment rate of a boundary condition was limited by the relaxation parameter $\alpha$. The fulfillment of particle and power balances and the agreement of the simulated profiles with the experimental data were manually assessed a few iterations after every new set of boundary values had been reached. \\
\begin{figure*}
    \centering
    \begin{subfigure}{0.43\textwidth}  
        \centering
        \includegraphics[width=\textwidth]{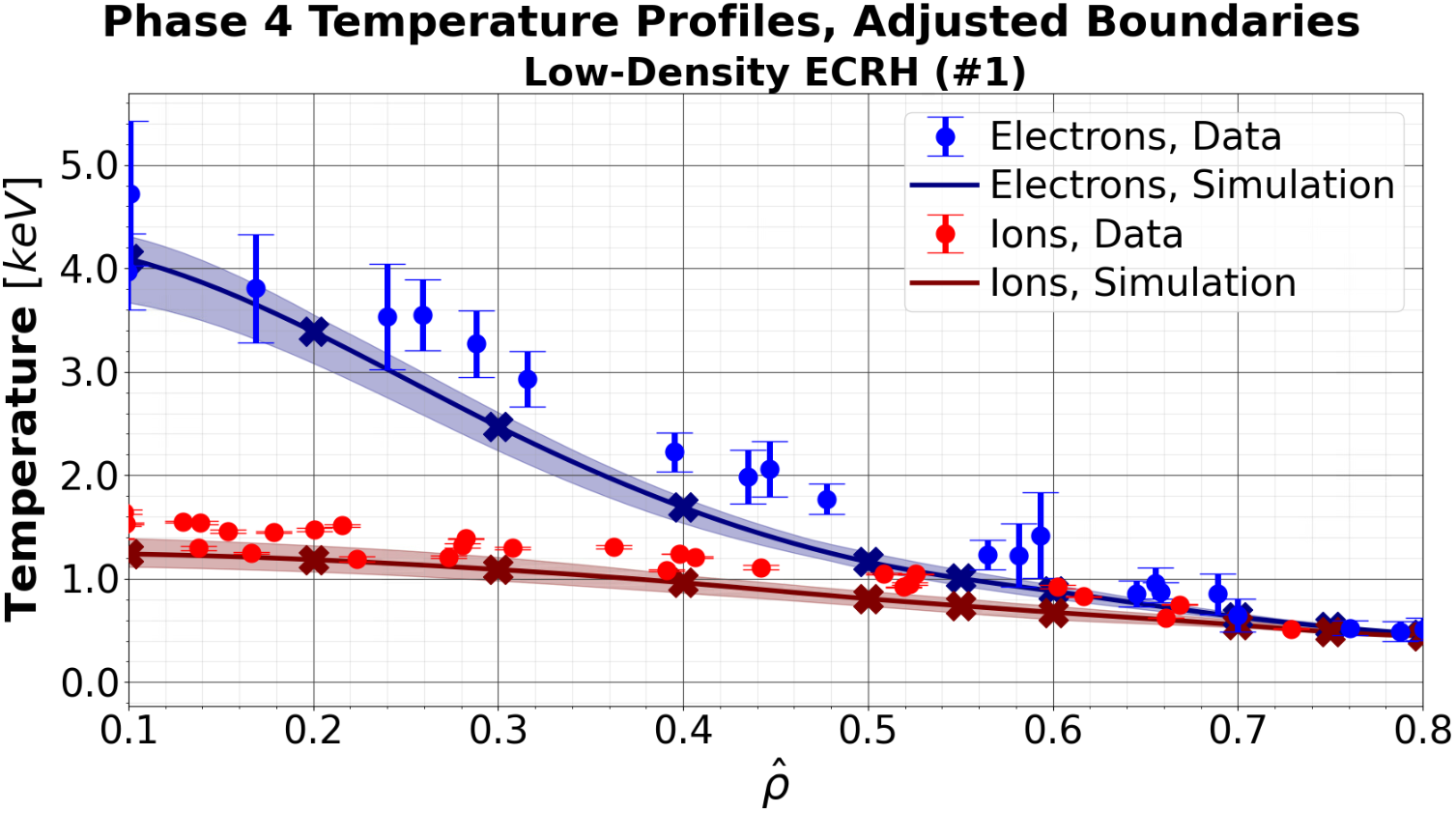} 
        \caption{}
        \label{fig:phase4_mod_T_Elow}
    \end{subfigure}
    \hspace{0.5cm}
    \begin{subfigure}{0.43\textwidth}  
        \centering
        \includegraphics[width=\textwidth]{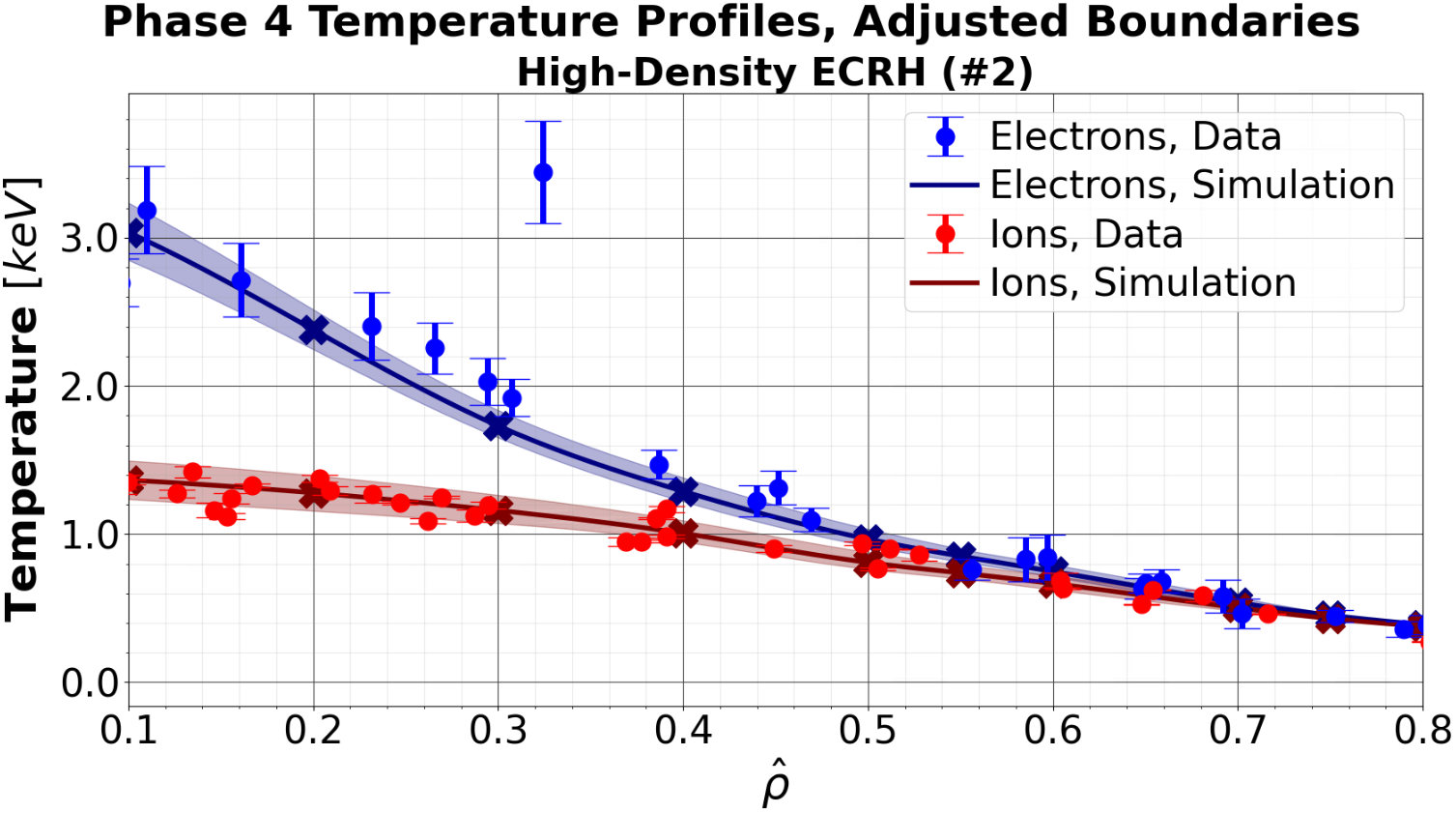}  
        \caption{}
        \label{fig:phase4_mod_T_Ehigh}
    \end{subfigure}
    \vspace{0.7cm}  
    \begin{subfigure}{0.43\textwidth}
        \centering
        \includegraphics[width=\textwidth]{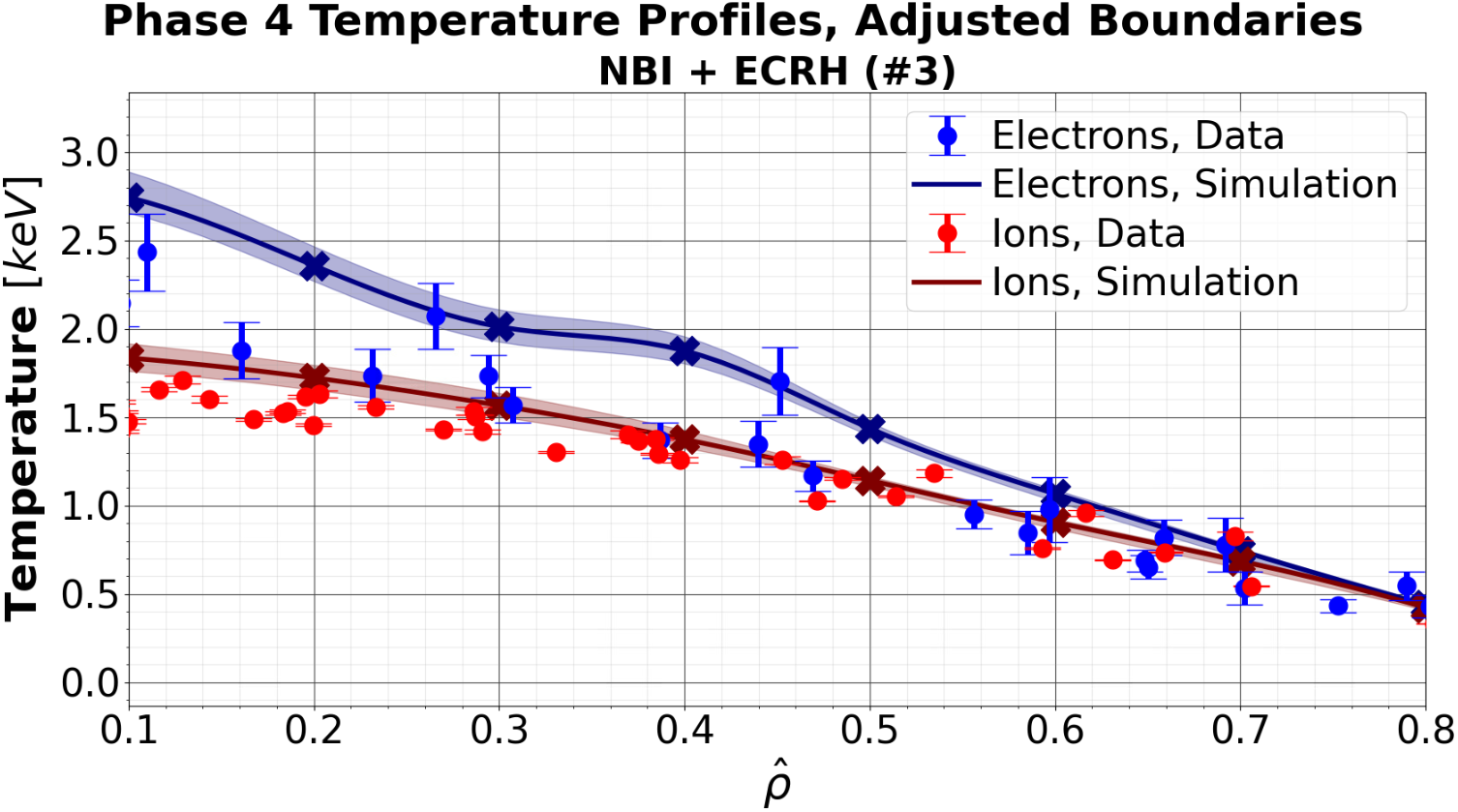}
        \caption{}
        \label{fig:phase4_mod_T_NII}
    \end{subfigure}
    \hspace{0.5cm}
    \begin{subfigure}{0.43\textwidth}
        \centering
        \includegraphics[width=\textwidth]{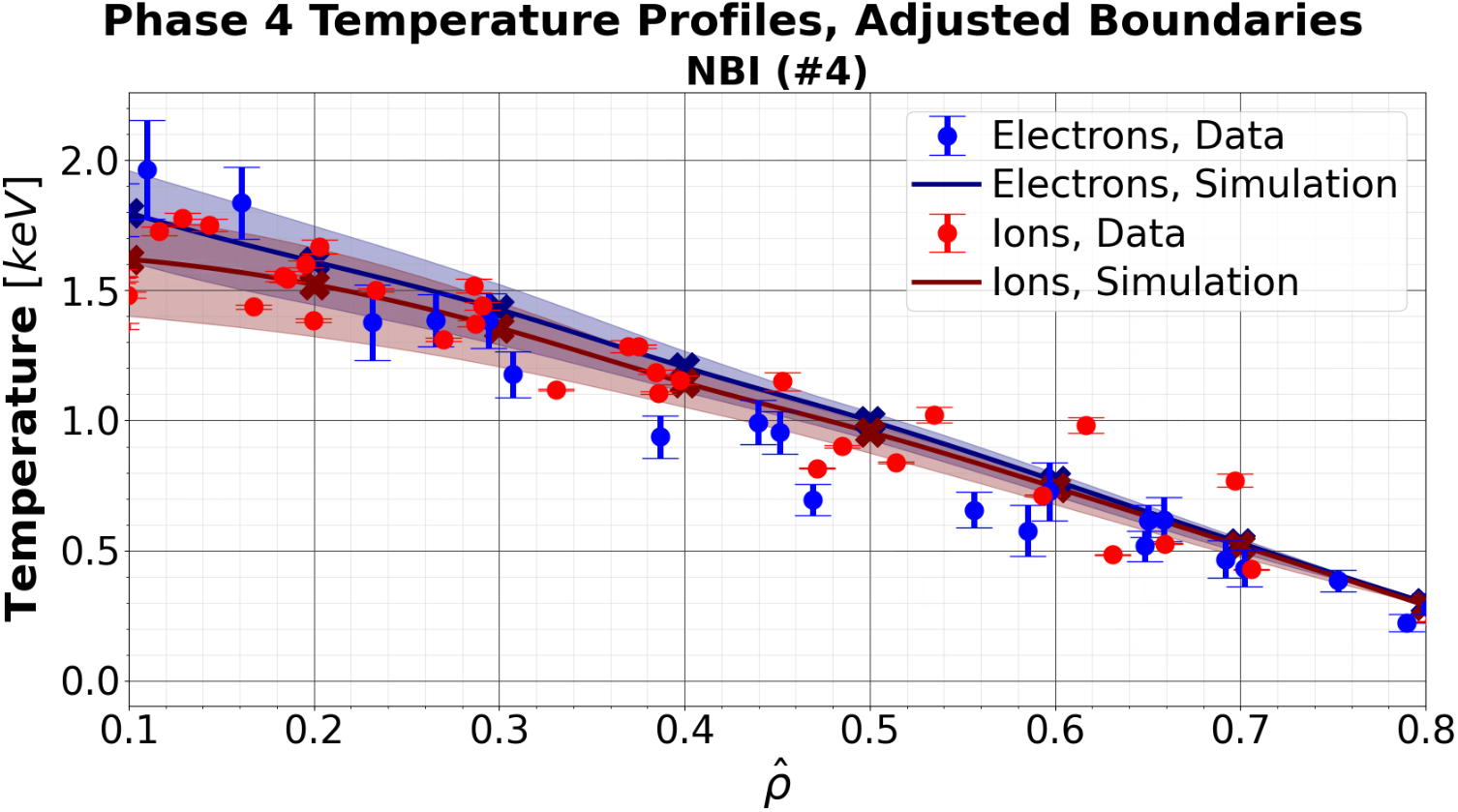}
        \caption{}
        \label{fig:phase4_mod_T_NIII}
    \end{subfigure}
    \vspace{0.7cm}
    \begin{subfigure}{0.43\textwidth}  
        \centering
        \includegraphics[width=\textwidth]{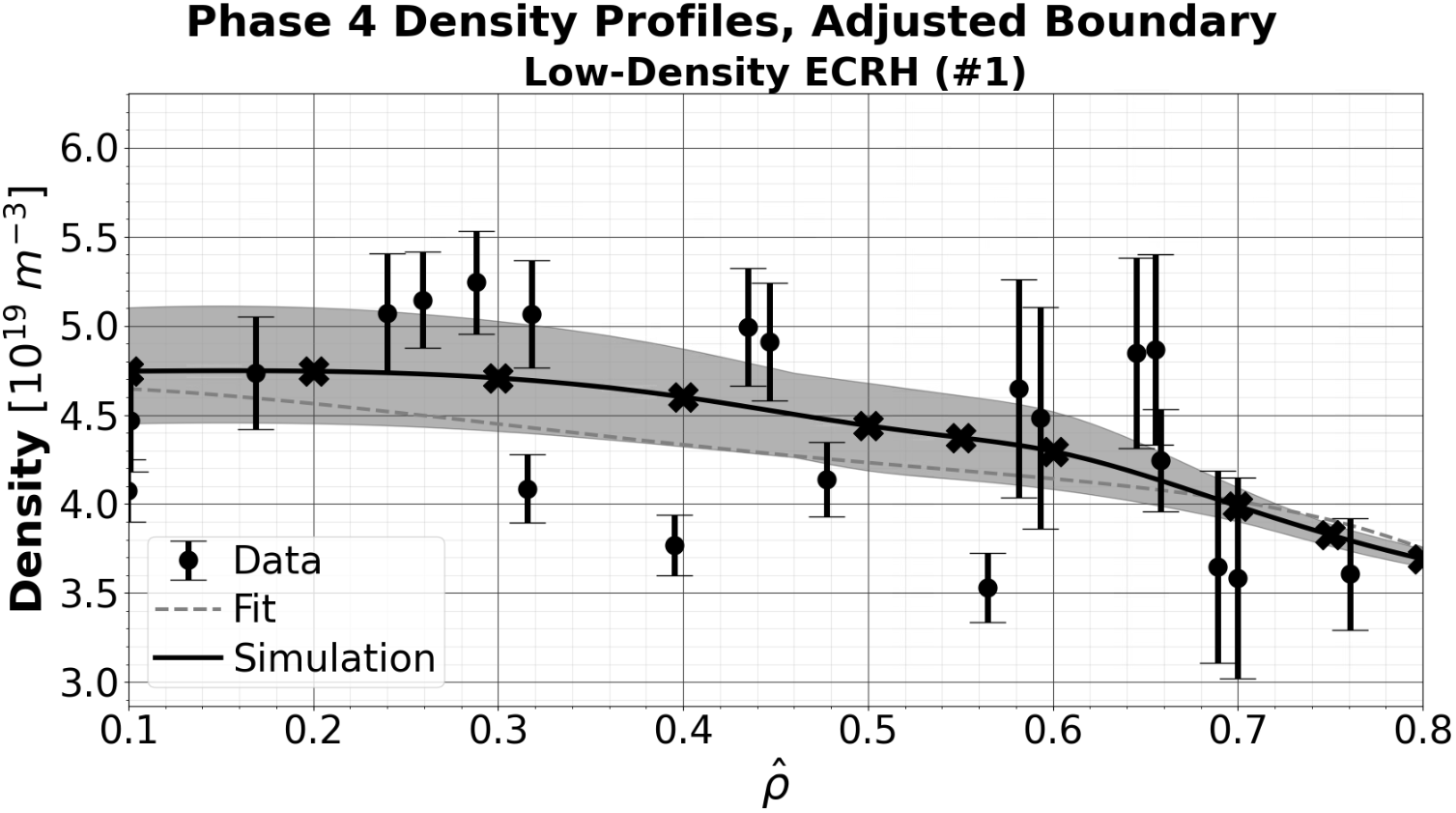} 
        \caption{}
        \label{fig:phase4_mod_n_Elow}
    \end{subfigure}
    \hspace{0.5cm}
    \begin{subfigure}{0.43\textwidth}  
        \centering
        \includegraphics[width=\textwidth]{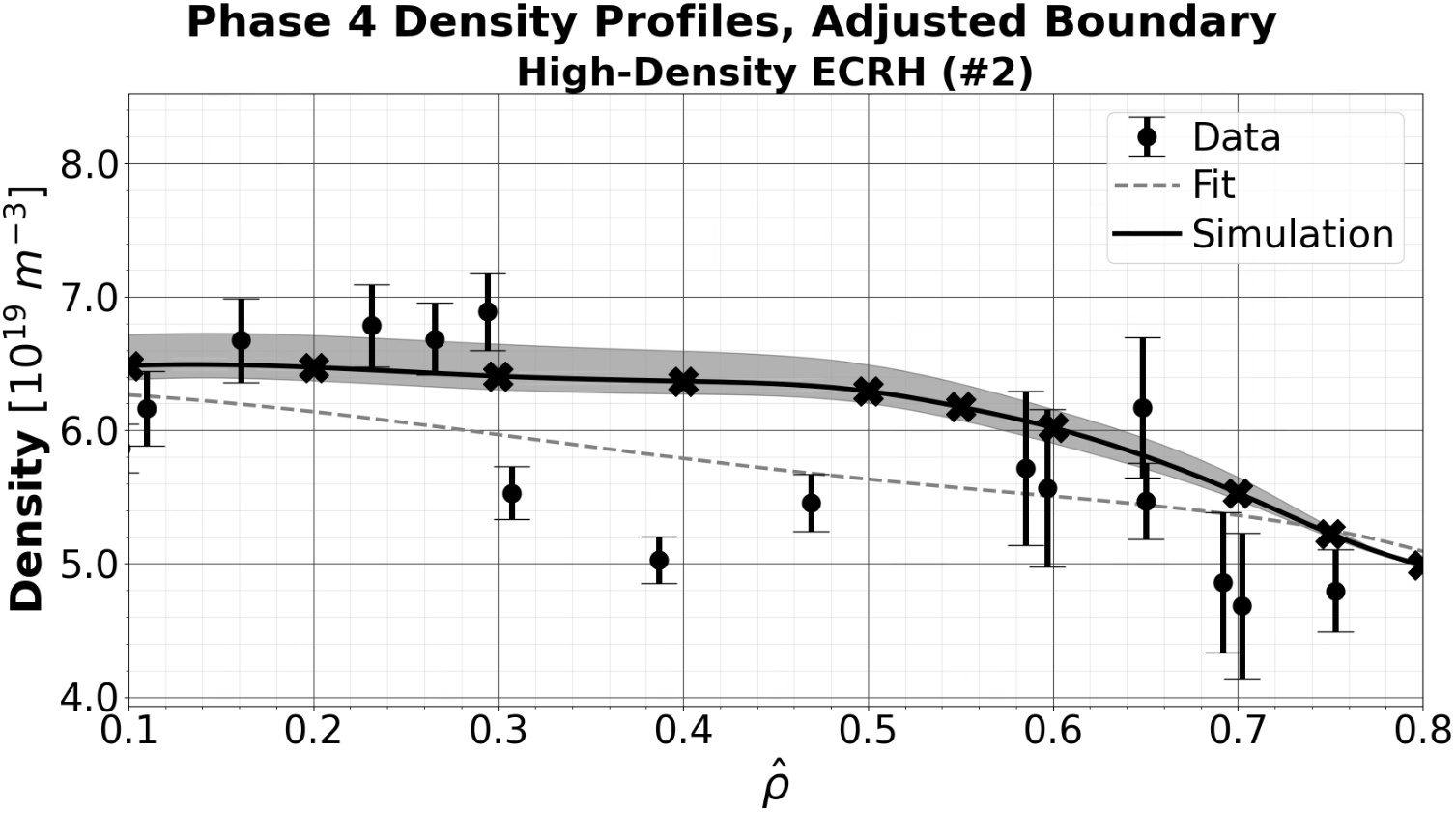}  
        \caption{}
        \label{fig:phase4_mod_n_Ehigh}
    \end{subfigure}
    \hspace{0.5cm}
    \begin{subfigure}{0.43\textwidth}
        \centering
        \includegraphics[width=\textwidth]{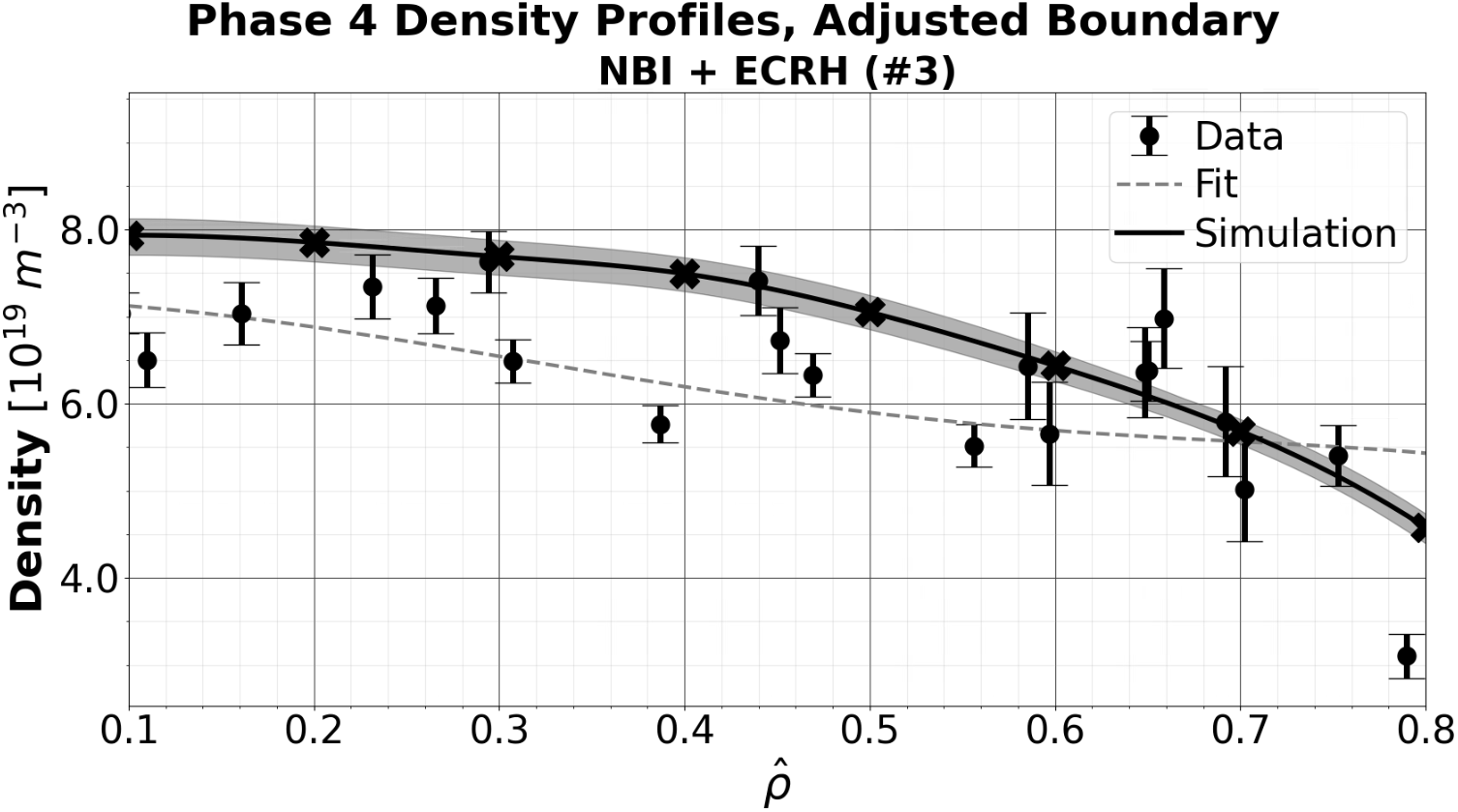}
        \caption{}
        \label{fig:phase4_mod_n_NII}
    \end{subfigure}
    \hspace{0.5cm}
    \begin{subfigure}{0.43\textwidth}
        \centering
        \includegraphics[width=\textwidth]{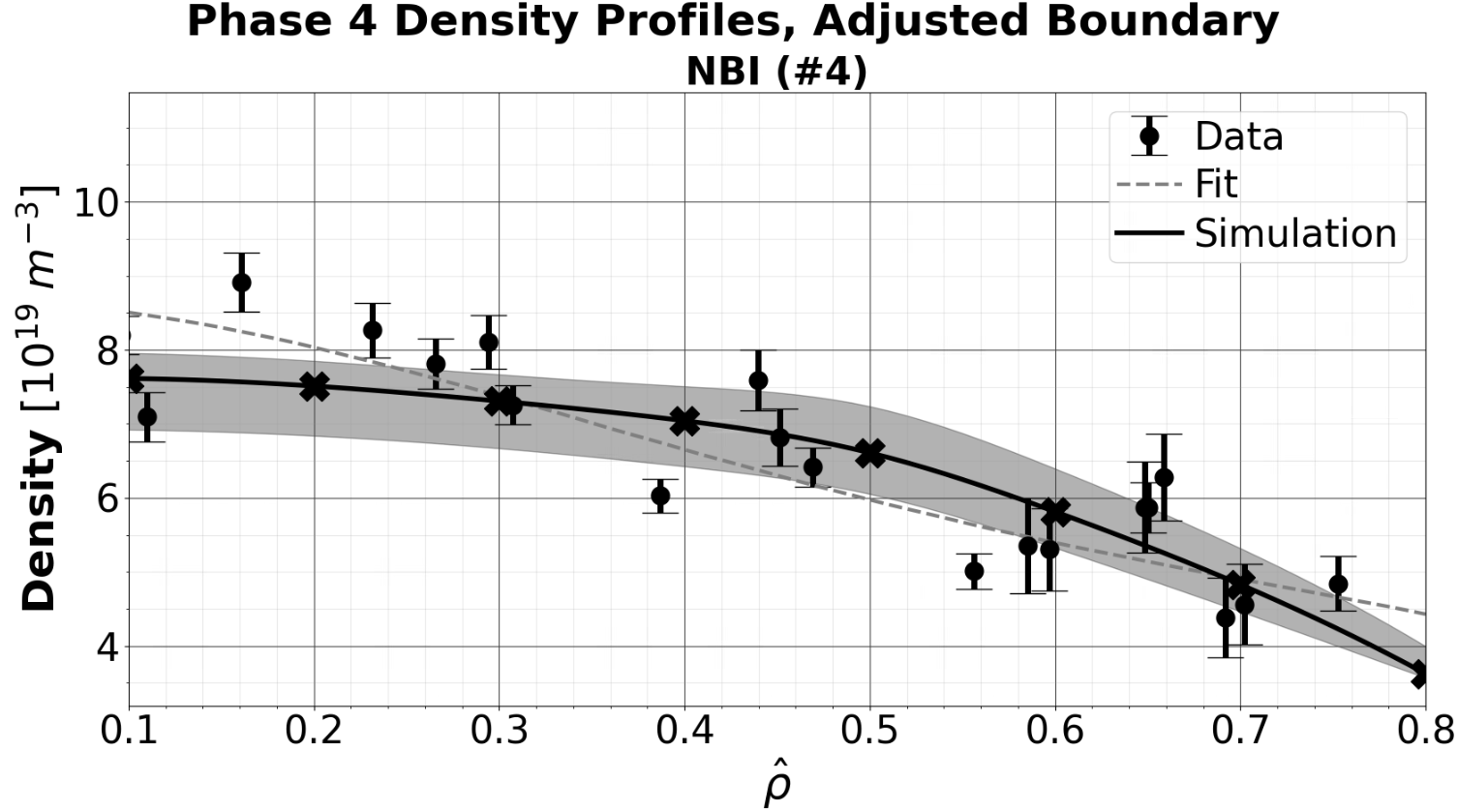}
        \caption{}
        \label{fig:phase4_mod_n_NIII}
    \end{subfigure}
    
    \caption{Converged temperature and density profiles of the low-density ECRH (\subref{fig:phase4_mod_T_Elow} and \subref{fig:phase4_mod_n_Elow}), high-density ECRH (\subref{fig:phase4_mod_T_Ehigh} and \subref{fig:phase4_mod_n_Ehigh}), NBI + ECRH (\subref{fig:phase4_mod_T_NII} and \subref{fig:phase4_mod_n_NII}), and NBI (\subref{fig:phase4_base_T_NIII} and \subref{fig:phase4_mod_n_NIII}) scenarios with adjusted temperature and density boundaries. Better agreement with the experimental data was achieved for all cases.}
    \label{fig:phase4_mod_Tn}
\end{figure*}

From Fig. \ref{fig:phase4_mod_Tn}, it can be seen that small adjustments within the error bars can have a significant effect on the inner core profiles, allowing for better matching of the experimental data points and compliance of power and particle balances. This emphasizes the importance of the plasma edge, which is unfortunately beyond the scope of this study. The fitted density profiles are shown in dotted gray, to facilitate comparison with the converged density profiles from the \texttt{GENE-KNOSOS-Tango} simulations. The earlier remark regarding the necessity of the larger density gradients at the boundary is clearly seen and confirmed by these second batch of phase 4 simulations.

\subsubsection{\label{sec:results_phase4_1e14}Adjusted Profile Boundaries with $\bm{n_{0,\text{edge}}\;=\;10^{14}}$~$\bm{m^{-3}}$}

\begin{figure}
    \centering
    \begin{subfigure}{0.22\textwidth}  
        \centering
        \includegraphics[width=\textwidth]{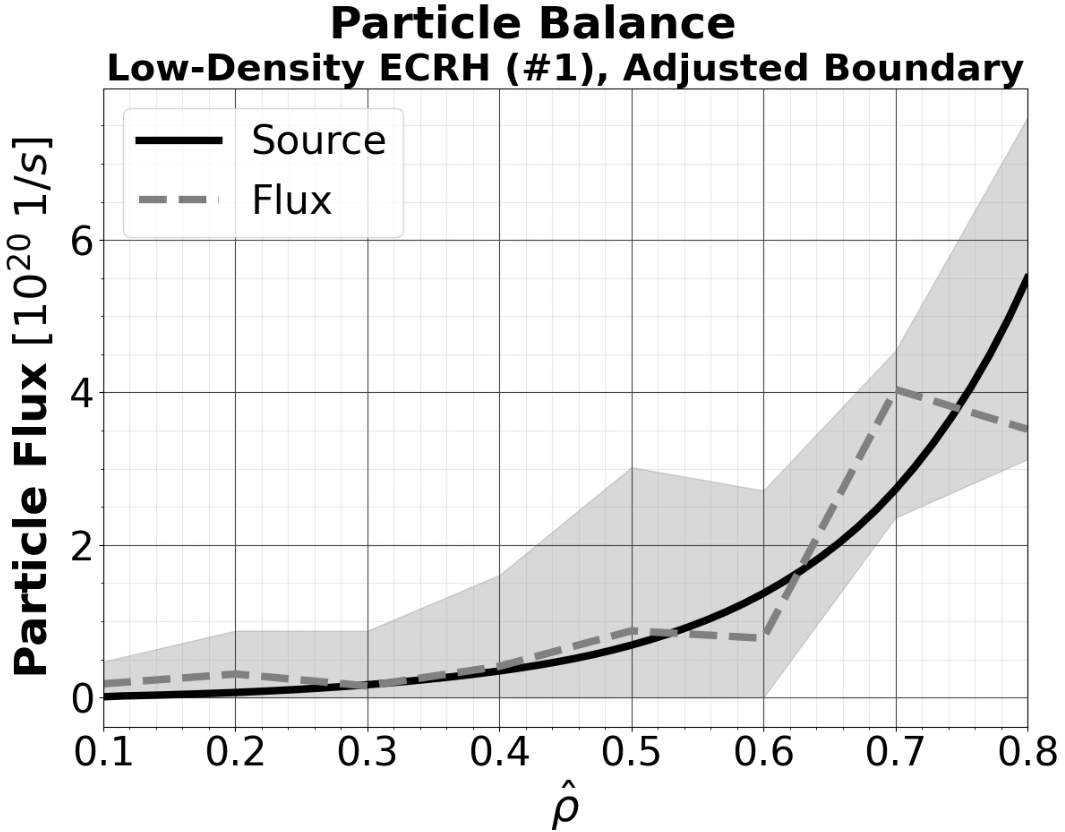} 
        \caption{}
        \label{fig:phase4_mod2_particle_Elow}
    \end{subfigure}
    \hspace{0.5cm}
    \begin{subfigure}{0.22\textwidth}  
        \centering
        \includegraphics[width=\textwidth]{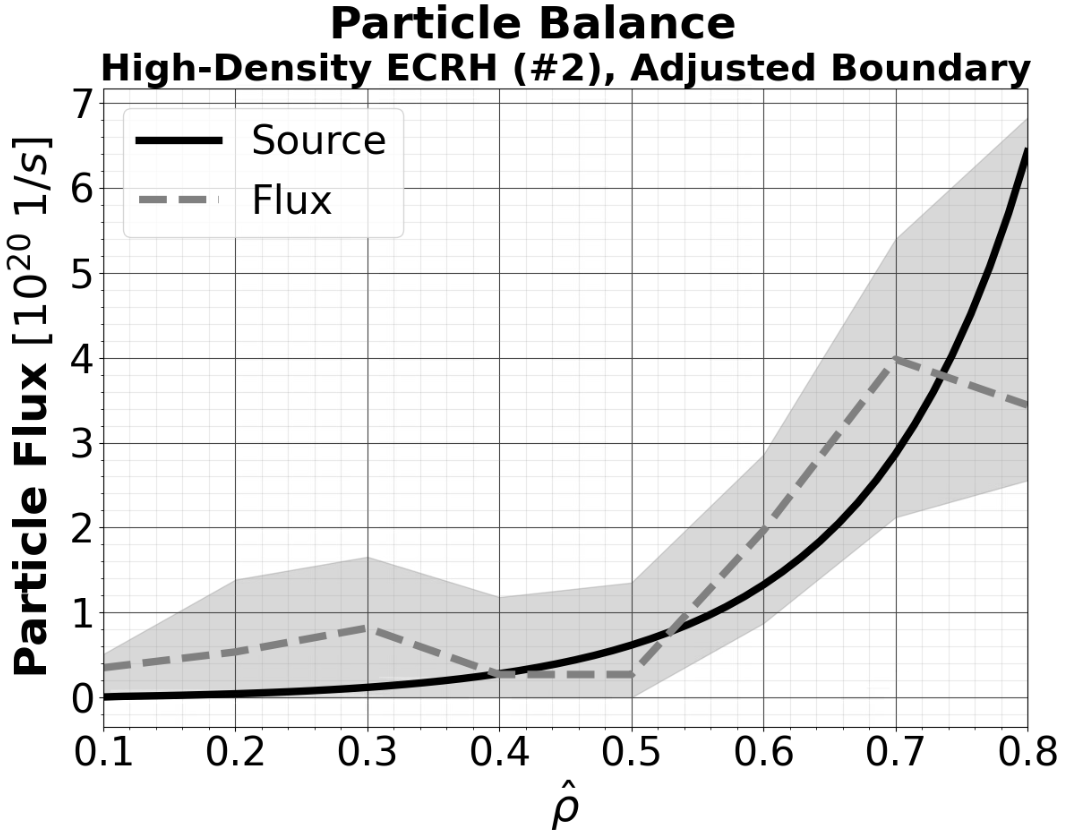}  
        \caption{}
        \label{fig:phase4_mod2_particle_Ehigh}
    \end{subfigure}
    
    \caption{Particle balances of the (\subref{fig:phase4_mod2_particle_Elow}) low- and (\subref{fig:phase4_mod2_particle_Ehigh}) high-density ECRH scenarios with adjusted boundaries for the plasma profiles and with an increased edge neutrals density of $\mathrm{10^{14}\;m^{-3}}$. Particle balances were more consistently matched, especially for $\hat{\rho} \ge 0.6$.}
    \label{fig:phase4_mod2_particle_ECRH}
\end{figure}

Despite these adjustments to the profiles, the sensitivity of $\Gamma$ to small changes in the profiles for the low- and high-density ECRH cases still remained. This is apparent from Figs. \ref{fig:phase4_mod_particle_Elow} and \ref{fig:phase4_mod_particle_Ehigh}, where $\Gamma$ still covered the whole region under the particle source curves rather than being confined to a small region close to the curves. In contrast, the two NBI cases had $\Gamma$ profiles that are consistently positive and matched the particle sources. It was hypothesized that this sensitivity stemmed from the difference in magnitudes of the particle sources, as the source terms of the ECRH cases were both approximately one order of magnitude smaller than those of the NBI cases. This motivated the third batch of phase 4 simulations, wherein $n_{0,edge}$ was increased to $10^{14}\;m^{-3}$. \\

The two ECRH cases underwent an order of magnitude increase for their particle sources, significantly larger than the 75-95\% increase observed in the two NBI cases. Fig. \ref{fig:phase4_mod2_particle_ECRH} shows the particle balances for cases 1 and 2, which were more adequately satisfied. Small adjustments were needed for the boundary density and temperatures but, in general, the plasma profiles stayed approximately the same relative to the $n_{0,edge} = 10^{14}\;m^{-3}$ simulations. \\

With both power and particle balances sufficiently matched, we now revisit the breakdown of heat fluxes by scale for each scenario. The low- and high-density ECRH scenarios had large electron-scale contributions to the total electron heat flux, peaking at about 60\% for both. As one moves radially outward, the electron-scale contribution first increased until $\hat{\rho} = 0.3$ followed by a gradual reduction toward the radial boundary. The upward trend is explained by the destabilization of the ETG mode by the increasing electron temperature gradient $\omega_{Te}$ and decreasing electron-ion temperature ratio $\tau = T_e / T_i$, \cite{Jenko2001} while the decreasing trend can be attributed to the steady increase of the density gradient $\omega_{n}$ with $\hat{\rho}$, which stabilizes the ETG mode. \cite{Jenko2001, Ren2011, Ruiz2015} \\

On the other hand, the two NBI cases showed relatively less electron-scale contributions to the total electron heat flux. The inverse dependence on $\hat{\rho}$ was still observed, but the peak percentage was reduced to approximately 40\% and 20\% for the NBI + ECRH and NBI scenarios respectively. For these two cases, where $T_i \sim T_e$ and $\omega_{Ti} \sim \omega_{Te}$, the ion-scale electron heat fluxes dominated; this is consistent with previous findings. \cite{Plunk2019} \\

\begin{figure}
    \centering
    \begin{subfigure}{0.48\textwidth}  
        \centering
        \includegraphics[width=\textwidth]{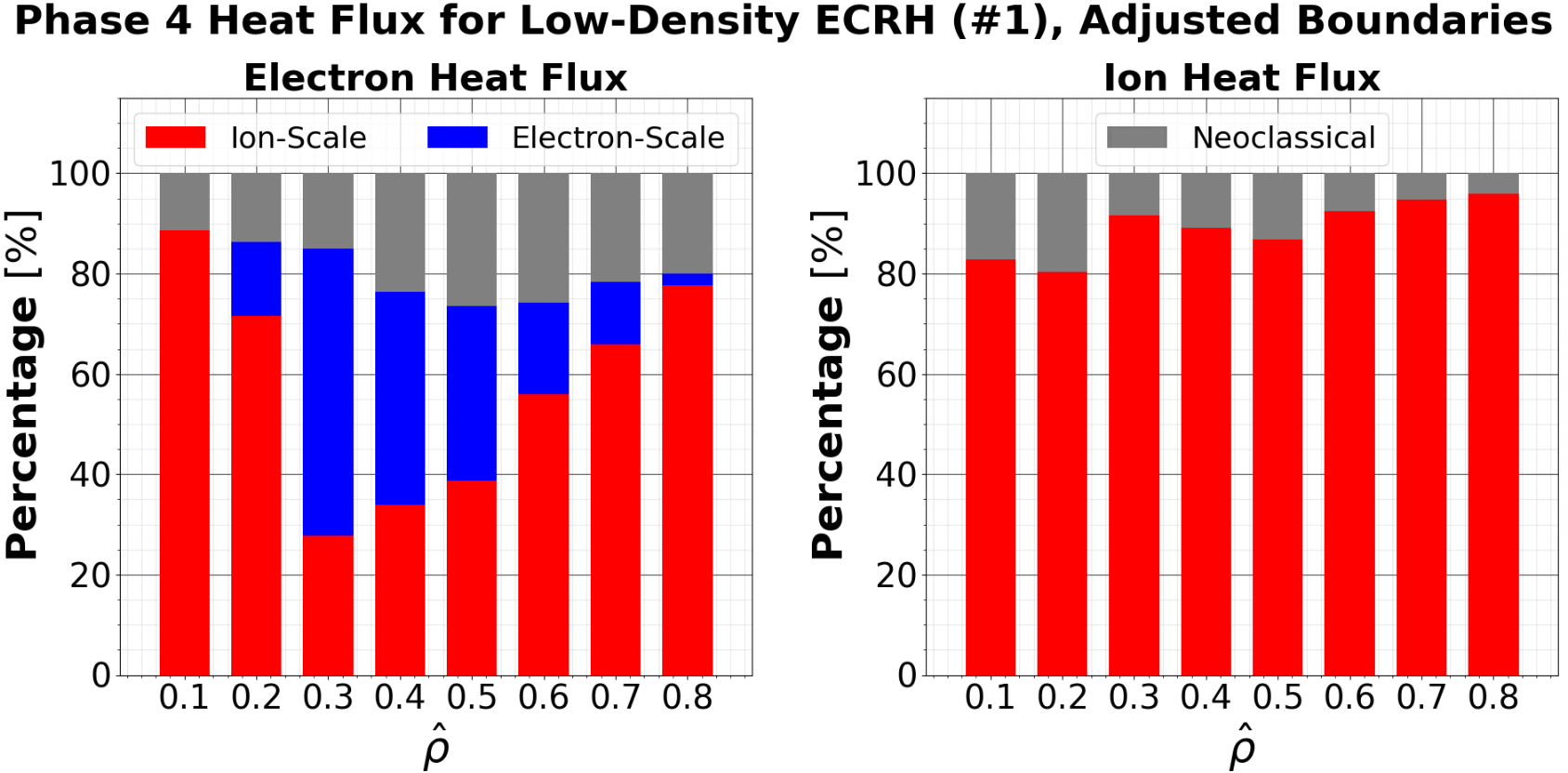} 
        \caption{}
        \label{fig:phase4_flux_breakdown_Elow}
    \end{subfigure}
    \begin{subfigure}{0.48\textwidth}  
        \centering
        \includegraphics[width=\textwidth]{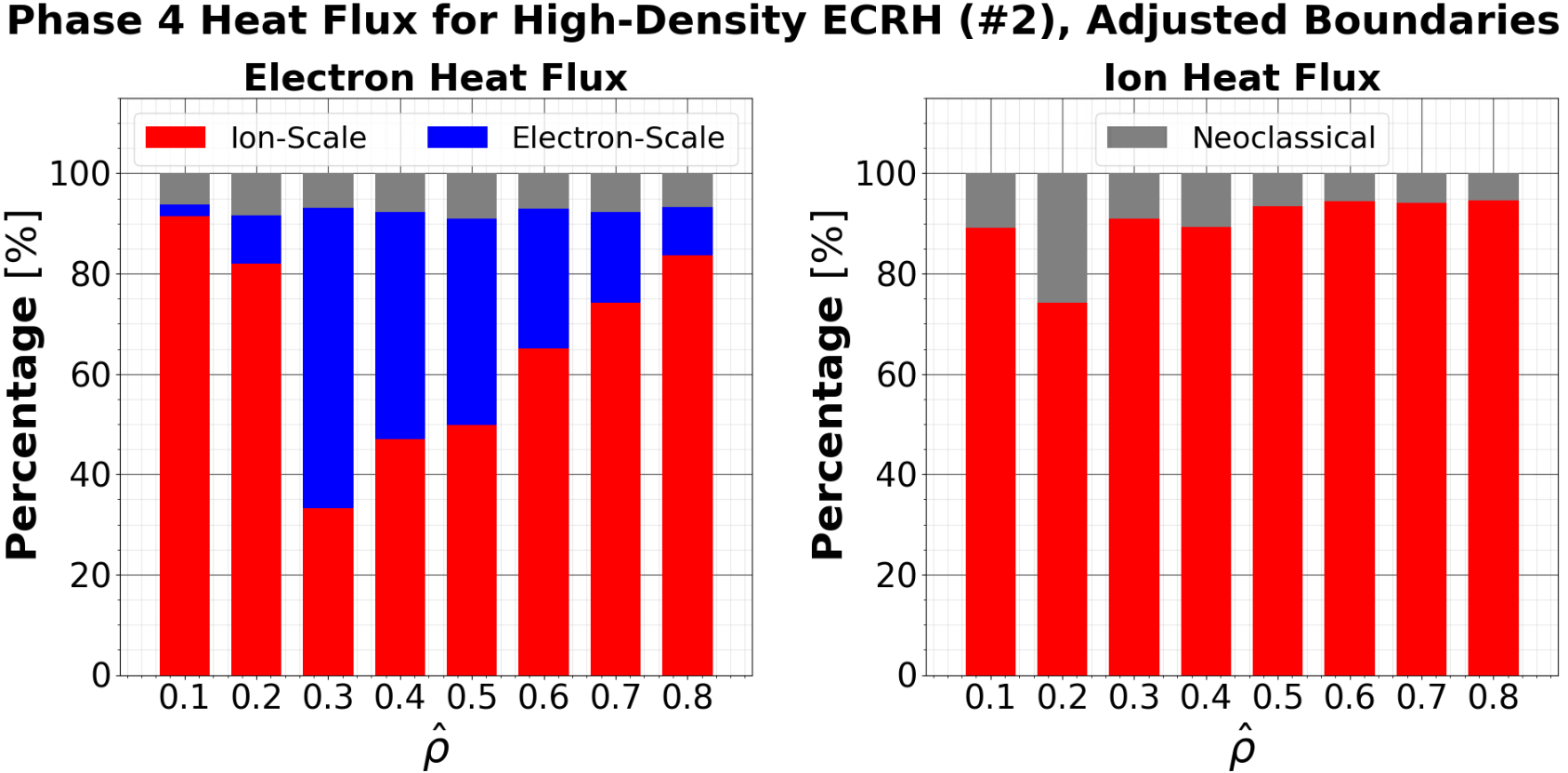}  
        \caption{}
        \label{fig:phase4_flux_breakdown_Ehigh}
    \end{subfigure}
    
    \caption{Breakdown of electron and heat fluxes for the (\subref{fig:phase4_flux_breakdown_Elow}) low- and (\subref{fig:phase4_flux_breakdown_Ehigh}) high-density ECRH scenarios. The electron-scale heat fluxes significantly contributed to the electron total.}
    \label{fig:phase4_flux_breakdown_ECRH}
\end{figure}

\begin{figure}
    \centering
    \begin{subfigure}{0.48\textwidth}  
        \centering
        \includegraphics[width=\textwidth]{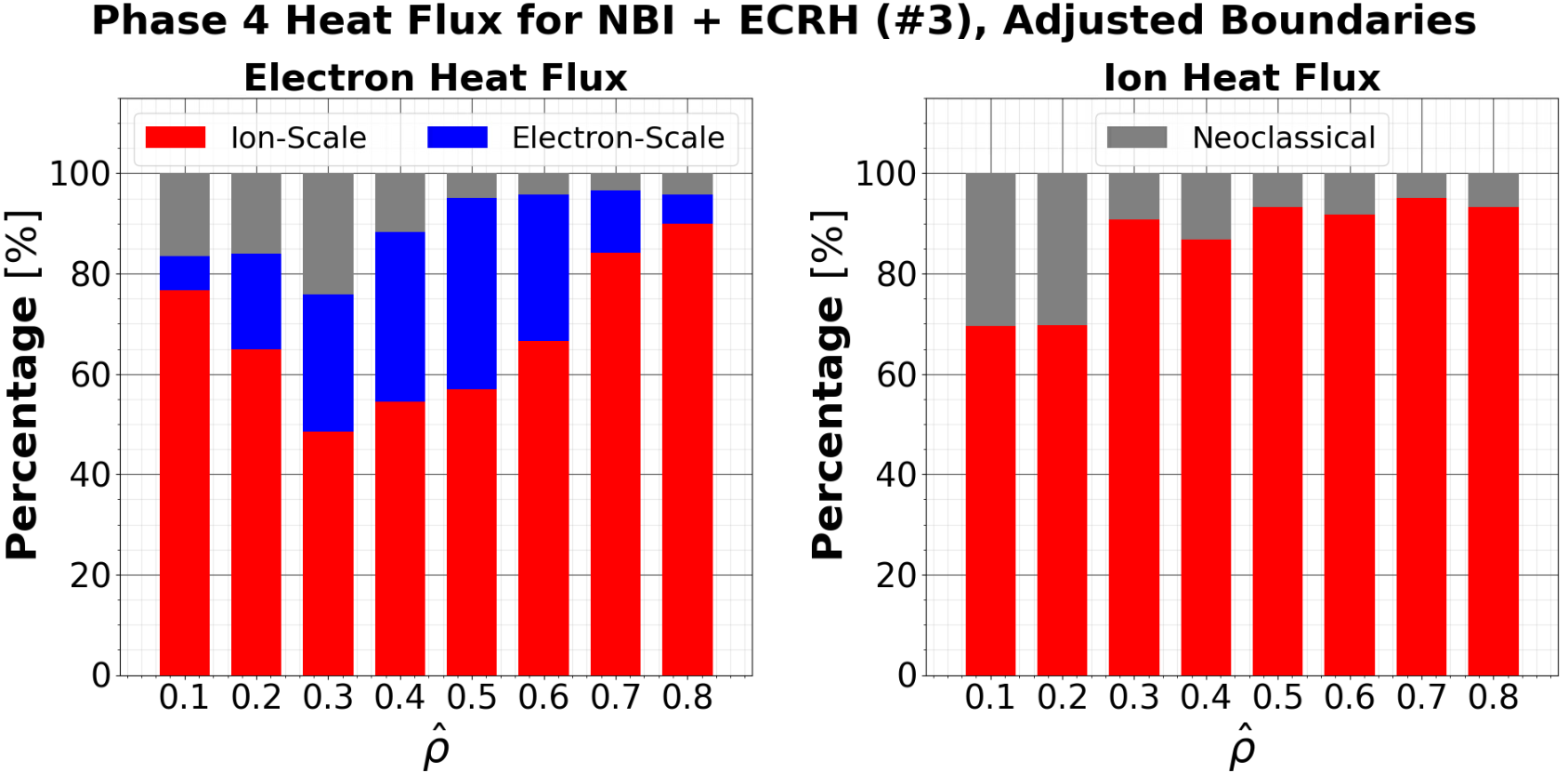} 
        \caption{}
        \label{fig:phase4_flux_breakdown_NII}
    \end{subfigure}
    \begin{subfigure}{0.48\textwidth}  
        \centering
        \includegraphics[width=\textwidth]{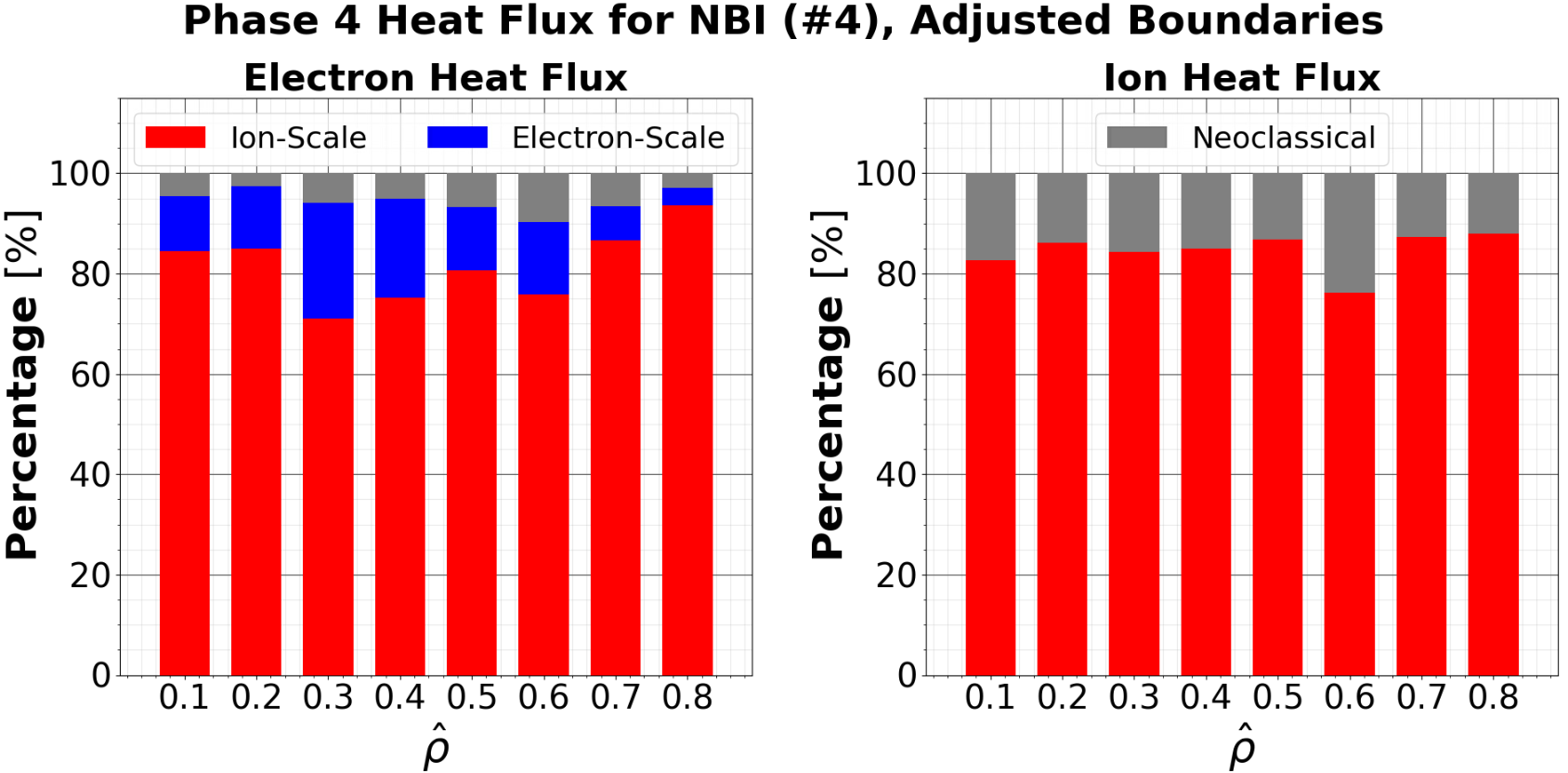}  
        \caption{}
        \label{fig:phase4_flux_breakdown_NIII}
    \end{subfigure}
    
    \caption{Breakdown of electron and heat fluxes for the (\subref{fig:phase4_flux_breakdown_NII}) NBI + ECRH and (\subref{fig:phase4_flux_breakdown_NIII}) NBI scenarios. The electron-scale heat fluxes accounted for relatively less when compared to the two ECRH scenarios.}
    \label{fig:phase4_flux_breakdown_NBI}
\end{figure}

After successfully reproducing the plasma profiles, the next step was to verify whether the turbulent properties derived from the simulations aligned with the observed experimental trends.

\subsection{\label{sec:results_expt_comparison}Comparison with Selected Experimental Trends}

It is important to note that the simulated turbulence trends were examined only at the end of each phase of the workflow. Therefore, the following results did not influence any of the simulations, especially for the final phase where the profile boundaries were adjusted. This approach ensured an unbiased check for the validation process to assess whether the combined results from the \texttt{GENE} and \texttt{KNOSOS} simulations could reproduce the turbulence trends observed in the experimental data. \\

\begin{figure}
    \centering
    \includegraphics[width=0.85\linewidth]{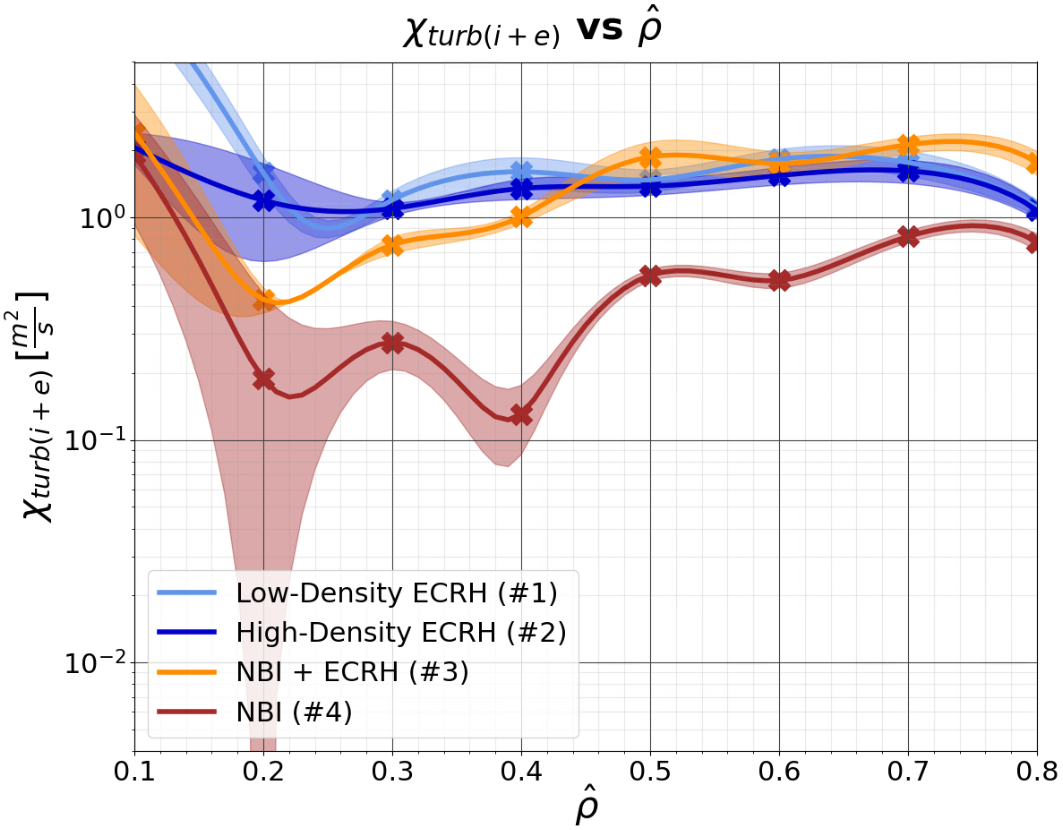}
    \caption{Radial profile of the total turbulent diffusivity $\chi$. The two ECRH and the NBI + ECRH cases have comparable $\chi$, while the NBI case has the lowest $\chi$ among all four scenarios.}
    \label{fig:carralero_plot1_combined}
\end{figure}

Three figures from Refs. \citenum{Carralero2021} and \citenum{Carralero2022} were chosen to be replicated using simulation results. First, in Fig. \ref{fig:carralero_plot1_combined}, the radial profiles of the total turbulent diffusivities $\chi_{(i + e)}$ for the four cases are illustrated. Good qualitative agreement with the experimental trend (Fig. 3b of Ref. \citenum{Carralero2022}) is evident. The two ECRH cases, the light and dark blue curves, indeed have roughly the same $\chi_{(i + e)}$ throughout the covered radial range. The NBI + ECRH scenario, the orange curve, has a cross-over point at about $\hat{\rho} = 0.4$, where its $\chi_{(i + e)}$ exceeds that of the ECRH cases. Finally, the NBI scenario, the red curve, exhibits the lowest value among the four. \\

\begin{figure}
    \centering
    \includegraphics[width=0.85\linewidth]{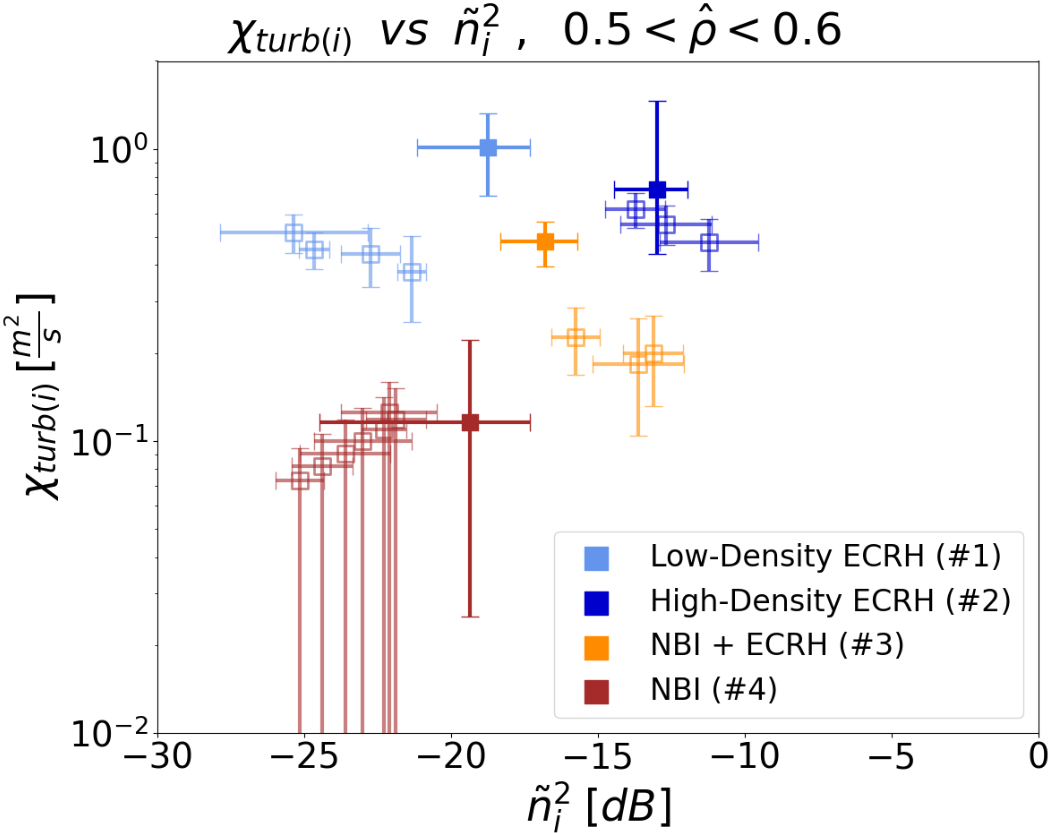}
    \caption{Dependence of $\chi_i$ on the squared ion density fluctuations $\tilde{n}^2_i$.\cite{Carralero2022} The hollow translucent points are the experimental data, which are also shown in Fig. 3c of Ref. \citenum{Carralero2022}. The $\nu$ and $\eta_i$ branches describe two turbulence trends that were successfully recovered by the simulations.}
    \label{fig:carralero_plot2_combined}
\end{figure}

\begin{figure}
    \centering
    \includegraphics[width=0.85\linewidth]{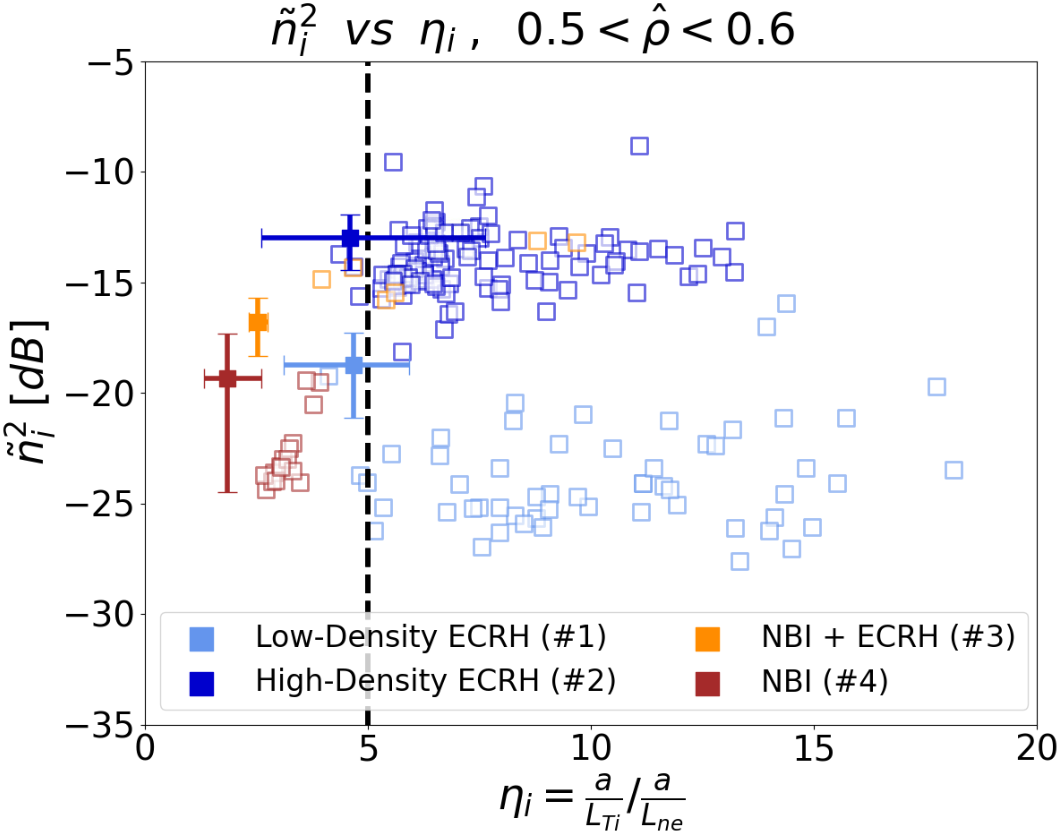}
    \caption{Relationship between the squared ion density fluctuations $\tilde{n}^2_i$ and $\eta_i$.\cite{Carralero2021} Good qualitative agreement is observed between the experiment and simulations. The hollow translucent points are the experimental data, which are also shown in Fig. 13 of Ref. \citenum{Carralero2021}. The quantitative discrepancy for the $\tilde{n}^2_i$ separation of the ECRH cases will be further studied.}
    \label{fig:carralero_plot3_combined}
\end{figure}

Fig. \ref{fig:carralero_plot2_combined} further builds upon this observation regarding the NBI scenario by comparing the ion turbulent diffusivity $\chi_i$ with the squared ion density fluctuations $\tilde{n}^2_i$. During the W7-X discharges, a Doppler reflectometry diagnostic measures the backscattered power at the cut-off layer, which is approximately proportional to the square of the density fluctuations at $0.5 \le \hat{\rho} \le 0.75$. \cite{Gusakov2004, Blanco2008} Spatial scales comparable to the ion gyroradius $\rho_i$, particularly $k_\perp \rho_i \simeq 1$ where $k_\perp$ is the mode wavenumber in the binormal direction, are probed. \cite{Carralero2022} To make the comparison with the experimental data equivalent, a wavenumber filter was applied to the $\tilde{n}^2_i$ values from the simulations. In Fig. \ref{fig:carralero_plot2_combined}, the data points for the four scenarios cluster within specific regions of the $\chi_i$ – $\tilde{n}^2_i$ phase space. Two specific branches for the collision frequency $\nu$ and for the parameter $\eta_i = L_n / L_{T_i}$, where $L_n$ and $L_{T_i}$ are the local gradient scales of $n_e$ and $T_i$ respectively, are highlighted. The parameter $\eta_i$ is used to gauge the onset of the ITG instability, the predominant form of turbulence in these scenarios. \cite{Horton1999, Riemann2016} First, in decreasing the $\nu$ value that is characteristic of the high-density ECRH case, we arrive at the low-density ECRH case. Notably, the turbulent transport level, as measured here with $\chi_i$, stays roughly the same. On the other hand, the $\eta_i$ branch illustrates that $\chi_i$ decreases as $\tilde{n}^2_i$ drops for the high-density ECRH, NBI + ECRH, and the NBI scenarios. \\

The simulations recovered these trends qualitatively. First, the results reinforce that varying $\nu$ between the two ECRH cases can have a negligible impact on turbulent transport, despite having very different $\tilde{n}^2_i$. The flux-tube simulations recovered a 5-db difference approximately between the $\tilde{n}^2_i$ magnitude of the two ECRH cases. This 5-dB disparity is a result of the differences in the plasma profiles of the two cases, which consequently gave way to a difference in the gyro-Bohm heat fluxes. While both scenarios have a heat flux of about 5 MW at $0.4 \le \hat{\rho} \le 0.6$, the ion gyro-Bohm heat flux of the high-density ECRH case is about thrice of the low-density ECRH scenario's. Since the ion-scale density fluctuations are directly proportional to the ion gyro-Bohm heat flux, this difference primarily explains the simulated 5-dB difference. \\

While this $\tilde{n}^2_i$ trend is present qualitatively, there is some quantitative disagreement. The experiments showed a 10-dB difference between the $\tilde{n}^2_i$ of the two ECRH cases, which is equivalent to an order of magnitude. Within the context of the experiments, the leading theory for this is that the total fluctuation level actually stayed the same between the low- and high-density ECRH scenarios, but they only moved to a different location away from the Doppler reflectometry measurement zone. \cite{Proll2013,Carralero2021,Carralero2022} This resulted in the reduction of the measured fluctuation amplitude. This theory is difficult to confirm here due to the limitations of the flux-tube approach and can only be understood with simulations of higher fidelity. Full flux-surface simulations are necessary to confirm this, and this will be done in a separate study using \texttt{GENE-3D}. \\

The $\eta_i$ branch reconstructed from the high-density ECRH, NBI + ECRH, and the NBI simulations mostly match well with the experimental $\chi_i$ values within the error bars. This branch is shown once again in Fig. \ref{fig:carralero_plot3_combined}, where $\tilde{n}^2_i$ is plotted against $\eta_i$. The simulations confirmed the experimental trend, where a reduction in $\eta_i$ decreased the fluctuation amplitude arising from turbulence. Increasingly steeper density gradients for these three cases led to progressively higher degrees of ITG turbulence stabilization.

\section{\label{sec:conclusions_outlook}Conclusions and Outlook}

In the present study, we validated the \texttt{GENE-KNOSOS-Tango} simulation framework against four W7-X experimental scenarios. A workflow was adapted to divide the validation study into four phases. In the first phase, only \texttt{GENE} and \texttt{Tango} were iterating at each step and the plasma density profile was kept fixed. Kinetic-electron ion-scale and adiabatic-ion electron-scale flux-tube simulations were each performed for eight radial positions to predict the plasma temperature profiles, given the heating power to the ions and electrons. In the second phase, neoclassical heat fluxes were included in the simulation loop through \texttt{KNOSOS}. In the third phase, the neoclassical $E_r \times B$ shear was calculated during each iteration and used as an additional input to \texttt{GENE}. Finally, in the fourth phase, the density profile was allowed to be varied. \\

In general, a good match between the experimental and simulated plasma profiles could be achieved. For the first three phases of the workflow, where the density profiles were kept fixed, the simulated temperature profiles remained within the experimental error bars while capturing the expected trends for the electron-scale heat flux contribution and neoclassical heat fluxes. The results of each phase also revealed some important results. The heat flux variation over several flux tubes on the same surface was moderate, and this variation could not explain notable differences between experimental and simulated profiles for the considered cases. Second, electron-scale turbulence can be important and needs to be taken into account to match experimental profiles, especially for electron-heated plasmas. Next, the inclusion of neoclassical $E_r \times B$ shear decreases plasma turbulence, especially on the ion scale. \\

For the fourth phase, where the density profiles were allowed to evolve, the importance of boundary conditions became more pronounced. While using boundary conditions from experimental data curve fits yielded a satisfactory match of the simulated profiles with the profile data for the two ECRH cases, the two NBI cases required adjustments in the boundary density and temperature values within the experimental error bars to allow the simulated profiles to align with the data. Another important parameter is the edge neutrals density $n_{0,edge}$, which dictates the particle source term via recycling and ionization. The work shows that more accurate measurements of the density profiles and its gradient and of the neutral species density near the plasma boundary are necessary for a more thorough assessment of the particle transport model. Furthermore, this shows that proper modeling of the edge conditions is necessary for predicting plasma profiles across the entire radial domain. \\

Lastly, the simulations qualitatively captured the experimental trends in turbulence characteristics, such as heat diffusivities $\chi$ and squared ion density fluctuations $\tilde{n}^2_i$. For instance, the simulated $\tilde{n}^2_i$ for the high-density ECRH, NBI + ECRH and NBI scenarios followed a decreasing trend as $\eta_i$ was reduced due to the stabilization of ITG turbulence. Though there are quantitative differences, the qualitative reproduction of experimentally observed trends using gyrokinetic simulations is already a promising result and an equally important aspect of the validation study.  \\

Looking ahead, we identify several potential directions for this study to pursue. First, we will perform radially global \texttt{GENE-3D} simulations with the converged \texttt{GENE} plasma profiles to check the impact of global effects. In this separate study, the kinetic-electron ion-scale simulations will be performed instead with \texttt{GENE-3D}. This study will also feature standalone full flux-surface simulations to delve deeper into the low- and high-density ECRH scenarios. Moreover, we will use synthetic diagnostics to study other turbulence properties, such as temperature fluctuations, derived from the global simulations to make the analysis more in-depth and the validation more robust. Next, performing multi-scale simulations for these four cases is a possible path to take, to determine if our findings regarding the persistence of electron-scale heat fluxes remains true and if there are interactions between the ion and electron scales. Finally, we will continue the validation study for other discharges beyond the parameter space covered here, such as the high-performance pellet fueling and high-beta scenarios.

\begin{acknowledgments}
This work has been carried out within the framework of the EUROfusion Consortium, funded by the European Union via the Euratom Research and Training Programme (Grant Agreement No 101052200 — EURO-fusion) and by the Spanish Ministry of Science, Innovation and Universities under grant PID2021-125607NB-I00. We acknowledge the EuroHPC Joint Undertaking for awarding this project access to the EuroHPC supercomputer LUMI, hosted by CSC (Finland) and the LUMI consortium through a EuroHPC Regular Access call. Views and opinions expressed are however those of the author(s) only and do not necessarily reflect those of the European Union or the European Commission. Neither the European Union nor the European Commission can be held responsible for them. Numerical simulations were performed at the Raven HPC system at the Max Planck Computing and Data Facility (MPCDF), Germany, the Marconi 100 \& Leonardo Fusion supercomputer at CINECA, Italy, and the LUMI supercomputer at the CSC data center, Finland.
\end{acknowledgments}

\section*{AUTHOR DECLARATIONS}
\subsection*{Conflict of Interest}
The authors have no conflicts to disclose.

\section*{Data Availability Statement}

The data that support the findings of this study are available from
the corresponding author upon reasonable request.

\section*{References}
\bibliography{aipsamp}

\end{document}